\documentclass[prb,aps,amssymb,twocolumn,superscriptaddress,notitlepage]{revtex4-1}
\usepackage{amsmath}
\usepackage{amssymb}
\usepackage{amsthm}
\usepackage{amsfonts}
\usepackage{listings}
\usepackage{enumerate}
\usepackage{latexsym}
\usepackage{graphicx}
\usepackage[caption=false]{subfig}
\usepackage{blkarray}
\usepackage{array}
\usepackage{color}
\usepackage[normalem]{ulem}
\usepackage{hyperref}
\hypersetup{colorlinks,allcolors=black} 
\usepackage{longtable} 

\newcolumntype{L}{>{$}l<{$}}

\newtheorem{defn}{Definition}
\newtheorem*{defn*}{Definition}

\newtheorem{cor}{Corollary}

\newtheorem{theorem}{Theorem}

\begin{document}
\title{Topology invisible to eigenvalues in obstructed atomic insulators}

\author{Jennifer Cano}
\affiliation{Department of Physics and Astronomy, Stony Brook University, Stony Brook, New York 11974, USA}
\affiliation{Center for Computational Quantum Physics, Flatiron Institute, New York, New York 10010, USA}
\author{L. Elcoro}
\affiliation{Department of Condensed Matter Physics, University of the Basque Country UPV/EHU, Apartado 644, 48080 Bilbao, Spain}
\author{M.~I.~Aroyo}
\affiliation{Department of Condensed Matter Physics, University of the Basque Country UPV/EHU, Apartado 644, 48080 Bilbao, Spain}
\author{B. Andrei Bernevig}
\affiliation{Department of Physics, Princeton University, Princeton, New Jersey 08544, USA}
\author{Barry Bradlyn}
\affiliation{Department of Physics and Institute for Condensed Matter Theory, University of Illinois at Urbana-Champaign, Urbana, IL, 61801-3080, USA}

\date{\today}
\begin{abstract}
We consider the extent to which symmetry eigenvalues reveal the topological character of bands.
Specifically, we compare \emph{distinct} atomic limit phases (band representations) that share the same irreducible representations (irreps) at all points in the Brillouin zone
and, therefore, appear equivalent in a classification based on eigenvalues.
We derive examples where such ``irrep-equivalent'' phases can be distinguished by a quantized Berry phase or generalization thereof.
These examples constitute a generalization of the Su-Schrieffer-Heeger chain: neither phase is topological, in the sense that localized Wannier functions exist, yet there is a topological obstruction between them.
We refer to two phases as ``Berry obstructed atomic limits'' if they have the same irreps, but differ by Berry phases. This is a distinct notion from eigenvalue obstructed atomic limits, which differ in their symmetry irreps at some point in the Brillouin zone.
We compute exhaustive lists of elementary band representations that are irrep-equivalent, in all space groups, with and without time-reversal symmetry and spin-orbit coupling, and use group theory to derive a set of necessary conditions for irrep-equivalence.
Finally, we conjecture, and in some cases prove, that irrep-equivalent elementary band representations that are not equivalent can be distinguished by a topological invariant.
\end{abstract}
\maketitle

\section{Introduction}

Topological band theory has revealed a subtle interplay between symmetry and topology.
Crystal symmetries can both identify and protect topological phases of matter.\cite{Fu2007,Teo08,Fu2010,Sato2010,Mong2010,Fu2011,Fang2012,Hsieh2012,Tanaka2012,Dziawa2012,Xu2012,Chiu2013,Morimoto2013,Jadaun2013,Fang2013,Slager2013,Hsieh2014,Liu2014,Shiozaki2014,Fang2014,Benalcazar2014,Yang2014,Koshino2014,Chiu2014,Bradlyn2016,Cano2017,Hourglass,Ma2017,Wieder17,Langbehn2017,Trifunovic2017,Benalcazar2017a,Benalcazar2017b,khalaf,Schindler2018,schindler2018higher,Khalaf2018,Geier2018,Trifunovic2019,Cano2019,robredo2019higher,elcoro2020magnetic,klemenz2020systematic,fang2020higher,Wieder2020,fang2021filling,velury2021topological}
However, the inherent challenge in identifying topological phases with crystal symmetry is that 
a different classification is needed for topological phases in each of the 230 space groups.
This challenge has only recently been overcome, through the theory of topological quantum chemistry\cite{NaturePaper,EBRTheory,GroupTheoryPaper,GraphTheoryPaper,GraphDataPaper,cano2020band,wieder2021topological} and, concurrently, the introduction of symmetry-based indicators.\cite{Kruthoff2016,Po2017,song2017,Ono2019}
Both theories make use of the symmetry of bands at high-symmetry points in the Brillouin zone (BZ) to identify topological phases.
This paradigm has been successful in predicting topological materials.\cite{vergniory2019complete,zhang2019catalogue}
In addition, it has led to the discovery of entirely new phases, such as higher order topological phases\cite{Benalcazar2017a,Benalcazar2017b,khalaf,Schindler2018,schindler2018higher} and fragile topological phases.\cite{PoFragile,NewTIs,Slager2018,hexagons,Liu2019,elcoro2020application}

However, not all topological phases can be determined by their irreducible representations (irreps) at high-symmetry points.
For example, a Chern insulator or a time-reversal invariant $\mathbb{Z}_2$ topological insulator can exist 
without any symmetry or with only translation symmetry.
In both cases, symmetry indicators that distinguish the topological from trivial phase do not exist.
It is not so surprising that these topological phases can be invisible to symmetry indicators, 
since they are not protected by crystal symmetry.
What is more surprising is that topological phases protected by crystal symmetry cannot always be distinguished by their symmetry irreps; mirror Chern insulators\cite{Teo08} and rotation anomaly insulators\cite{fang2019new} are examples. 
Thus, while symmetry indicators are a powerful tool to identify topological bands, they render certain topological phases invisible.

The theory of topological quantum chemistry\cite{NaturePaper,EBRTheory,GroupTheoryPaper,GraphTheoryPaper,GraphDataPaper,cano2020band} is based on band representations\cite{Zak1980,Zak1981}, and thus incorporates information about Bloch states beyond just symmetry indicators.
Band representations exactly span the space of topologically trivial (atomic limit) phases. Therefore, topological bands are those that do not transform as band representations.
Since distinct band representations can have the same symmetry indicators, band representations refine the classification of symmetry-indicated topological phases.
This refinement has given rise to the discovery of fragile topological phases, which are in the trivial class of symmetry indicators, but can be detected via band representations.\cite{PoFragile,NewTIs,Slager2018,hexagons,Liu2019,elcoro2020application}

In this work, we further refine the classification of symmetry indicators by
comparing pairs of bands that have 
the same irreps at every high-symmetry point in the BZ; we refer to such a pair as irrep-equivalent.
Irrep-equivalent bands exhibit Bloch wave functions that transform the same way as each other under symmetry at each point in the BZ.
However, as was pointed out in Refs.~\onlinecite{Bacry1988b,Michel1992}, it is possible that two bands that transform identically under all symmetries at each point in the BZ differ by a topologically non-trivial global gauge transformation, thus rendering them distinct.
These bands do not need to be topological: in two and three dimensions, distinct trivial phases can be irrep-equivalent, but distinguished by topological invariants. 
(In one dimension, the only crystal symmetry operation is inversion, which completely distinguishes distinct phases.)
This is exactly the study of the current work: we will show that two distinct but irrep-equivalent band representations can be distinguished by topological invariants, despite both being trivial band insulators.
We will further classify all such irrep-equivalent band representations.

We start in Sec.~\ref{sec:notation} with a self-contained review of band representations.
In Sec.~\ref{sec:f222}, we review the earliest example of two irrep-equivalent atomic limit phases, in space group $F222$ (No. 22).\cite{Bacry1988b,Michel1992,Bacry1993}
We derive a Berry phase invariant to prove that the two phases are topologically distinct.
We then introduce in Sec.~\ref{sec:P2} a second example of irrep-equivalent atomic limit phases with time-reversal (TR) symmetry and spin-orbit coupling (SOC) in space group $P112$ (No. 3), for which we also derive a Berry phase invariant that distinguishes the phases.

In Sec.~\ref{sec:irrep-equiv}, using the results of topological quantum chemistry,
we enumerate \textit{all} of the irrep-equivalent elementary band representations (EBRs) and use group theory to derive general conditions that explain the tables.
This builds on earlier work by Bacry, Michel, and Zak (BMZ)\cite{Bacry1988}; importantly, our analysis reveals one set of cases missed by BMZ.
In addition, our tables are the first list of all of the irrep-equivalent atomic limit phases with TR and/or SOC.
Finally, we conjecture that all irrep-equivalent -- but distinct -- atomic limit phases differ by a topological invariant derived from Berry phases.
The problem of finding an invariant that distinguishes them in each case remains outstanding.


\section{Review of Band Representations}
\label{sec:notation}

A set of orbitals from atoms residing at specific positions in a particular space group defines a band representation in direct (real) space, i.e., an atomic limit. Fourier transforming the band representation completely determines the irreps that appear at each point in the BZ, independently of energetics.
The concept of a band representation
was introduced by Zak\cite{Zak1980,Zak1981} to understand how ``$\mathbf{k}\cdot \mathbf{p}$'' representations at different points in the BZ connect to each other.
A modern interpretation and extension of Zak's theory was introduced in the theory of topological quantum chemistry\cite{NaturePaper,EBRTheory,GroupTheoryPaper,GraphTheoryPaper,GraphDataPaper,cano2020band}.
In order to make the present work self-contained, we review the notation for band representations established in Ref.~\onlinecite{EBRTheory}, emphasizing the parts of the theory most relevant for this work.

Let $\mathbf{q}$ be one particular site (which we sometimes label by the Wyckoff position to which it belongs) in the lattice of a crystal invariant under the symmetries of a space group, $G$.
The site-symmetry group, $G_\mathbf{q}$, consists of the symmetry operations in $G$ that leave the site $\mathbf{q}$ invariant:
\begin{equation}
G_\mathbf{q} \equiv \{ g \in G | g\mathbf{q} = \mathbf{q} \}
\end{equation}
Since $G_\mathbf{q}$ is a subgroup of $G$, one can choose a set of coset representatives, $g_\alpha$, for $G_\mathbf{q}$ in $G$:
\begin{equation}
G=\bigcup_{\alpha=1}^{n}g_\alpha \left(G_\mathbf{q}\ltimes\mathbb{Z}^3\right),
\label{eq:coset}
\end{equation}
where $\ltimes$ denotes the semi-direct product, $g_1$ is the identity and $g_{\alpha\neq 1} \notin G_\mathbf{q}$.
Each coset representative defines a site related by symmetry to $\mathbf{q}$:
\begin{equation}
\mathbf{q}_\beta \equiv g_\beta \mathbf{q} ,
\end{equation}
such that the site-symmetry group of $\mathbf{q}_\beta$ is conjugate to that of $\mathbf{q}$: $G_{\mathbf{q}_\beta} = g_\beta G_\mathbf{q} g_\beta^{-1}$.
The coset decomposition in Eq.~(\ref{eq:coset}) effectively maps each space group element to an element in the site-symmetry group. Specifically, for each $h\in G$ and each $g_\alpha$, there is a unique coset representative $g_\beta$ and site-symmetry group element $g\in G_\mathbf{q}$ that satisfies
\begin{equation}
hg_\alpha =\{E|\mathbf{t}_{\beta\alpha}(h)\}g_{\beta}g,
\label{eq:defbeta} 
\end{equation}
where
\begin{equation}
\mathbf{t}_{\beta\alpha}(h) \equiv h\mathbf{q}_\alpha - \mathbf{q}_\beta
\label{eq:deftbetaalpha}
\end{equation}
is a lattice translation.
We will use Eq.~(\ref{eq:defbeta}) to build a representation of $G$ given a representation of $G_\mathbf{q}$, which amounts to determining the symmetry of a band from the symmetry of an orbital.

Let $\rho$ be a representation of $G_\mathbf{q}$.
Following Ref.~\onlinecite{EBRTheory}, $\rho$ induces a band representation\cite{Zak1980} of $G$, denoted $\rho \uparrow G$ or $\rho_G$.
According to Eq (5) of Ref.~\onlinecite{EBRTheory}, the matrix form of $\rho_G(h)$ consists of infinitely many blocks, labelled by pairs $(\mathbf{k}',\mathbf{k})$, where $\mathbf{k}'$ is a row index and $\mathbf{k}$ is a column index.
For each symmetry element, $h=\{R|\mathbf{v}\} \in G$ (the notation denotes a point group operation, $R$, followed by a translation, $\mathbf{v}$), and each set of columns corresponding to $\mathbf{k}$, there is exactly one non-zero block, which corresponds to $\mathbf{k}' = R\mathbf{k}$.
We denote this block by $\rho_G^\mathbf{k}(h)$ (as in Ref.~\onlinecite{NaturePaper}, although there the block structure was not explicitly emphasized); 
its matrix elements are given by,
\begin{equation}
\rho_G^\mathbf{k}(h)_{j\beta, i\alpha} = 
e^{-i(R\mathbf{k})\cdot\mathbf{t}_{\beta\alpha}(h)} \tilde{\rho}_{ji}( g_{\beta}^{-1}\{E|-\mathbf{t}_{\beta\alpha}(h)\}hg_\alpha),
\label{eq:blocks}
\end{equation}
where we have defined
\begin{equation}
\tilde{\rho}_{ij}(a) = \begin{cases} \rho_{ij}(a) & a\in G_\mathbf{q} \\
0 & a\notin G_\mathbf{q} 
\end{cases}
\label{eq:tilderho}
\end{equation}
Eq.~(\ref{eq:blocks}) warrants some unpacking. The left-hand-side (LHS) of Eq.~(\ref{eq:blocks}) is a matrix that
specifies how wave functions at $h\mathbf{k}$ are related to those at $\mathbf{k}$.
The subscripts on the LHS run over the bands at $\mathbf{k}$. 
The symmetry of wave functions at $\mathbf{k}$ is determined by the representation $\rho$ in direct space (real space), which is the content of the right-hand-side (RHS) of Eq.~(\ref{eq:blocks}).
The first term on the RHS is a $\mathbf{k}$-dependent phase determined by positions of atoms in the unit cell.
The second term is equal to an element of $\rho$ when $h$ can be mapped to an element of $G_\mathbf{q}$ using 
Eq.~(\ref{eq:defbeta}): specifically, this term is non-zero if and only if $g_\beta^{-1} h g_\alpha$ is equal to an element of $G_\mathbf{q}$ up to a translation determined by $\mathbf{t}_{\beta\alpha}(h)$.

The little group, $G_\mathbf{k}$, of a point, $\mathbf{k}$, in the BZ, is the set of symmetry operations whose rotational part leaves the point $\mathbf{k}$ invariant modulo a reciprocal lattice vector:
\begin{equation}
G_\mathbf{k} \equiv \{ g=\{R|\mathbf{v} \} \in G | R\mathbf{k} \equiv \mathbf{k} \},
\label{eq:deflittlegroup}
\end{equation}
where the equivalence relation $R\mathbf{k} \equiv \mathbf{k}$ is defined by equivalence up to a reciprocal lattice vector, i.e. $\mathbf{k} \equiv\mathbf{k+K}$ if and only if $\mathbf{K}$ is a reciprocal lattice vector.
When Eq~(\ref{eq:blocks}) is restricted to elements in $G_\mathbf{k}$, it furnishes a representation of $G_\mathbf{k}$, which we denote $\rho_G \downarrow G_\mathbf{k}$.
We define the characters of $\rho_G \downarrow G_\mathbf{k}$:
\begin{align}
\chi_G^{\mathbf{k}}(h) &\equiv  \sum_{i,\alpha} \rho_G^{\mathbf{k}}(h)_{i\alpha,i\alpha}
= \sum_{i,\alpha}  e^{-i\mathbf{k}\cdot \mathbf{t}_{\alpha\alpha}(h)}\tilde{\rho}_{ii}(h_\alpha) \nonumber\\
&= \sum_{\alpha} e^{-i\mathbf{k}\cdot \mathbf{t}_{\alpha\alpha}(h)}\tilde{\chi}(h_\alpha),
\label{eq:chars}
\end{align}
where we have defined
\begin{equation}
\tilde{\chi}(h) =  \sum_i \tilde{\rho}(h)_{ii}
\label{eq:tildechi}
\end{equation}
and defined the shorthand,
\begin{equation}
h_\alpha \equiv g_\alpha^{-1} \{ E|-\mathbf{t}_{\alpha\alpha}(h)\} h g_\alpha
\label{eq:halpha}
\end{equation}
If two band reps share the same little group irreps, (i.e., yield the same characters $\chi_G^\mathbf{k}$ in Eq~(\ref{eq:chars}) for all $\mathbf{k}$), then we refer to them as irrep-equivalent.

Equivalent band representations are necessarily irrep-equivalent, where equivalence is defined as follows (Def. 5 of Ref.~\onlinecite{EBRTheory}):
two band representations $\rho_G$ and $\sigma_G$ are equivalent iff there exists a unitary matrix-valued function $S(\mathbf{k},\tau,g)$ smooth in $\mathbf{k}$ and continuous in $\tau$ such that for all $g\in G$, $\tau \in [0,1]$, 
$S(\mathbf{k},\tau,g)$ is a band representation and
\begin{equation}
S(\mathbf{k},0,g) = \rho_G^\mathbf{k}(g) \text{ and } S(\mathbf{k},1,g) = \sigma_G^\mathbf{k}(g)
\label{eq:defequivalence}
\end{equation}
An EBR is a band representation that is not equivalent to a direct sum of other band representations.

We will now refer to the definition of equivalence in Eq.~(\ref{eq:defequivalence}) as \textbf{homotopic equivalence}, to distinguish it from irrep-equivalence.
As pointed out in Ref.~\onlinecite{EBRTheory}, homotopic equivalence implies irrep-equivalence, since the deformation provided by $S$ does not change the characters of the little group irreps.
However, the reverse is not true.\cite{NaturePaper,Bacry1988b,Bacry1993,Michel1992}
The purpose of this manuscript is to enumerate irrep-equivalent EBRs, study how to distinguish them by examples,  and present a general conjecture.

One of the key tools that we will use is the Wilson loop,\cite{ArisCohomology,Soluyanov2011,Wieder17,Zakphase,Fu06,Ryu10,Yu11,Taherinejad14,Alexandradinata14,Hourglass,Benalcazar2017a,Holler2017,aris18-2} which allows for a non-Abelian generalization of the Berry phase.
Given a Hamiltonian where the (cell-periodic parts of the) Bloch wave functions are denoted by $|u_i(\mathbf{k})\rangle$, we define the Wilson loop matrix of a set of bands, $\mathcal{B}$, over a closed path, $l$, in the BZ by:
\begin{equation}
	W_l \equiv \mathcal{P}e^{i\int_l dl \cdot \mathbf{A}(\mathbf{k}) },
	\label{eq:wilson}
\end{equation}
where $\left[ \mathbf{A}(\mathbf{k})\right]_{ij} \equiv i\langle u_i(\mathbf{k}) | \nabla_\mathbf{k}| u_j(\mathbf{k}) \rangle$, $i,j\in \mathcal{B}$, is the Berry connection, and $\mathcal{P}$ denotes that the exponential is path-ordered.
Eq.~(\ref{eq:wilson}) is well-defined as long as the bands in $\mathcal{B}$ do not touch any bands  in the complement of $\mathcal{B}$.
When the path $l$ is defined by a reciprocal lattice vector, $\mathbf{K}$, such that $l = x\mathbf{K}$, $0\leq x \leq 1$, we will call $W_l$ the $\mathbf{K}$-directed Wilson loop and denote it $W_\mathbf{K}$.

As we will show, sometimes symmetry forces the Wilson loop eigenvalues of the bands transforming as an EBR to be quantized, even without specifying the Hamiltonian.
In particular, if $\mathcal{B}$ includes all of the bands in the Hamiltonian,
it has been proven\cite{ArisCohomology} that the eigenvalues of $W_\mathbf{K}$ are given by $e^{i\mathbf{K} \cdot \mathbf{r}_i}$,
where $\mathbf{r}_i$ is the real space position of the $i^{\rm th}$ degree of freedom in the Hamiltonian.
(This connection between the Wilson loop eigenvalue, which is a Berry phase, and the position of charge in real space, illustrates the ``Modern Theory of Polarization.''\cite{ksv,ksv2,RestaReview})

More generally, we will give examples where the eigenvalues of $W_\mathbf{K}$ remain quantized even when $\mathcal{B}$ is a subset of bands in the Hamiltonian.
Examples of this phenomenon exist for topological crystalline insulators protected by inversion symmetry\cite{Alexandradinata14,Hughes2011inversion},
but, in that case, bands with distinct Wilson loop eigenvalues necessarily exhibit distinct inversion eigenvalues at high-symmetry points, making them irrep-\emph{in}equivalent. Similarly, EBRs with distinct rotational symmetry eigenvalues at high symmetry points can also have quantized Wilson loop eigenvalues\cite{Holler2017}.
In contrast, the examples that we show here distinguish bands that transform as irrep-equivalent EBRs.
Consequently, the Wilson loop eigenvalues along particular high-symmetry lines serve to distinguish the EBRs in cases where the symmetry eigenvalues cannot.


\section{Example: $F222$. EBRs that are irrep-equivalent but not equivalent}
\label{sec:f222}

As an example of irrep-equivalent EBRs that are not equivalent, we show that in space group $F222$, for each EBR induced from the $4a$ position at $(0,0,0)$, there is an irrep-equivalent EBR induced from the $4b$ position at $(0,0,\frac{1}{2})$.
This example was studied in earlier works;\cite{Bacry1988b,Michel1992,Bacry1993,Zeinerf2221,Zeinerf2222} here we reprove existing results in modern language, establish a more general Berry phase invariant, and introduce a ``generalized obstructed atomic limit'' that cannot be distinguished by little group irreps.

\begin{figure}
\centering
\includegraphics[width=2.5in]{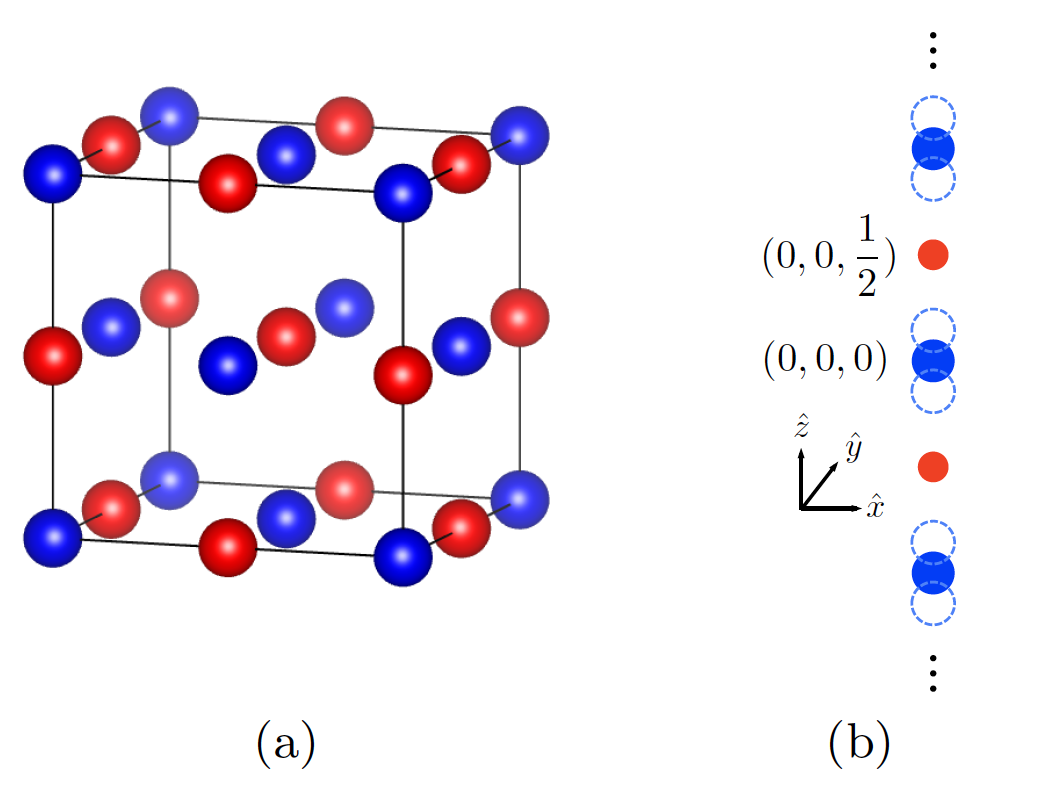}
\caption{(a) The black lines outline the conventional unit cell of $F222$ (No. 22).
The blue atoms are in the $4a$ $(0,0,0)$ position, while the red atoms are at the $4b$ $(0,0,1/2)$ position.
(b) The $(0,0,z)$ (and $(1/2,1/2,z)$) lines each separately implement the Rice-Mele chain, where $C_{2,100}$ or $C_{2,010}$ play the role of the inversion symmetry operation in 1D.
Since the $4a$ Wyckoff position has no free parameter, it is impossible to continuously deform
a single Wannier function centered on a blue lattice site to be centered on a red lattice site while preserving $C_{2,100}$ symmetry, or vice versa.
However, two Wannier functions both centered on a blue lattice site could be deformed to be centered on red lattice sites by moving pairwise, as indicated by the blue dashed circles. (Similarly, two Wannier functions both centered on a red lattice site could be deformed to be centered on blue lattice sites by moving pairwise).
Figure reproduced from Ref.~\onlinecite{EBRTheory}.}
\label{fig:F222-lattice}
\end{figure}

The space group  $F222$ contains the symmetry operations $\{ C_{2,100} | \mathbf{0} \}$, $\{ C_{2,010} | \mathbf{0} \}$, $\{ C_{2,001} | \mathbf{0} \}$ (point group operations are written in Sch\"onflies notation, following Ref.~\onlinecite{PointGroupTables}) and the face-centered lattice translations; we denote the primitive lattice vectors:
\begin{equation}
\mathbf{t}_1 = \frac{1}{2}(0,b,c), \,
\mathbf{t}_2 = \frac{1}{2}(a,0,c), \,
\mathbf{t}_3 = \frac{1}{2}(a,b,0) 
\label{eq:vectors222}
\end{equation}
The reciprocal lattice vectors are given by:
\begin{equation}
\mathbf{g}_1= 2\pi(-\frac{1}{a},\frac{1}{b},\frac{1}{c}), \, \mathbf{g}_2 = 2\pi(\frac{1}{a},-\frac{1}{b},\frac{1}{c}), \, \mathbf{g}_3 =2\pi(\frac{1}{a},\frac{1}{b},-\frac{1}{c})
\label{eq:reciprocalvectors222}
\end{equation}
which are related by the rotations as follows:
\begin{align}
C_{2,100}:& \,\,\,\, \mathbf{g}_3 \leftrightarrow \mathbf{g}_2, \,\, \mathbf{g}_1 \leftrightarrow -(\mathbf{g}_1 + \mathbf{g}_2 + \mathbf{g}_3) \label{eq:symmf222-x}\\
C_{2,010}:& \,\,\,\, \mathbf{g}_3 \leftrightarrow \mathbf{g}_1, \,\, \mathbf{g}_2 \leftrightarrow  -(\mathbf{g}_1 + \mathbf{g}_2 + \mathbf{g}_3) \label{eq:symmf222-y}\\
C_{2,001}:& \,\,\,\, \mathbf{g}_2 \leftrightarrow \mathbf{g}_1, \,\, \mathbf{g}_3 \leftrightarrow  -(\mathbf{g}_1 + \mathbf{g}_2 + \mathbf{g}_3)
\label{eq:symmf222-z}
\end{align}
We are interested in band representations induced from irreps on the sites  (in conventional coordinates) $\mathbf{q} = (0,0,0)$ and $\mathbf{q}' = (0,0,1/2)$, shown in Fig.~\ref{fig:F222-lattice}, which correspond to the $4a$ and $4b$ Wyckoff positions, respectively.
(The Wyckoff multiplicity of 4 indicates that there are four sites in the conventional unit cell, although there is only one site in the primitive unit cell.)
The site-symmetry group $G_\mathbf{q}$ is generated by $\{ C_{2,110} | \mathbf{0} \}, \{ C_{2,010} | \mathbf{0} \} $ and $\{ C_{2,001} | \mathbf{0} \}$, while the site-symmetry group $G_{\mathbf{q}'}$ is generated by $\{ C_{2,100} | \mathbf{t}_z\}, \{C_{2,010} | \mathbf{t}_z\} $ and $\{ C_{2,001} | \mathbf{0} \}$, where 
\begin{equation} \mathbf{t}_z = \mathbf{t}_1 + \mathbf{t}_2 - \mathbf{t}_3 
\label{eq:deftz}
\end{equation}
is an integer linear combination of the primitive lattice vectors defined in Eq.~(\ref{eq:vectors222}). 
$G_\mathbf{q}$ and $G_{\mathbf{q}'}$ are isomorphic to $D_2$, whose character table is in Table~\ref{table:d2chars}.

\begin{table}
\begin{tabular}{c|cccc}
$\rho$ & $[E]$ & $[C_{2,001}]$ & $[C_{2,010}]$ & $[C_{2,100}]$ \\
\hline
$A$ & $1$ & $1$ & $1$ & $1$ \\
$B_1$ & $1$ & $1$ &$-1$ &$-1$ \\
$B_2$ & $1$ & $-1$ &$1$ &$-1$ \\
$B_3$ & $1$ & $-1$ &$-1$ &$1$ 
\end{tabular}
\caption{Character table for the irreducible representations of the point group $D_2$.}\label{table:d2chars}
\end{table}

In $F222$, each element in the space group can be written as $\{ E|\mathbf{t} \}g$, where $g\in G_\mathbf{q}$ and $\mathbf{t}$ is a lattice vector.
Thus, a single-valued representation $\rho$ of $G_\mathbf{q}$ induces an elementary band representation, $\rho_G$, given by Eq.~(\ref{eq:blocks}):
\begin{equation}
\rho_G^\mathbf{k}(\{ E | \mathbf{t} \} g) = e^{-i(g\mathbf{k})\cdot \mathbf{t} }\rho(g),
\label{eq:bandrep222-4a}
\end{equation}
where the indices $i,j,\alpha,\beta$ are absent because all single-valued representations of $G_\mathbf{q}$ are one-dimensional (the group is Abelian) and there is only one site in the primitive unit cell corresponding to the $4a$ position.

We now consider a band representation induced from a representation of $G_{\mathbf{q}'}$.
In this case, a generic space group element can be written as $\{ E|\mathbf{t} \} g = \{ E|\mathbf{t}'\} g'$, where $g' \in G_{\mathbf{q}'}$ and $\mathbf{t}'$ is a lattice vector.
Specifically, if $g=\{ C_{2,100}|\mathbf{0}\}$ or $\{ C_{2,010} | \mathbf{0}\}$, then $\mathbf{t}' = \mathbf{t} - \mathbf{t}_z$, while if $g=\{ C_{2,001} | \mathbf{0} \}$, then $\mathbf{t}' = \mathbf{t}$.
Let $\rho'$ be a representation of the site-symmetry group $G_{\mathbf{q}'}$, defined by 
\begin{align}
\rho'( \{ C_{2,100(010)} | \mathbf{t}_z \}) &= \rho( \{ C_{2,100(010)}|\mathbf{0} \} ), \nonumber\\
\rho'( \{ C_{2,001} | \mathbf{0} \}) &= \rho ( \{ C_{2,001} | \mathbf{0} \})
\label{eq:defrho222}
\end{align}
Then the induced band representation is given by:
\begin{equation}
(\rho')_G^\mathbf{k} (\{ E | \mathbf{t} \} g) = e^{-i(g\mathbf{k})\cdot \mathbf{t}' }\rho(g) 
\label{eq:bandrep222-4b}
\end{equation}
We showed in Appendix D of Ref.~\onlinecite{EBRTheory} that the EBRs defined by Eqs.~(\ref{eq:bandrep222-4a}) and (\ref{eq:bandrep222-4b}) for the same choice of $\rho$ are irrep-equivalent because $e^{-i(g\mathbf{k})\cdot \mathbf{t}} = e^{-i(g\mathbf{k})\cdot \mathbf{t}'}$ when $\mathbf{k}$ is a high-symmetry point.
We now show that despite being irrep-equivalent, the two EBRs are not related by a gauge transformation that respects the periodicity of the BZ and hence are not homotopically equivalent.\cite{Bacry1988b}
(See Appendix~\ref{sec:equivalence} for a proof that homotopic equivalence implies the existence of a BZ-periodic gauge transformation.)

Since the band representation $\rho_G^\mathbf{k}(h)$ acting on a state at $\mathbf{k}$ yields a state at $h\mathbf{k}$,  $\rho_G^\mathbf{k}(h)$ transforms under a gauge transformation $M_\mathbf{k}$ as $\rho_G^\mathbf{k}(h) \rightarrow M_{h\mathbf{k}}^\dagger \rho_G^\mathbf{k}(h) M_\mathbf{k}$.
The band representations in Eqs.~(\ref{eq:bandrep222-4a}) and (\ref{eq:bandrep222-4b}) are related by the gauge transformation:
\begin{equation}
(\rho')_G^\mathbf{k}(h) =  M_{h\mathbf{k}}^\dagger  \rho_G^\mathbf{k}(h) M_{\mathbf{k}},
\label{eq:gauge222}
\end{equation}
where $M_{(k_x,k_y,k_z)} = e^{-ik_z/2}$.
Since there is only one band, it is clear that $M_\mathbf{k}$ is the unique gauge transformation that relates $\rho_G^\mathbf{k}$ and $(\rho')_G^\mathbf{k}$.
However, since $M_\mathbf{k}$ does not respect the periodicity of the BZ, we conclude that the band representations $\rho_G^\mathbf{k}$ and $(\rho')_G^\mathbf{k}$ are not related by any BZ-periodic gauge transformation and hence are distinct EBRs.

We can further show that the EBRs defined by Eqs.~(\ref{eq:bandrep222-4a}) and (\ref{eq:bandrep222-4b}) are distinct by comparing the values of particular Berry phases computed within each EBR.\cite{Michel1992,EBRTheory}
Following Eq.~(\ref{eq:wilson}),
let $W_\mathbf{K}$ denote the Berry phase acquired when the wavefunction is transported from the origin, $\Gamma$, to $\mathbf{K}$, where $\mathbf{K}$ is any reciprocal lattice vector; to be concrete, we take the path from $\Gamma$ to $\mathbf{K}$ to be the path of shortest length between the two points.
When the Hilbert space includes only a single orbital on either $\mathbf{q}$ or $\mathbf{q}'$, $W_{\mathbf{g}_j} = e^{i\mathbf{g}_j\cdot \mathbf{q}}=1$ or $e^{i\mathbf{g}_j \cdot \mathbf{q}'}= -1$, respectively.\cite{ArisCohomology} 
Physically, this phase corresponds to the polarization along the $\mathbf{g}_j$ direction relative to the center of the unit cell.
Note that changing the unit cell center would change the polarization, but not the relative difference between the two polarizations.

Generically, the Hilbert space of a real material includes more than a single orbital.
Combined with the fact that there is no symmetry that transforms $\mathbf{g}_i \rightarrow -\mathbf{g}_i$, the Berry phase, $W_{\mathbf{g}_i}$, will cease to be quantized (as noted for this example in Refs.~\onlinecite{Zeinerf2221,Zeinerf2222}).
Thus, we are motivated to develop an invariant that goes beyond the Berry phase studied in Ref.~\onlinecite{Michel1992}.

\begin{figure}[t]
\centering
\includegraphics[width=1.5in]{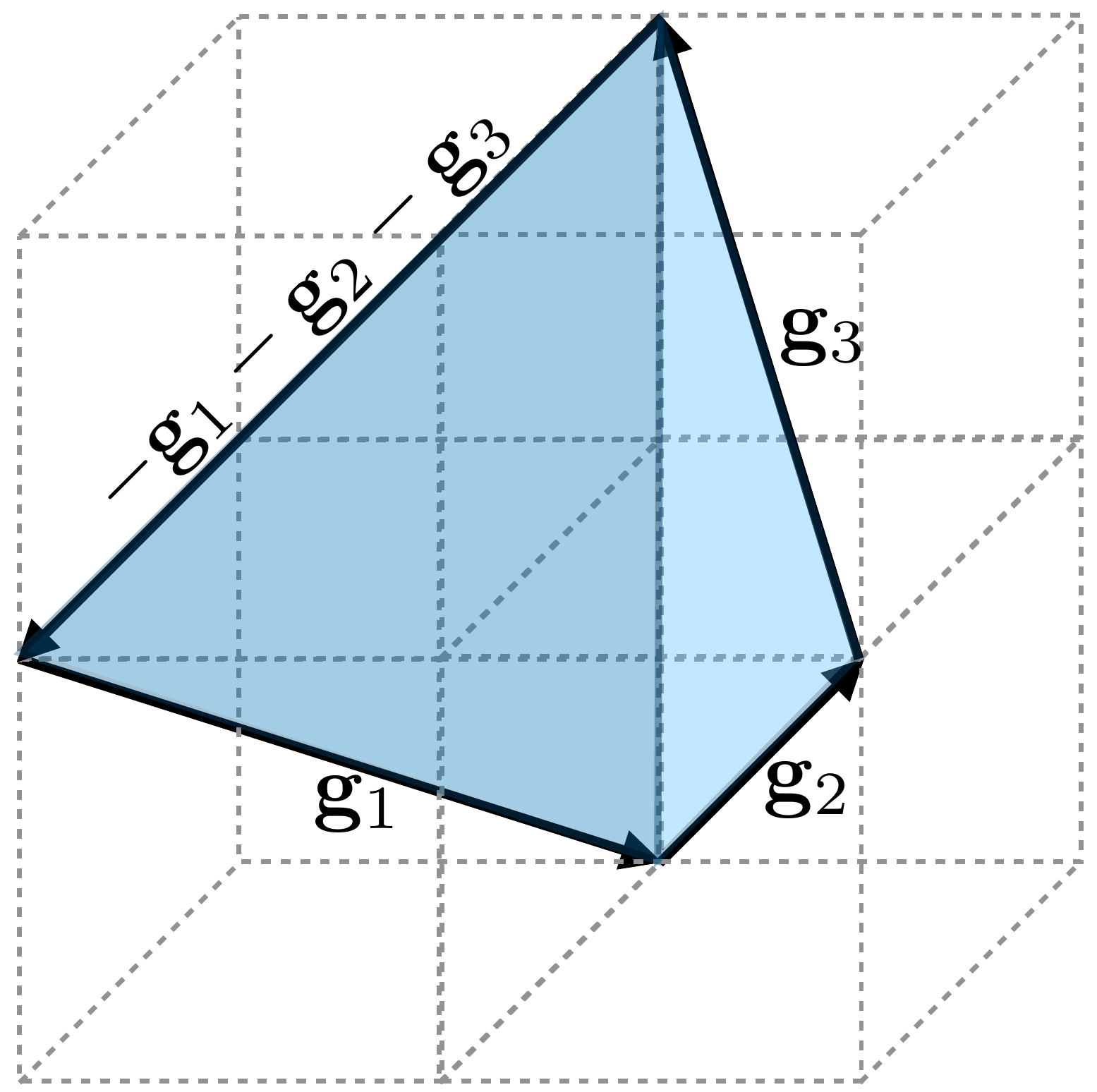}
\caption{Reciprocal lattice vectors defining the topological invariant in Eq.~(\ref{eq:22-Berry}). Light gray dotted lines outline cubes in the BZ with sides of length $2\pi$. The vectors $\mathbf{g}_1$, $\mathbf{g}_2$, $\mathbf{g}_3$ and $-\mathbf{g}_1-\mathbf{g}_2-\mathbf{g}_3$ outline a loop, $\ell$. The surface $\Sigma$ that appears in Eq.~(\ref{eq:222-berry}) can be any surface whose boundary is $\ell$; one such surface consists of the two blue shaded triangles.}
\label{fig:F222invariant}
\end{figure}

We derive an invariant that will distinguish the two EBRs in Eqs.~(\ref{eq:bandrep222-4a}) and (\ref{eq:bandrep222-4b}) but does not rely on them comprising the entire Hilbert space.
We utilize the action of the crystal symmetry operations on the reciprocal lattice vectors (Eqs.~(\ref{eq:symmf222-x}) -- (\ref{eq:symmf222-z})), which enforces:
\begin{equation}
W_{\mathbf{g}_1} = W_{\mathbf{g}_2} = W_{\mathbf{g}_3} = W^{-1}_{\mathbf{g}_1+\mathbf{g}_2+\mathbf{g}_3} 
\label{eq:222-symm-Wilson}
\end{equation}
Let $\ell$ be the loop traced by putting the vectors $\mathbf{g}_i$ and $-\mathbf{g}_1 - \mathbf{g}_2 - \mathbf{g}_3 $ end-to-end (see Fig.~\ref{fig:F222invariant}).
By Stoke's theorem, the Berry phase acquired upon traversing $\ell$ is equal to the flux of Berry curvature through any surface, $\Sigma$, whose boundary is $\ell$:
\begin{equation}
W_{\mathbf{g}_1} W_{\mathbf{g}_2} W_{\mathbf{g}_3} W^{-1}_{\mathbf{g}_1+\mathbf{g}_2+\mathbf{g}_3}  = e^{i \int_\Sigma \Omega \cdot d\Sigma},
\label{eq:222-berry}
\end{equation}
where $\Omega = \nabla \times \mathbf{A}$ is the Berry curvature and $\Sigma$ is a region whose boundary is $\ell$; an example is shown in Fig.~\ref{fig:F222invariant}.
(Note that Eq.~(\ref{eq:222-berry}) does not hold in general for non-Abelian Wilson loops, but is valid for a one-dimensional band representation.)
Combining Eqs.~(\ref{eq:222-symm-Wilson}) and (\ref{eq:222-berry}) yields a topological invariant, $n$, defined mod 4, by:
\begin{equation}
e^{\frac{2\pi i n}{4}} = W_{\mathbf{g}_1}e^{-\frac{i}{4}\int_\Sigma \Omega \cdot d\Sigma}
\label{eq:22-Berry}
\end{equation}
Since bands that transform as an EBR induced from a representation of $G_{\mathbf{q}}$ 
can be continuously deformed to have $\Omega =0$ and $W_{\mathbf{g}_1} = 1$, these bands must have $n=0$ in Eq.~(\ref{eq:22-Berry}), 
while bands that transform as an EBR induced from a representation of $G_{\mathbf{q}'}$ 
can be continuously deformed to have $\Omega = 0$ and $W_{\mathbf{g}_1} = -1$ and hence have $n=2$ in Eq.~(\ref{eq:22-Berry}).
Thus, the two EBRs can be distinguished by knowledge of the Berry curvature, $\Omega$, and the Berry phase $W_{\mathbf{g}_1}$ by computing the RHS of Eq.~(\ref{eq:22-Berry}) for a given band to obtain $n\in \mathbb{Z}_4$ on the LHS, despite the fact that $W_{\mathbf{g}_1}$ itself is not quantized.
Thus, Eq.~(\ref{eq:22-Berry}) serves to distinguish the two irrep-equivalent EBRs $\rho_G$ and $\rho'_G$, for any irrep $\rho$ of $G_{\mathbf{q}}$ and corresponding irrep $\rho'$ of $G_{\mathbf{q'}}$ defined by Eq.~(\ref{eq:defrho222}).
We further show in the Supplemental Material\cite{SMF222-cd} that the values of $n=1$ and $n=3$ on the LHS of Eq.~(\ref{eq:22-Berry}) distinguish the irrep-equivalent EBRs induced from the $4c$ position $(\frac{1}{4}, \frac{1}{4}, \frac{1}{4})$ and $4d$ position $(\frac{1}{4},\frac{1}{4},\frac{3}{4})$, expressed in conventional basis coordinates.
(Note: EBRs induced from the $4c$ and $4d$ positions are not irrep-equivalent to those induced from the $4a$ and $4b$ positions).
Thus, Eq.~(\ref{eq:22-Berry}) serves to distinguish all pairs of irrep-equivalent single-valued EBRs in $F222$.

One physical consequence of this irrep-equivalence is that it allows for obstructed atomic limit\cite{NaturePaper} (also called frozen-polarization\cite{Po2017,turner-frozenref}) phases in $F222$ that cannot be diagnosed by their little group representations; we refer to these as Berry obstructed atomic limits. 
The most well-known example of an obstructed atomic limit phase is the Su-Schrieffer-Heeger\cite{ssh1979} (SSH) or Rice-Mele\cite{RiceMele} chain protected by inversion symmetry: the two phases of the SSH model each correspond to an atomic limit phase -- that is, exponentially localized and symmetry-preserving Wannier functions exist -- but the two phases are separated by a gap-closing phase transition.
Hence, there is an obstruction to continuously deforming a band in one phase to a band in the other phase.
However, in the SSH model, the two phases can be distinguished by the product of their inversion eigenvalues at $\Gamma$ and $X$, as well as by their polarization.\cite{Alexandradinata14}
What is unique about the Berry obstructed atomic limit phases in $F222$ that we will shortly present is that they cannot be distinguished by either their symmetry eigenvalues or their polarizations.

To study the transition between the two Berry obstructed atomic limit phases in $F222$, we utilize the intermediate site $\mathbf{q}'' = (0,0,z)$ (in conventional coordinates), with $0<z<\frac{1}{2}$, that interpolates between $\mathbf{q}$ and $\mathbf{q}'$. (A similar construction for the SSH or Rice-Mele chain is shown in the Supplemental Material.\cite{SMRiceMele}) 
Since $G_{\mathbf{q}''} \subset (G_\mathbf{q} \cap G_{\mathbf{q}'})$, $\mathbf{q}''$ is part of the non-maximal $8g$ Wyckoff position.
The site-symmetry group $G_{\mathbf{q}''}$ is generated by $\{ C_{2,001} | \mathbf{0} \}$ and thus isomorphic to $C_2$; the character table for $C_2$ is shown in Table~\ref{table:c2chars}.

\begin{table}[t]
\begin{tabular}{c|cccc}
$\rho$ & $[E]$ & $[C_{2}]$ & $[^dE]$ & $[^dC_{2}]$  \\
\hline
$A$ & $1$ & $1$ & $1$ & $1$ \\
$B$ & $1$ & $-1$ & $1$ & $-1$ \\
$^1\bar{E}$ & $1$ & $-i$ & $-1$ & $i$\\
$^2\bar{E}$ & $1$ & $i$ & $-1$ & $-i$
\end{tabular}
\caption{Character table for the double group $^d C_2$. 
The superscript $d$ indicates elements that arise due to the double cover of $SO(3)$ by $SU(2)$ \cite{GroupTheoryPaper}.
$A$ and $B$ are single-valued irreps, while $^1\bar{E}$ and $^2\bar{E}$ are double-valued (spinor) irreps.}\label{table:c2chars}
\end{table}

Consider the trivial representation of $G_{\mathbf{q}''}$, denoted $A_{\mathbf{q}''}$; the subscript indicates the site.
This irrep induces representations of $G_{\mathbf{q}}$ and $G_{\mathbf{q}'}$ with the same labels, i.e.,
\begin{align}
A_{\mathbf{q}''} \uparrow G_{\mathbf{q}} &= A_{\mathbf{q}} \oplus B_{1,\mathbf{q}},\nonumber\\
A_{\mathbf{q}''} \uparrow G_{\mathbf{q}'} &= A_{\mathbf{q}'} \oplus B_{1,\mathbf{q}'},
\label{eq:22-induce8g}
\end{align}
where the irreps on the right-hand sides of Eq.~(\ref{eq:22-induce8g}) are defined in Table~\ref{table:d2chars}.
Eq.~(\ref{eq:22-induce8g}) shows that bands derived from $A_{\mathbf{q}''}$ orbitals at the $8g$ position  will transform as either a sum of EBRs induced from the $4a$ position or a sum of EBRs induced from the $4b$ position; the two sums of EBRs on the right-hand sides of Eq.~(\ref{eq:22-induce8g}) are homotopically equivalent in the sense of Eq.~(\ref{eq:defequivalence}) and the $8g$ position furnishes an equivalence.
Consequently, given a system whose valence band transforms as the EBR $A_\mathbf{q} \uparrow G$ and whose conduction band transforms as the EBR $B_{1,\mathbf{q}} \uparrow G$, it is possible to drive a quadratic gap-closing phase transition at a high-symmetry point such that when the gap re-opens, the valence band transforms as the EBR $A_{\mathbf{q}'} \uparrow G$  and the conduction band transforms as $B_{1,\mathbf{q}'} \uparrow G$ (the quadratic dispersion is required by symmetry).
The gap \textit{must} close during the transition because the valence band in each phase can be distinguished by the quantized invariant $n\in \mathbb{Z}_4$ defined in Eq.~(\ref{eq:22-Berry}), despite being indistinguishable by their little group irreps.
We construct such a model and comment on the phase transition in the Supplemental Material.\cite{SMf222}

This construction is analogous to the obstructed atomic limit transition in the Rice-Mele model.\cite{SMRiceMele}
The $A_\mathbf{q} \oplus B_{1,\mathbf{q}}$ representation describes $s$ and $p_z$ orbitals at the $4a$ position, which, in the 3D crystal, can be considered as an array of Rice-Mele chains oriented in the $\hat{\mathbf{z}}$ direction, and stacked according to the face-centered lattice vectors of Eq.~(\ref{eq:vectors222}) (see Fig.~\ref{fig:F222-lattice}). 
The (spinless) $C_{2,100}$ and $C_{2,010}$ symmetry operations act on these chains exactly as the inversion symmetry operation acts in the Rice-Mele model, while $C_{2,001}$ acts as the identity. 
Thus, our Berry obstructed atomic limit phase transition corresponds to the polarization transition in the Rice-Mele model. 
However, unlike the Rice-Mele chain, our Berry obstructed atomic limit cannot be diagnosed by a single Berry phase because, as discussed above, the Berry phase $W_{\mathbf{g}_i}$ need not be quantized.

Notice that the polarization along the $\mathbf{z}$-direction is quantized because the $\mathbf{z}$-axis maps into itself under the rotation $C_{2,100}$, but this polarization cannot distinguish the two phases.  This is because in $F222$, $\hat{\mathbf{z}}$ is neither a direct nor reciprocal primitive lattice vector. To compute the polarization in the $\mathbf{z}$-direction, we must integrate the Berry connection from $\mathbf{k}=0$ to $\mathbf{k}=\mathbf{g}_1+\mathbf{g}_2 = 4\pi  c(0,0,1).$\cite{ksv}A Wannier center at $\mathbf{q}'$ thus yields the Berry phase $W_{\mathbf{g}_1+\mathbf{g}_2} = e^{i\mathbf{q}' \cdot (\mathbf{g}_1+\mathbf{g}_2)} = 1,$ corresponding to a polarization of $e=0\mod e$, which is indistinguishable from the polarization corresponding to a Wannier center at $\mathbf{q}$.

\section{Example: $P112$. Berry phase distinguishing irrep-equivalent EBRs with time-reversal symmetry}
\label{sec:P2}

\begin{figure}
\centering
\includegraphics[height=.8in]{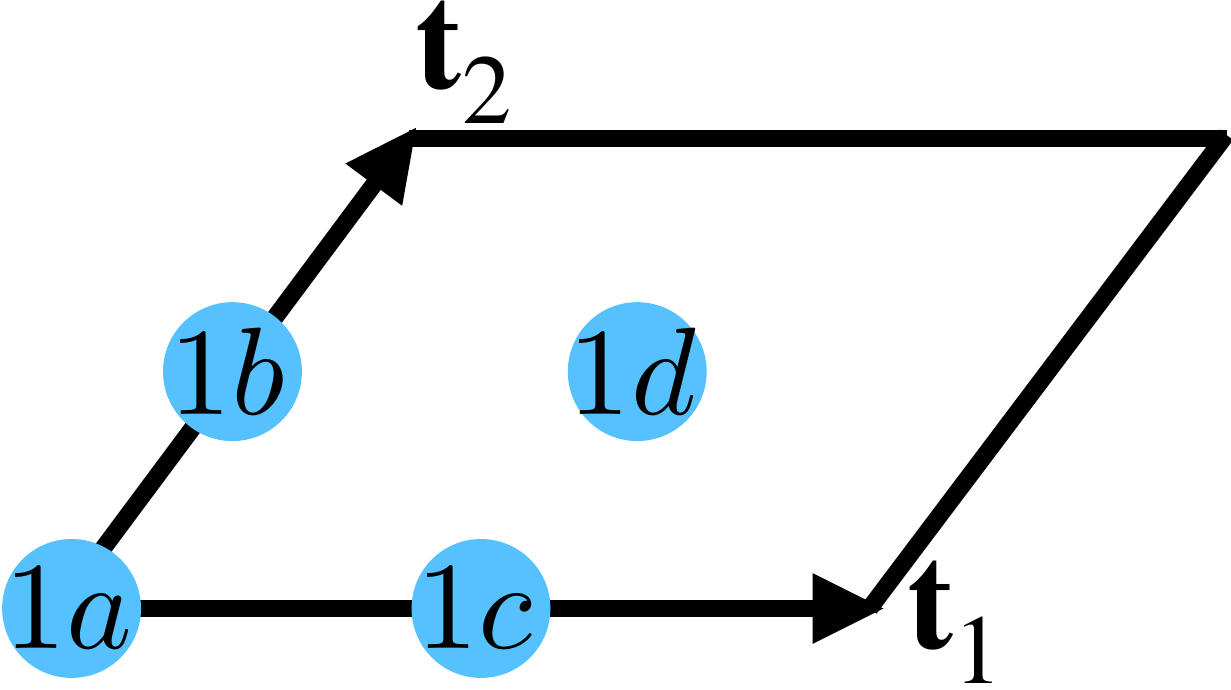}
\caption{Unit cell, lattice vectors, and maximal Wyckoff positions in $P112$ in the $z=0$ plane.}
\label{fig:P2-Wyckoff}
\end{figure}

We consider a second example of irrep-equivalent -- but distinct -- EBRs in space group $P112$ ($P2$, No. 3), which is generated by $\{ C_{2,001} | \mathbf{0} \}$ and primitive lattice translations. 
There are four maximal Wyckoff positions:
\begin{align}
&\mathbf{q}_a = \left(0,0,z \right),\,\, \mathbf{q}_b = \left(0,\frac{1}{2},z \right), \nonumber\\
&\mathbf{q}_c = \left(\frac{1}{2},0,z\right), \mathbf{q}_d = \left(\frac{1}{2},\frac{1}{2},z\right) ,
\label{eq:P2-Wyckoff}
\end{align}
shown in Fig.~\ref{fig:P2-Wyckoff}.
The site-symmetry group, $G_{\mathbf{q}_i}$, of each $\mathbf{q}_i$ is generated by a $C_{2,001}$ rotation about an axis that goes through $\mathbf{q}_i$. Hence, $G_{\mathbf{q}_i}$ is isomorphic to the point group $C_2$, which has two single-valued and two double-valued irreps, as shown in Table~\ref{table:c2chars}.
(The parameter $z$ in Eq.~(\ref{eq:P2-Wyckoff}) is free because the site-symmetry group $G_{\mathbf{q}_i}$ is independent of $z$.)
The EBRs induced from these irreps are all distinguishable by their irreps at high-symmetry points, as can be verified using the BANDREP tool on the Bilbao Crystallographic Server (BCS).\cite{NaturePaper}

When TR is present, the EBRs induced from the single-valued irreps remain distinguishable by their irreps at high-symmetry points, because TR does not impose extra constraints on the high-symmetry points.
However, TR is important for the \textit{double}-valued irreps and their induced EBRs because it enforces a Kramers degeneracy.
Specifically, since the pair of complex conjugate irreps irreps $^1\bar{E}$ and $^2\bar{E}$ of $^d C_2$ (the superscript $d$ indicates the double group; see Table~\ref{table:c2chars}) are exchanged under TR, the point group generated by $^d C_{2}$ and TR has a unique double-valued irrep, which we denote $^1\bar{E}^2\bar{E}$.
In real space, then, $G_{\mathbf{q}_i}$ has a unique double-valued irrep, which we denote $\rho_i$.
In reciprocal space (momentum space), the four EBRs $\rho_{i} \uparrow G$ are irrep-equivalent because the little group of each high-symmetry point, which is generated by $C_{2,001}$ and TR, has only one irrep, $^1\bar{E}^2\bar{E}$.

We now show that despite being irrep-equivalent, each of the four EBRs $\rho_{i} \uparrow G$ is distinguishable by a combination of two Berry phases.
Specifically, we will consider the eigenvalues of $W_{\mathbf{g}_1}$ and $W_{\mathbf{g}_2}$, 
which are the $\mathbf{g}_1$- and $\mathbf{g}_2$-directed Wilson loop matrices defined in Eq.~(\ref{eq:wilson}) by transporting the wave functions along the path $k\mathbf{g}_{1}$ or $k\mathbf{g}_2$ as $k$ goes from $0$ to $2\pi$. (The reciprocal lattice vectors are defined by the usual relation $\mathbf{t}_i \cdot \mathbf{g}_j = 2\pi \delta_{ij}$, where $\mathbf{t}_{1,2}$ are shown in Fig.~\ref{fig:P2-Wyckoff} and $\mathbf{t}_3$ is the lattice vector in the $\mathbf{z}$ direction.)
Since each EBR consists of two bands with time-reversal symmetry, $W_{\mathbf{g}_j}$ is a $2\times 2$ matrix.
Following the discussion below Eq.~(\ref{eq:wilson}),
when the Hilbert space includes only the two orbitals transforming as $\rho_i$, the two eigenvalues of $W_{\mathbf{g}_j}$ are both $e^{i\mathbf{g}_j \cdot \mathbf{q}_i}$. As shown in Table~\ref{tab:P2Wilson}, the combination of eigenvalues of $W_{\mathbf{g}_1}$ and $W_{\mathbf{g}_2}$ uniquely determines the EBR.

\begin{table}
\begin{tabular}{c|cc}
Site & $W_{\mathbf{g}_1}$ eig. & $W_{\mathbf{g}_2}$ eig. \\
\hline
$\mathbf{q}_a$ & $1$ & $1$ \\
$\mathbf{q}_b$ & $1$ & $-1$ \\
$\mathbf{q}_c$ & $-1$ & $1$ \\
$\mathbf{q}_d$ & $-1$ & $-1$
\end{tabular}
\caption{Wilson loop eigenvalues of the EBRs induced from $\rho_i$ on site $\mathbf{q}_i$ with time-reversal symmetry in space group $P112$. The second(third) column lists the eigenvalue of $W_{\mathbf{g}_1}(W_{\mathbf{g}_2})$. As explained in the text, the two eigenvalues of $W_{\mathbf{g}_1}(W_{\mathbf{g}_2)}$ are degenerate; hence, only one number is listed in each column even though the Wilson loop matrices are $2\times 2$.
The combination of the eigenvalues of $W_{\mathbf{g}_1}$ and $W_{\mathbf{g}_2}$ uniquely determine the EBR.}
\label{tab:P2Wilson}
\end{table}

However, as we discussed in Sec.~\ref{sec:f222}, the Hilbert space generically includes other orbitals.
We now prove that, unlike in the example in Sec.~\ref{sec:f222}, the Wilson loop eigenvalues remain fixed at the values in Table~\ref{tab:P2Wilson} even in the presence of other orbitals.
The proof is as follows: TR requires the two Wilson loop eigenvalues to be degenerate,\cite{ArisCohomology} which forces $W_{\mathbf{g}_{i}} = e^{i\theta_{i}} \mathbb{I}$, where $\theta_{i}$ is real and $\mathbb{I}$ indicates the $2\times 2$ identity matrix.
Since $\{ C_{2,001} | \mathbf{0} \}$ reverses the orientation of the Wilson loop, it forces $W_{\mathbf{g}_{i}}$ to also be particle-hole symmetric\cite{ArisCohomology}, which requires $\theta_{i} \in \{ 0,\pi \} $.
Thus, if other bands are introduced into the Hilbert space, the Wilson loop eigenvalues in Table~\ref{tab:P2Wilson} remain fixed to $\pm 1$, and the four EBRs remain distinguishable.

Notice that the same argument applies in space group $P11m$ ($Pm$, No. 6), which is generated by a mirror reflection through the $z=0$ plane and translations.
In this group, there are only two maximal Wyckoff positions, $1a$ $(x,y,0)$ and $1b$ $(x,y,\frac{1}{2})$.
When spin-orbit coupling is included and time-reversal symmetry is enforced, there is a unique double-valued irrep of each site-symmetry group and the EBRs induced from different sites are irrep-equivalent.
Since there are only two maximal Wyckoff positions, the irrep-equivalent EBRs can be distinguished by a single Wilson loop, $W_{\mathbf{g}_3}$, defined in Eq.~(\ref{eq:wilson}) by transporting the wavefunctions along the path $(0,0,k)$ as $k$ goes from $0$ to $2\pi$, because the combination of mirror and TR forces $W_{\mathbf{g}_3}=\pm \mathbb{I}$.

As in Sec.~\ref{sec:f222}, one can construct an obstructed atomic limit transition between two phases, where the valence bands in the two phases are irrep-equivalent, but distinguishable by the combination of $W_{\mathbf{g}_{1,2}}$ eigenvalues.


\section{Irrep-equivalent EBRs}
\label{sec:irrep-equiv}

We now generalize the examples in Secs.~\ref{sec:f222} and \ref{sec:P2} to other space groups.
Specifically, we answer the question: when are the EBRs induced from representations of the site-symmetry groups $G_\mathbf{q}$ and $G_{\mathbf{q}'}$ irrep-equivalent?
The case of single-valued representations without time-reversal symmetry was considered by Bacry, Michel, and Zak (BMZ)\cite{Bacry1988}.
In this paper, we present a complete answer to this question for EBRs with and without TR and SOC by a computational search; the results are listed in Appendix~\ref{sec:irrep-equiv-tables}.
In addition to expanding to include TR and SOC, our results also reveal two cases missed by BMZ,\cite{Bacry1988} which we analyze in Sec.~\ref{sec:differences}.

To better understand our computational results, in this section we use group theory to derive a set of necessary (but not sufficient) conditions for irrep-equivalence.
For simplicity, we apply them to the case without TR or SOC to see how the conditions limit the space groups that can contain irrep-equivalent EBRs.

\subsection{Deriving irrep-equivalence from characters of the site-symmetry group}
\label{sec:irrep-equiv-chars}

Although irrep-equivalence is defined by the representations at each high-symmetry point, it is elegant and useful to consider the full band representation, $\rho_G$, whose blocks $\rho_G^\mathbf{k}$ we explicitly constructed in Eq.~(\ref{eq:blocks}) for all $\mathbf{k}$ in the BZ.
To utilize this formalism, we prove in Appendix~\ref{app:unitary}:
\begin{theorem}
Two band representations, $\rho_G$ and $\rho'_G$, of a space group, $G$, are irrep-equivalent if and only if they are related by a unitary transformation, $U$, such that:
\begin{equation}
U \left[ \rho_G(g) \right] U^\dagger =  \rho'_G(g) \text{ for all }g\in G
\end{equation} \label{th:unitary}
\vspace{-15pt}
\end{theorem}
\noindent 
Theorem~\ref{th:unitary} eliminates the need to consider individual $\mathbf{k}$ points; instead, we consider the unitary equivalence of entire band representations at once. 
Further, Theorem~\ref{th:unitary} does not place any constraints on the $\mathbf{k}$-dependence of $U$ such as continuity or BZ periodicity.
Thus, irrep-equivalence is weaker than the homotopic definition of equivalence in Eq.~(\ref{eq:defequivalence}), which requires that the unitary matrix be smooth and periodic in $\mathbf{k}$.
This is exactly the point of this work: as was illustrated by the examples in Secs.~\ref{sec:f222} and \ref{sec:P2}, band representations can be unitarily equivalent -- and thus display all the same representations at high-symmetry points -- without being homotopically equivalent. When this happens, the unitary transformation is not smooth and periodic in the BZ, as we saw in Eq.~(\ref{eq:gauge222}) where $M_\mathbf{k}$ was not BZ-periodic.

We now seek conditions for when two EBRs are unitarily equivalent.
Since EBRs are induced representations, in the language of group theory, we want to know when two induced representations are unitarily equivalent.
The question is well-defined for \textit{finite} groups using character theory (we review the representation theory of finite groups in Appendix~\ref{app:reptheory}).
However, difficulties arise due to the infinite nature of space groups (which have an infinite set of translation symmetries).
This motivates us, following BMZ\cite{Bacry1988} to define the finite ``Born von Karman'' space groups,
$G_{N} \equiv G/T_N$, where $T_N$ is the subgroup of $G$ 
generated by $\mathbf{t}_i^N$ and $\mathbf{t}_i$ are the primitive lattice vectors (notice that $T_N$ is infinite and, as a result, $G_N = G/T_N$ is finite). These are the symmetry groups of finite-sized crystals with periodic boundary conditions.

We restrict ourselves to choices of $N$ such that the high-symmetry points in the BZ of $G_N$ are identical to those of $G$;\cite{Evarestov85}
with this constraint on $N$, EBRs will be irrep-equivalent in $G$ if and only if they are also irrep-equivalent in $G_N$.
(For example, in the case of inversion symmetry in one dimension, the high-symmetry points are $k=0$ and $k=\pi$, so we require $N$ to be even.)
Thus, our search for irrep-equivalent EBRs of $G$ is identical to searching for irrep-equivalent EBRs of $G_N$.
With this understanding, we will drop the subscript $N$ and proceed to use the representation theory of finite groups, reviewed in Appendix~\ref{app:reptheory}.

Using the character theory of finite groups, we derive in Appendix~\ref{app:irrep-equiv} a necessary and sufficient condition for two induced representations to be unitarily equivalent. This result, which we will utilize extensively in the following, is expressed
in the language of band representations as:
\begin{theorem}
Given a representation of $G_\mathbf{q}$ with characters $\chi$ and a representation of $G_{\mathbf{q}'}$ with characters $\chi'$, 
the induced representations with characters $\chi\uparrow G$ and $\chi' \uparrow G$ will be irrep-equivalent if and only if, for every $g\in G_\mathbf{q} \cup G_{\mathbf{q}'}$:
\begin{equation}
\frac{1}{|G_\mathbf{q}|} \sum_{h\in G_\mathbf{q} \cap [g]_G } \chi(h) = 
\frac{1}{|G_{\mathbf{q}'}|} \sum_{h' \in G_{\mathbf{q}'} \cap [g]_G } \chi' (h'),
\label{eq:irrep-equiv}
\end{equation}
where $[g]_G \equiv \{ (g')^{-1} g g' | g'\in G \}$ denotes the conjugacy class of $g$ in $G$. 
\label{th:irrep-equiv}
\end{theorem}
\noindent
Theorem~\ref{th:irrep-equiv} provides an algorithm to make a complete list of irrep-equivalent EBRs in all space groups by evaluating Eq.~(\ref{eq:irrep-equiv}) for all pairs of sites $\mathbf{q}, \mathbf{q}'$, characters $\chi, \chi'$ and space group elements $g\in G_\mathbf{q} \cup G_{\mathbf{q}'}$.
We have compiled tables of irrep-equivalent EBRs (described in Appendix~\ref{sec:irrep-equiv-tables}) in a different way, by explicitly comparing the irreps of each EBR at high-symmetry momenta.
Thus, Theorem~(\ref{th:irrep-equiv}) serves as an independent check of the tables.

In the remainder of this manuscript, we will use Eq.~(\ref{eq:irrep-equiv}) to derive constraints on which site-symmetry groups can induce irrep-equivalent EBRs.
Ultimately, we find two main results: first, in Sec.~\ref{sec:samesite}, we derive that only seven point groups permit irrep-equivalent EBRs induced from the same Wyckoff position.
Second, in Sec.~\ref{sec:differentsite}, we show that there are only 29 pairs of point groups that permit irrep-equivalent EBRs induced from different Wyckoff positions.

\subsection{Examples}

Before deriving more general constraints, we provide two examples of how to use Theorem~\ref{th:irrep-equiv} by applying it to the irrep-equivalent EBRs studied in Secs.~\ref{sec:f222} and \ref{sec:P2}.

\subsubsection{Application of Eq.~(\ref{eq:irrep-equiv}) to $F222$}
\label{sec:F222-ex}

We showed in Sec.~\ref{sec:f222} that the EBRs $\rho \uparrow G$ and $\rho'\uparrow G$, defined in Eqs.~(\ref{eq:bandrep222-4a}) and (\ref{eq:bandrep222-4b}), are irrep-equivalent for any irrep $\rho$ of $G_{\mathbf{q}=(0,0,0)}$ and the corresponding irrep $\rho'$ of the isomorphic group $G_{\mathbf{q}'=(0,0,\frac{1}{2})}$, where $\rho'$ is defined in Eq.~(\ref{eq:defrho222}).
Recall from Sec.~\ref{sec:f222} that both $G_{\mathbf{q}}$ and $G_{\mathbf{q}'}$ contain $\{ C_{2,001} | \mathbf{0}\}$, while each other nontrivial element of $G_{\mathbf{q}}$ is in the same conjugacy class (with respect to the space group $G$)
as an element in $G_{\mathbf{q}'}$, specifically,
\begin{align}
 \{ C_{2,100} | \mathbf{t}_z \} &= \{ E | \mathbf{t}_2 \} \{ C_{2,100} | \mathbf{0} \}\{ E | \mathbf{t}_2 \}^{-1} \nonumber\\
 \{ C_{2,010} | \mathbf{t}_z \} &= \{ E | \mathbf{t}_3 \} \{ C_{2,010} | \mathbf{0} \}\{ E | \mathbf{t}_3 \}^{-1},
 \label{eq:f222conjugate}
\end{align}
where $\{ C_{2,100(010)}|\mathbf{t}_z \} \in G_{\mathbf{q}'}$, while $\{ C_{2,100(010)} \mathbf{0} \}\in G_\mathbf{q}$; the lattice vectors $\mathbf{t}_{1,2,3}$ are defined in Eq.~(\ref{eq:vectors222}); and $\mathbf{t}_z$ is defined in Eq.~(\ref{eq:deftz}).

We now show explicitly that Eq.~(\ref{eq:irrep-equiv}) is satisfied (utilizing $|G_\mathbf{q} | = |G_{\mathbf{q}'} |$ and taking $\chi$ and $\chi'$ to denote the characters of $\rho$ and $\rho'$, respectively):
\begin{description}
\item[$g=E$] In this case, $G_\mathbf{q}\cap [g]_G = G_{\mathbf{q}'} \cap [g]_G = E$. Thus, Eq.~(\ref{eq:irrep-equiv}) yields $ \chi(E) = \chi'(E) = 1$.
\item[$g=\{ C_{2,001} | \mathbf{0} \}$] In this case, 
\begin{align}
G_\mathbf{q}\cap [g]_G = G_{\mathbf{q}'} \cap [g]_G = \{ C_{2,001} | \mathbf{0} \} 
\end{align}
Eq.~(\ref{eq:irrep-equiv}) yields $\chi( \{ C_{2,001} | \mathbf{0} \} ) = \chi'( \{ C_{2,001} | \mathbf{0} \} )$, which is satisfied by the definition of $\rho'$ in Eq.~(\ref{eq:defrho222}).
\item[$g=\{ C_{2,100(010)} | \mathbf{0} \}$ or $\{ C_{2,100(010)} | \mathbf{t}_z \}$] We showed in Eq.~(\ref{eq:f222conjugate}) that $ \{ C_{2,100(010)} | \mathbf{0} \}$ and $\{ C_{2,100(010)} | \mathbf{t}_z \}$ are conjugate in $G$. Since $g$ only enters Eq.~(\ref{eq:irrep-equiv}) through $[g]_G$, the equation is the same for either choice of $g$. Since
\begin{align}
G_\mathbf{q}\cap [g]_G &= \{ C_{2,100(010)} | \mathbf{0} \} \nonumber\\
G_{\mathbf{q}'} \cap [g]_G &= \{ C_{2,100(010)} | \mathbf{t}_z \} 
\end{align}
Eq.~(\ref{eq:irrep-equiv}) yields 
$ \chi( \{ C_{2,100(010)} | \mathbf{0} \} ) = \chi'( \{ C_{2,100(010)} | \mathbf{t}_z \} ),$
which is satisfied by the definition of $\rho'$ in Eq.~(\ref{eq:defrho222}).
\end{description}
Thus, the example in Sec.~\ref{sec:f222} of irrep-equivalent EBRs in $F222$ induced from the $1a$ and $1b$ position satisfy Eq.~(\ref{eq:irrep-equiv}) -- as of course they must, since we already showed that they are irrep-equivalent.

\subsubsection{Application of Eq.~(\ref{eq:irrep-equiv}) to $P112$}
\label{sec:P2-ex}

We showed in Sec.~\ref{sec:P2} that in the double SG $P112$, the EBRs induced from the double-valued $^1\! \bar{E}\,^2\!\bar{E}$ representation (which is irreducible with respect to time reversal symmetry) of the site-symmetry groups $G_{\mathbf{q}_i}$ are irrep-equivalent for the four sites $\mathbf{q}_{i=a,b,c,d}$ defined in Eq.~(\ref{eq:P2-Wyckoff}). The site-symmetry group for each site is generated by a two-fold rotation:
\begin{align}
G_{\mathbf{q}_a} &= \langle \{ C_{2,001} | 000 \} \rangle \nonumber\\
G_{\mathbf{q}_b} &= \langle \{ C_{2,001} | 010 \} \rangle \nonumber\\
G_{\mathbf{q}_c} &= \langle \{ C_{2,001} | 100 \} \rangle \nonumber\\
G_{\mathbf{q}_d} &= \langle \{ C_{2,001} | 110\} \rangle,
\label{eq:P2-ss}
\end{align}
where the angled brackets enclose the site-symmetry group generator.

It is straight-forward to check that, unlike in the previous example in $F222$ (Sec.~\ref{sec:F222-ex}), none of the two-fold rotations that appear in Eq.~(\ref{eq:P2-ss}) are in the same conjugacy class with respect to the space group $P112$.
For example, the conjugacy class $\left[ \{ C_{2,001} | 000 \} \right]_G$ only contains elements of the form:
\begin{multline}
\{ E | n_1 n_2 0 \}^{-1}  \{ C_{2,001} | 000 \}\{ E | n_1 n_2 0\} \\ = \{ C_{2,001} | -2n_1, - 2n_2 ,0 \},
\label{eq:P2-conj-c2z}
\end{multline}
where $n_{1,2} \in \mathbb{Z}$ so that $\{ E | n_1 n_2 0 \} \in G$.
Therefore, none of the rotations in $G_{\mathbf{q}_b}$, $G_{\mathbf{q}_c}$ or $G_{\mathbf{q}_d}$, as defined in Eq.~(\ref{eq:P2-ss}), are conjugate to $\{C_{2,001} | 000 \}$, due to the factors of two on the RHS of Eq.~(\ref{eq:P2-conj-c2z}).

We now check that Eq.~(\ref{eq:irrep-equiv}) is satisfied when $\mathbf{q} = \mathbf{q}_a$ and $\mathbf{q}' = \mathbf{q}_b$:
\begin{description}
\item[$g=E$] Since $G_{\mathbf{q}_a} \cap [g]_G = G_{\mathbf{q}_b} \cap [g]_G = E$, Eq.~(\ref{eq:irrep-equiv}) yields $ \chi(E) = \chi'(E) = 2$.
\item[$g=\{C_{2,001} | 000 \}$] Eq.~(\ref{eq:P2-conj-c2z}) showed that $g$ is not conjugate to any element of $G_{\mathbf{q}_b}$. Thus, the RHS of Eq.~(\ref{eq:irrep-equiv}) is zero, which requires on the LHS $\chi(g) = 0$. The characters in Table~\ref{table:c2chars} confirm this is satisfied for the $^1\!\bar{E}\,^2\!\bar{E}$ irrep.
\item[$g=\{C_{2,001} | 010 \} $] Eq.~(\ref{eq:P2-conj-c2z}) showed that $g$ is not conjugate to any element of $G_{\mathbf{q}_a}$, so that the LHS of Eq.~(\ref{eq:irrep-equiv}) is zero. The characters in Table~\ref{table:c2chars} show that the RHS is also zero because $\chi(g)=0$ for the $^1\!\bar{E}\,^2\!\bar{E}$ irrep.
\end{description}
The same arguments hold for the other pairs of $\mathbf{q}_{a,b,c,d}$.
Thus, the irrep-equivalent EBRs in $P112$ satisfy Eq.~(\ref{eq:irrep-equiv}), as expected.

\subsection{Merging conjugacy classes}
\label{sec:theorems}

We now use Eq.~(\ref{eq:irrep-equiv}) to constrain which site-symmetry groups can induce irrep-equivalent EBRs.
It will be useful to introduce the notion of merging conjugacy classes: two distinct conjugacy classes, $[g_1]_{G_\mathbf{q}}\neq [g_2]_{G_\mathbf{q}}$, defined with respect to the site-symmetry group, $G_\mathbf{q}$, are said to \textbf{merge} in the full space group, $G$, if $[g_1]_G = [g_2]_G$.
We use the concept or merging conjugacy classes to rewrite the LHS of Eq.~(\ref{eq:irrep-equiv}) when $g\in G_\mathbf{q}$:
\begin{theorem}
Given $g\in G_\mathbf{q}$, if the conjugacy class $[g]_{G_\mathbf{q}}$ does not merge with any distinct conjugacy class $[g']_{G_\mathbf{q}}$, where $g'\in G_\mathbf{q}$, $[g']_{G_\mathbf{q}} \neq [g]_{G_\mathbf{q}}$, then the LHS of Eq.~(\ref{eq:irrep-equiv}) is given by:
\begin{equation} 
\frac{1}{|G_\mathbf{q}| }\sum_{h\in  [g]_{G_\mathbf{q}} } \chi(h) = \frac{| [g]_{G_\mathbf{q} } | }{ |G_\mathbf{q}| } \chi(g),
\label{eq:nomerge}
\end{equation}
\label{th:nomerge}
\end{theorem}
\noindent Proof: Let $g\in G_\mathbf{q}$. 
If $[g]_{G_\mathbf{q}}$ does not merge with any distinct conjugacy class of $G_\mathbf{q}$ in $G$, then $[g]_{G_\mathbf{q}}=G_\mathbf{q} \cap [g]_G $ 
(for otherwise, there exists $h\in \left( G_\mathbf{q} \cap [g]_G \right)$, such that $h\notin [g]_{G_\mathbf{q}}$, which means the conjugacy classes $[h]_{G_\mathbf{q}}$ and $[g]_{G_\mathbf{q}}$ are distinct and merge in $G$, violating the hypothesis.)
Thus, when $[g]_{G_\mathbf{q}}$ does not merge with any distinct conjugacy classes in $G$, the LHS of Eq.~(\ref{eq:irrep-equiv}) can be rewritten as the LHS of Eq.~(\ref{eq:nomerge}).
The equality in Eq.~(\ref{eq:nomerge}) follows because all elements in the same conjugacy class have the same character, which completes the proof.

We now establish two theorems about merging conjugacy classes that we will use in the following sections to show that
many site-symmetry groups cannot induce
irrep-equivalent EBRs, following BMZ\cite{Bacry1988}.

The first theorem pertains to crystallographic classes: two symmetry operations $g_1$ and $g_2$ 
are part of the same \textbf{crystallographic class} if and only if there exists a crystallographic symmetry operation $g$ such that $g_1 = g^{-1} g_2 g$.
For example, $\{ C_{2,100} | \mathbf{0} \}$ and $\{ C_{2,010}|\mathbf{0} \}$ are in the same crystallographic class because they are conjugate by $\{ C_{4,001} | \mathbf{0} \}$, but
$\{ C_{2,100} | \mathbf{0} \}$ and $\{ C_{2,110}|\mathbf{0} \}$ are not in the same crystallographic class because they are conjugate by the rotation $\{ C_{8,001} | \mathbf{0} \}$, which is not a crystallographic symmetry operation.

By the definition of a crystallographic class, we deduce:
\begin{theorem}
If $G_\mathbf{q}$ is isomorphic to one of the following point groups:
\begin{gather}
C_1, C_i, C_2, C_s, C_{2h}, C_{4v}, D_{2d}^*, D_3, C_{3v}, \nonumber\\  D_{3d}, D_6,  C_{6v}, D_{3h}, D_{6h}, T_d, O, O_h,
\label{eq:ptgrpsnomerge}
\end{gather}
which do not have any distinct conjugacy classes with elements in the same crystallographic class, 
then no two conjugacy classes of $G_\mathbf{q}$ merge in any space group, $G$.
The asterisk ($^*$) indicates the $\bar{4}m2$ orientation of $D_{2d}$, discussed below.
\label{th:ptgrpsnomerge}
\end{theorem}
\noindent
Proof: suppose that $G_\mathbf{q}$ is isomorphic to a point group listed in (\ref{eq:ptgrpsnomerge}) and that two distinct conjugacy classes $[g_1]_{G_\mathbf{q}}$ and $[g_2]_{G_\mathbf{q}}$ merge in $G$.
Then, by definition, $g_1$ and $g_2$ are in the same crystallographic class. 
This completes the proof by contradiction, since the groups listed in (\ref{eq:ptgrpsnomerge}) do not have any distinct conjugacy classes with elements in the same crystallographic class.

We now make a few comments on the list (\ref{eq:ptgrpsnomerge}). 
First, as an example, $C_2$ appears on this list because its two conjugacy classes are $[C_2]$ and $[E]$, and $C_2$ and $E$ are not in the same crystallographic class; on the other hand, $D_2 = \{ E, C_{2,100}, C_{2,010}, C_{2,001} \}$ is not on this list because each element in $D_2$ is in its own conjugacy class, but all the $C_2$ rotations in $D_2$ are in the same crystallographic class.
Second, $D_{2d}$ in the $\bar{4}m2$ orientation, which contains $C_{2,001}$ and $C_{2,110}$ in separate conjugacy classes, appears on this list because the two operations are not conjugated by a space group symmetry operation.
On the other hand $D_{2d}$ in the $\bar{4}2m$ orientation contains $C_{2,001}$ and $C_{2,100}$ in separate conjugacy classes; this group does not appear on the list because $C_{2,001}$ and $C_{2,100}$ are conjugated by $C_{4,010}$, which is a space group symmetry operation.
Third, we note that one can make a similar list for the double crystallographic point groups.
The list will be different because the double groups have an extra generator -- due to the double cover of $SO(3)$ by $SU(2)$ \cite{GroupTheoryPaper} -- that changes the distribution of symmetry elements into conjugacy classes.
Here, for simplicity, we exclude the double groups from our analysis and also ignore time reversal symmetry.
However, the tables in Appendix~\ref{sec:irrep-equiv-tables} are listed for both point groups and double point groups, with and without time-reversal symmetry.

The second theorem results from considering the conjugacy class of a \textbf{point-fixing} symmetry, which is a symmetry operation that has exactly one fixed point, such as the rotoreflections $S_2$ (inversion), $S_3$, $S_4$ and $S_6$:
\begin{theorem}
Let $g_{1,2} \in G_\mathbf{q}$ be point-fixing symmetry operations.
If $[g_1]_G = [g_2]_G$ then $[g_1]_{G_\mathbf{q}} = [g_2]_{G_\mathbf{q}}$.
\label{th:nonpolarnomerge}
\end{theorem}
\noindent  This is clear when $g_1$ is an inversion operation, since there can only be one inversion operation in a site-symmetry group.
The general proof is as follows:
suppose $g_{1,2}\in G_\mathbf{q}$ are point-fixing symmetry operations and $[g_1]_G = [g_2]_G$.
Then there exists a $g\in G$ such that $g^{-1} g_1 g = g_2$, which implies $g_1 (g \mathbf{q} ) = gg_2\mathbf{q} = g\mathbf{q}$, i.e., $g\mathbf{q}$ is a fixed point of $g_1$. By hypothesis, $g_1$ only has one fixed point, $\mathbf{q}$; thus, it must be that $g\mathbf{q} = \mathbf{q}$. Consequently, $g\in G_\mathbf{q}$.
Since $g^{-1} g_1 g = g_2$, this means $[g_1]_{G_\mathbf{q}} = [g_2]_{G_\mathbf{q}}$.

We now utilize the theorems in this section to restrict which site-symmetry groups can host irrep-equivalent EBRs.
We first consider EBRs induced from the same site-symmetry group, i.e. $\mathbf{q} = \mathbf{q}'$ in Eq.~(\ref{eq:irrep-equiv}), and then consider the case $\mathbf{q} \neq \mathbf{q}'$.
In the former case, we prove that irrep-equivalent EBRs are only possible when $G_\mathbf{q}$ is isomorphic to one of seven possible point groups (out of the 32 point groups that occur in crystals.)
Within the second case, we narrow down the possible pairs of $G_\mathbf{q}$ and $G_{\mathbf{q}'}$ to 29 possible pairs and find examples that were missed by BMZ.\cite{Bacry1988}
This provides sufficient conditions for irrep-equivalent EBRs. The Tables in Appendix~\ref{sec:irrep-equiv-tables} provide an exhaustive list of all examples.

\subsection{Same site: $\mathbf{q} = \mathbf{q}'$}
\label{sec:samesite}

We first consider the case where $\mathbf{q} = \mathbf{q}'$ in Eq.~(\ref{eq:irrep-equiv}): we prove using the 
theorems in Secs.~\ref{sec:theorems} that irrep-equivalent EBRs are only possible when $G_\mathbf{q}$ is isomorphic to one of the following seven point groups:
\begin{equation}
C_{2v}, C_3, C_4, C_6, D_2, D_{2h}, T
\label{eq:samesiteptgrps}
\end{equation}
We then prove by explicit computation (Table~\ref{tab:same_q}) that this list is necessary and sufficient.
Recall from the discussion below (\ref{eq:ptgrpsnomerge}) that this is a list of single groups;
there is a different list for double groups that we do not derive.

We now establish the list in (\ref{eq:samesiteptgrps}), following BMZ\cite{Bacry1988}.
When $\mathbf{q} = \mathbf{q}'$, Eq.~(\ref{eq:irrep-equiv}) gives a necessary condition for two band representations induced from the same site to be irrep-equivalent:
\begin{cor}
A necessary condition for two band representations induced from distinct representations of the same site-symmetry group, $G_\mathbf{q}$, to be irrep-equivalent, is that two distinct conjugacy classes of $G_\mathbf{q}$ merge with respect to $G$.
\label{cor:mergesamesite}
\end{cor}
\noindent We prove Corollary~\ref{cor:mergesamesite} by contradiction:
suppose that no distinct conjugacy classes of $G_\mathbf{q}$ merge in $G$ and let $\chi$ and $\chi'$ be characters of two inequivalent representations of $G_\mathbf{q}$ such that $\chi\uparrow G$ and $\chi'\uparrow G$ are irrep-equivalent.
Since no distinct conjugacy classes of $G_\mathbf{q}$ merge in $G$, 
both sides of Eq.~(\ref{eq:irrep-equiv}) simplify according to Theorem~\ref{th:nomerge}:
\begin{equation} 
\frac{|[g]_{G_\mathbf{q} } |}{|G_\mathbf{q}|} \chi(g) = \frac{|[g]_{G_\mathbf{q} } |}{|G_\mathbf{q}|} \chi'(g) \Rightarrow \chi(g) = \chi'(g),
\label{eq:sumconjugacyclass}
\end{equation}
for all $g\in G_\mathbf{q}$.
Since two representations with the same character are equivalent, our assumption that $\chi$ and $\chi'$ are characters of inequivalent representations of $G_\mathbf{q}$ is contradicted, which completes the proof.

Theorem~\ref{th:ptgrpsnomerge} established that for the point groups listed in (\ref{eq:ptgrpsnomerge}), no two conjugacy classes of $G_\mathbf{q}$ merge in $G$, for any choice of $G$.
Combined with Corollary~\ref{cor:mergesamesite}, it follows that when $G_\mathbf{q}$ is isomorphic to a group in (\ref{eq:ptgrpsnomerge}), distinct irreps of $G_\mathbf{q}$ will not yield irrep-equivalent EBRs.
This conclusion rules out 16 of the 32 point groups.
(Notice that although 17 point groups are listed in (\ref{eq:ptgrpsnomerge}), $D_{2d}$ in the $\bar{4}2m$ orientation is not ruled out.)

We rule out five additional groups by the following:
\begin{cor}
If $G_\mathbf{q}$ is an Abelian group generated by a set of point-fixing symmetry operations, then no two distinct representations of $G_\mathbf{q}$ will induce irrep-equivalent EBRs.
\label{cor:nonpolarAbelian}
\end{cor}
\noindent Proof: suppose that $\chi$ and $\chi'$ are characters of two distinct representations of $G_\mathbf{q}$ and that $\chi \uparrow G$ and $\chi'\uparrow G$ are irrep-equivalent.
Then for each point-fixing generator, $g_i$, of $G_\mathbf{q}$, Theorem~\ref{th:nonpolarnomerge} guarantees that $[g_i]_G \cap G_\mathbf{q} = [g_i]_{G_\mathbf{q}}$.
Thus, both sides of the sum in Eq.~(\ref{eq:irrep-equiv}) simplify according to Theorem~\ref{th:nomerge},
yielding exactly Eq.~(\ref{eq:sumconjugacyclass}), which implies $\chi(g_i) = \chi'(g_i)$ for each point-fixing generator, $g_i$, of $G_\mathbf{q}$.
By hypothesis, $G_\mathbf{q}$ is generated by the set of $g_i$;
hence, each element, $g\in G_\mathbf{q}$, can be written as $g=\prod_i g_i^{n_i}$.
Since, also by hypothesis, $G_\mathbf{q}$ is Abelian, the order of the $g_i$ in the product does not matter.
Thus, the character of $g$ can be expressed as the product: 
\begin{equation}
\chi(g) = \prod_i \chi(g_i)^{n_i} = \prod_i \left[ \chi'(g_i)\right]^{n_i} = \chi'(g),
\end{equation}
where we have used $\chi(g_1g_2) = \chi(g_1)\chi(g_2)$ for an Abelian group (which does not necessarily hold in a non-Abelian group) and
the middle equality follows because we proved $\chi(g_i) = \chi'(g_i)$.
But $\chi(g) = \chi'(g)$ for all $g\in G_\mathbf{q}$ violates our hypothesis that $\chi$ and $\chi'$ are characters of distinct representations of $G_\mathbf{q}$, which completes the proof.

There are five Abelian crystallographic point groups that do not appear in (\ref{eq:ptgrpsnomerge}) and can be generated by only point-fixing symmetry operations:
\begin{gather}
S_4 = \langle S_4^+ \rangle , \, \, C_{4h} = \langle S_4^+ , i \rangle ,\,\, S_6 = \langle S_6^+ \rangle, \nonumber\\
C_{3h} = \langle S_3^+ \rangle , \,\,  C_{6h} = \langle S_3^+, i \rangle,
\label{eq:nonpolarAbelian}
\end{gather}
where the angled brackets enclose the group generators, $S_n^+$ indicates an $n$-fold rotoreflection and $i$ indicates inversion, following Sch\"onflies notation. \cite{PointGroupTables}
Corollary~\ref{cor:nonpolarAbelian} proves that if $G_\mathbf{q}$ is isomorphic to one of the point groups in (\ref{eq:nonpolarAbelian}), then no two distinct irreps of $G_\mathbf{q}$ will induce irrep-equivalent EBRs.

Finally, by using the POINT application on the BCS (see end of Appendix~\ref{app:chars-gamma}), one finds that if $G_\mathbf{q}$ is one of the following four point groups:
\begin{equation}
D_4, D_{2d}^{**}, D_{4h}, T_h,
\label{eq:conjmergenoequiv}
\end{equation}
conjugacy classes with respect to $G_\mathbf{q}$ can merge in $G$, but no distinct irreps of $G_\mathbf{q}$ induce irrep-equivalent EBRs. 
The double asterisk ($^{**}$) in (\ref{eq:conjmergenoequiv}) indicates the $\bar{4}2m$ orientation of $D_{2d}$, as explained below Theorem ~\ref{th:ptgrpsnomerge}.
We show in Appendix~\ref{app:chars-gamma} that for these groups (as well as the groups listed in (\ref{eq:ptgrpsnomerge})), distinct irreps of the site-symmetry group induce band representations with different little group representations at $\Gamma$, which, consequently, are not irrep-equivalent.

Thus, there are only seven possible choices of $G_\mathbf{q}$, listed in (\ref{eq:samesiteptgrps}), from which distinct irreps can induce irrep-equivalent EBRs.
This list is not only sufficient, but also necessary: Table~\ref{tab:same_q} reveals that for each point group in (\ref{eq:samesiteptgrps}), there exists at least one space group with a site whose site-symmetry group is isomorphic to one of the groups in (\ref{eq:samesiteptgrps}) and for which two distinct irreps induce irrep-equivalent EBRs.
Table~\ref{tab:same_q} contains the complete list of irrep-equivalent EBRs induced from the same site, without time-reversal symmetry.
The table contains both single-valued and double-valued EBRs: while the former were enumerated in Tables 2 and 3 of BMZ\cite{Bacry1988}, the latter list is new to this work.
Also new to this work is the analogous list for irrep-equivalent time-reversal symmetric EBRs induced from the same site in Table~\ref{tab:TRsame_q}.

\subsection{Irrep-equivalent EBRs when $\mathbf{q}$ and $\mathbf{q}'$ are not part of the same Wyckoff position}
\label{sec:differentsite}

We now consider the case that $\mathbf{q}$ and $\mathbf{q}'$ are not part of the same Wyckoff position.
We use the theorems in Sec.~\ref{sec:theorems} to prove that certain pairs of point groups $G_\mathbf{q}$ and $G_{\mathbf{q}'}$ cannot have irreps that induce irrep-equivalent EBRs.
We limit the total number of possible cases to 29 $(G_\mathbf{q}, G_{\mathbf{q}'})$ pairs, out of $528$ possible pairs of crystallographic point groups.
The possible pairs are indicated by empty squares in Table~\ref{tab:pointgrouppairs}, which we now derive.

We first limit the possible cases of irrep-equivalence by proving that
representations whose character of a point-fixing symmetry operation is non-zero do not induce an EBR that is irrep-equivalent to any distinct EBR:
\begin{cor}
Given a site-symmetry group, $G_\mathbf{q}$, a point-fixing symmetry operation $g\in G_\mathbf{q}$,
and an irrep of $G_\mathbf{q}$ with character $\chi$, if $\chi(g) \neq 0$, then $\chi\uparrow G$ is not irrep-equivalent to any  EBR induced from the site-symmetry group of a site that is not part of the same Wyckoff position as $\mathbf{q}.$
\label{cor:pointgroupsnonpolarnonzero}
\end{cor}
\noindent Proof: let $g\in G_\mathbf{q}$ be a point-fixing symmetry and let $\chi$ be the character of an irrep of $G_\mathbf{q}$ such that $\chi(g) \neq 0$.
Theorem~\ref{th:nonpolarnomerge} says that $[g_i]_G \cap G_\mathbf{q} = [g_i]_{G_\mathbf{q}}$; therefore, the sum on the LHS of Eq.~(\ref{eq:irrep-equiv}) is over the conjugacy class $[g]_{G_\mathbf{q}}$. Then, using Eq.~(\ref{eq:nomerge}),
the LHS of Eq.~(\ref{eq:irrep-equiv}) is proportional to $\chi(g)$.
Since $\chi(g)\neq 0$ by hypothesis, the LHS of Eq.~(\ref{eq:irrep-equiv}) is non-zero.
Now suppose there is a site $\mathbf{q}'$ and an irrep of $G_{\mathbf{q}'}$ with character $\chi'$ such that $\chi\uparrow G$ and $\chi'\uparrow G$ are irrep-equivalent.
Since we have established that the LHS of Eq.~(\ref{eq:irrep-equiv}) is nonzero, it must also be that the RHS of Eq.~(\ref{eq:irrep-equiv}) is nonzero.
Thus, the sum on the RHS of Eq.~(\ref{eq:irrep-equiv}) must be over a nonempty set, i.e., $g$ is conjugate in $G$ to some element, $g'$, of $G_{\mathbf{q}'}$.
Then there exists $h\in G$ such that $h^{-1} g h = g'$.
Consequently, $g(h\mathbf{q}') = hg'\mathbf{q}' = h\mathbf{q}'$ (the last equality follows because $g'\in G_{\mathbf{q}'}$), i.e., $h\mathbf{q}'$ is a fixed point of $g$.
Since, by hypothesis, $g$ has a single fixed point, $h\mathbf{q}'= \mathbf{q}$, which, by definition, means that $\mathbf{q}$ and $\mathbf{q}'$ are part of the same Wyckoff position.
This completes the proof.

There are 14 point groups that have a point-fixing symmetry operation whose character is non-zero in all irreps:
\begin{gather}
C_i, C_{2h}, D_{2h}, S_4, C_{4h}, D_{4h}, S_6, D_{3d}, \nonumber\\
C_{3h}, C_{6h}, D_{3h}, D_{6h}, T_h, O_h
\label{eq:pointgroupsnonpolarnonzero}
\end{gather}
Corollary~\ref{cor:pointgroupsnonpolarnonzero} guarantees that if $G_\mathbf{q}$ is isomorphic to one of the point groups in the list in (\ref{eq:pointgroupsnonpolarnonzero}), then no EBR induced from an irrep of $G_\mathbf{q}$ will be irrep-equivalent to an EBR induced from an irrep of $G_{\mathbf{q}'}$ when $\mathbf{q}'$ is not part of the same Wyckoff position as $\mathbf{q}$.

We further rule out the case where $G_\mathbf{q}$ is the trivial group ($C_1$)
because if $G$ has the general Wyckoff position (which, by definition, has a trivial site-symmetry group) as a maximal Wyckoff position, then it has no special Wyckoff positions.
(Such groups are called fixed-point-free space groups or Bieberbach groups.)

There are 17 remaining choices for $G_\mathbf{q}$:
\begin{gather}
C_2, C_s, D_2, C_{2v}, C_4, D_4, C_{4v}, D_{2d}, C_3, \nonumber\\ D_3, 
C_{3v}, C_6, D_6, C_{6v}, T, O, T_d
\label{eq:pointgroupsnonpolarzero}
\end{gather}
All these groups except for $D_{2d}$ and $T_d$ lack point-fixing symmetry operations.
While $D_{2d}$ and $T_d$ have point-fixing symmetry operations, they also have irreps where the character of the point-fixing operation is zero; hence none of the groups listed in (\ref{eq:pointgroupsnonpolarzero}) are ruled out by Corollary~\ref{cor:pointgroupsnonpolarnonzero}.
Thus, there are $(17\cdot 16/2) +17 = 153$ pairs of crystallographic groups that could correspond to $G_\mathbf{q}$ and $G_{\mathbf{q}'}$
(including the 17 pairs where $G_\mathbf{q}$ and $G_{\mathbf{q}'}$ are isomorphic even though $\mathbf{q}$ and $\mathbf{q}'$ are not part of the same Wyckoff position).
These $153$ pairs are shown boxed in Table~\ref{tab:pointgrouppairs}.

\begin{table*}[t]
\begin{tabular}{c|c|c|c|c|c|c|c|c|c|c|c|c|c|c|c|c|c|}
  &  $C_2$ & $C_s$ & $D_2$ & $C_{2v}$ & $C_4$ & $D_4$ & $D_{2d}$ & $C_{4v}$ & $C_3$ & $D_3$ & $C_{3v}$ & $C_6$ & $D_6$ & $C_{6v}$ & $T$ & $O$  & $T_d$ \\
\hline 
$ C_2$ & \\ \cline{1-3}
$C_s $ & X  &  \\ \cline{1-4}
$D_2$ & D & X & \\ \cline{1-5}
$C_{2v}$ & D & D &    & \\ \cline{1-6}
$C_4$ & D  & X & &  & \\ \cline{1-7}
$D_4$ & D & X & & W & W & \\ \cline{1-8}
$D_{2d}$ & D & D &  & W & W & & \\ \cline{1-9}
$C_{4v}$ & D& D & W & & W & W  &  & \\ \cline{1-10}
$C_3$ & X & X & X & X & X & X & D &  X & \\  \cline{1-11}
$D_3$& X  & X & X & X & X & X & X & X & & \\ \cline{1-12}
$C_{3v}$& X & X & X & X & X & X & X & X & & W  & \\ \cline{1-13}
$C_6$ & X & X & X & X & X & X & X & X & X & X & X & W \\ \cline{1-14}
$D_6$ & X & X & X & X & X & X & X & X & X & X & X &  & \\ \cline{1-15}
$C_{6v}$ & X & X & X & X & X & X & X & X & X & X & X & W & W & W\\ \cline{1-16}
$T$ & D & X &  & W & W & W & W & W & X & D & X & X  & X  & X & \\ \cline{1-17}
$O$ & D & X & D & D & D& & D' & W & D& D & X & X & X & X & & \\ \cline{1-18}
$T_d$  &D & X & D& D& D & D'  & D' & D' & X & D & X & X & X & X & W & W &  \\ \cline{1-18}
\end{tabular}
\caption{The possible pairs of site-symmetry groups $G_\mathbf{q}$ and $G_{\mathbf{q}'}$ (rows and columns) that can induce irrep-equivalent EBRs according to the list in (\ref{eq:pointgroupsnonpolarzero}) are indicated by empty boxes.
An X indicates that a pair $(G_\mathbf{q}, G_{\mathbf{q}'})$ is ruled out because $G_\mathbf{q}$ has an element $g$ in Table~\ref{tab:crys-zero-ind} that guarantees the LHS of Eq.~(\ref{eq:irrep-equiv}) is non-zero, while $G_{\mathbf{q}'}$ has no element in the same crystallographic class as $g$, thus guaranteeing that the RHS of Eq.~(\ref{eq:irrep-equiv}) is zero.
Of the remaining pairs, those that are ruled out because they do not have irreps that satisfy the dimensionality constraint in Eq.~(\ref{eq:irrep-equiv-E}) are marked with a D or D' for dimension (see text for distinction).
Those that are ruled out because there is no space group with both point groups as maximal site-symmetry groups are marked with a W for Wyckoff.
}
\label{tab:pointgrouppairs}
\end{table*}

We will eliminate $77$ of the $153$ possible pairs of $(G_\mathbf{q}, G_{\mathbf{q}'})$ (marked with an X in Table~\ref{tab:pointgrouppairs}) 
in the following way: suppose $G_\mathbf{q}$ has an element $g$ such that the LHS of Eq.~(\ref{eq:irrep-equiv}) is non-zero for any $\chi$ (we will explain below how this can happen).
If, further, $[g]_G \cap G_{\mathbf{q}'} = \emptyset$, then the RHS of Eq.~(\ref{eq:irrep-equiv}) will be zero for any $\chi'$.
Therefore, Eq.~(\ref{eq:irrep-equiv}) is not satisfied for any irreps $\chi, \chi'$ of $G_\mathbf{q}, G_{\mathbf{q}'}$, respectively.
Consequently, the pair $(G_\mathbf{q},G_{\mathbf{q}'})$ can be ruled out as a candidate for irrep-equivalence.

To this end, for each point group listed in~(\ref{eq:pointgroupsnonpolarzero}), we list in Table~\ref{tab:crys-zero-ind} the crystallographic classes for which the LHS of Eq.~(\ref{eq:irrep-equiv}) is necessarily non-zero, for any $\chi$.
We now explain how to find the entries in Table~\ref{tab:crys-zero-ind}.
For the following point groups:
\begin{gather}
C_2, C_s, C_{2v}, C_4, C_{4v}, D_3, C_{3v}, \nonumber\\
C_6 \text{ (when $g$ is $C_2$)}, 
D_6, C_{6v}, T, O, T_d
\label{eq:crys-zero-ind-list-1}
\end{gather}
the element $g$ listed in Table~\ref{tab:crys-zero-ind} meets two conditions:
\begin{enumerate}
\item All of the elements of $G_\mathbf{q}$ in the crystallographic class of $g$ are in the conjugacy class $[g]_{G_\mathbf{q}}$ and hence the conjugacy class $[g]_{G_\mathbf{q}}$ does not merge with any distinct conjugacy classes in $G$; and
\item For all irreps of $G_\mathbf{q}$, $\chi(g) \neq 0$.
\end{enumerate}

\noindent
For example, if $G_\mathbf{q}$ is isomorphic to the point group $C_2$, and $g$ indicates the two-fold rotation in $G_\mathbf{q}$, then for both irreps of $C_2$, $\chi(g)\neq 0$.
Since $G_\mathbf{q}$ does not have any other conjugacy class with a $C_2$ rotation, it follows from Theorem~\ref{th:nomerge} that the LHS of Eq.~(\ref{eq:irrep-equiv}) is given by
$\frac{| [g]_{G_\mathbf{q} } | }{ |G_\mathbf{q}| } \chi(g) = \frac{1}{2}\chi(g) \neq 0$,
for any space group $G$. 
We then deduce that if $G_{\mathbf{q}'}$ does not have a two-fold rotation, $G_{\mathbf{q}'} \cap [g]_G=\emptyset$ and therefore the RHS of Eq.~(\ref{eq:irrep-equiv}) will be zero for all irreps of $G_\mathbf{q}'$.
Hence there will never be an irrep-equivalence between $G_{\mathbf{q}}$ and $G_{\mathbf{q}'}$.

\begin{table}[h]
\begin{tabular}{cc}
$G_\mathbf{q}$ & $g$ \\
\hline
$C_2$ & $C_2$\\
$C_s$ & $m$\\
$D_2$ & \\
$C_{2v}$ & $C_2$ \\
$C_4$ & $C_2$\\
$D_4$ & \\
$C_{4v}$ & $C_2$\\
$C_3$ & $C_3$ \\
$D_3$ & $C_3$\\
\end{tabular}
\quad\quad
\begin{tabular}{cc}
$G_\mathbf{q}$ & $g$ \\
\hline
$C_{3v}$ & $C_3$\\
$C_6$ & $C_2, C_3, C_6$\\
$D_6$ & $C_{2,001}, C_3, C_6$ \\ 
$C_{6v}$ & $C_2, C_3, C_6$\\
$T$ & $C_2$\\
$O$ & $C_{2,001}$ \\
$D_{2d}$ & \\
$T_d$ & $C_2$\\
& 
\end{tabular}
\caption{For each point group, $G_\mathbf{q}$, in (\ref{eq:pointgroupsnonpolarzero}), the elements $g$ are listed for which the LHS of Eq.~(\ref{eq:irrep-equiv}) will necessarily be nonzero, for any space group $G$ and any irrep, $\chi$, of $G_\mathbf{q}$.
(No entry means that there is no such $g\in G_\mathbf{q}$ with this property).
The orientation of the axis of rotation is only specified when there is more than one axis of the same order in different conjugacy classes.}
\label{tab:crys-zero-ind}
\end{table}


We now explain the point groups in Table~\ref{tab:crys-zero-ind} that do not appear in (\ref{eq:crys-zero-ind-list-1}), namely $C_3$ and $C_6$ (when $g=C_3,C_6$).
Consider the case when $G_\mathbf{q}$  is isomorphic to $C_3$ and generated by $g=\{ C_{3,001} | \mathbf{0} \}$.
Table~\ref{table:c3chars} shows that $g$ and $g^{-1}$ are in different conjugacy classes with respect to $G_\mathbf{q}$.
If they remain in different conjugacy classes with respect to $G$, then the LHS of Eq.~(\ref{eq:irrep-equiv}) will be non-zero according to Eq.~(\ref{eq:nomerge}), since $\chi(g), \chi(g^{-1}) \neq 0$.
If, on the other hand, their conjugacy classes merge in $G$ (which would happen if, for example, $G$ contained $\{ C_{2,100} | \mathbf{0} \}$, since $\{ C_{2,100} | \mathbf{0} \}^{-1} \{ C_{3,001} | \mathbf{0} \} \{ C_{2,100} | \mathbf{0} \} = \{ C_{3,001}^{-1} | \mathbf{0} \} $), then the LHS of Eq.~(\ref{eq:irrep-equiv}) would be proportional to $\chi(g) + \chi(g^{-1}) \neq 0$.
Thus, whether or not the conjugacy classes of $g$ and $g^{-1}$ merge in $G$, the LHS of Eq.~(\ref{eq:irrep-equiv}) is always non-zero when applied to $g$.
This explains why $C_3$ is in Table~\ref{tab:crys-zero-ind} with $g=C_3$.
The same logic applies to $C_6$ for $g=C_3,C_6$, which can be verified by the characters in Table~\ref{table:c6chars}.

\begin{table}
\begin{tabular}{c|ccc}
$\rho$ & $[E]$ & $[C_{3}]$ & $[C_{3}^{-1}]$ \\
\hline
$A$ & $1$ & $1$ & $1$ \\
$^1E$ & $1$ & $\omega^2$ & $\omega$\\
$^2E$ & $1$ & $\omega$ & $\omega^2$
\end{tabular}
\caption{Character table for $C_3$; $\omega = e^{2\pi i/3}$.}\label{table:c3chars}
\end{table}

\begin{table}
\begin{tabular}{c|cccccc}
$\rho$ & $[E]$ & $[C_{6}]$ & $[C_{3}]$ & $[C_{2}]$ & $[C_{3}^{-1}]$ & $[C_{6}^{-1}]$ \\
\hline
$A$ & $1$ & $1$ & $1$ & $1$ & $1$ & $1$ \\
$B$ & $1$ & $-1$ & $1$ & $-1$ & $1$ & $-1$ \\
$^1E_2$ & $1$ & $\omega$ & $\omega^2$ & $1$ & $\omega$ & $\omega^2$ \\
$^2E_2$ & $1$ & $\omega^2$ & $\omega$ & $1$ & $\omega^2$ & $\omega$ \\
$^1E_1$ & $1$ & $-\omega$ & $\omega^2$ & $-1$ & $\omega$ & $-\omega^2$ \\
$^2E_1$ & $1$ & $-\omega^2$ & $\omega$ & $1$ & $\omega^2$ & $-\omega$ \\
\end{tabular}
\caption{Character table for $C_6$; $\omega = e^{2\pi i/3}$.}\label{table:c6chars}
\end{table}


Constraints from dimensionality further restrict the pairs of point groups in Table~\ref{tab:pointgrouppairs}.
Taking $g$ to be the identity element in Eq.~(\ref{eq:irrep-equiv}) yields a necessary condition for $\chi$ and $\chi'$ to induce irrep-equivalent EBRs:
\begin{equation}
\chi(E)/|G_\mathbf{q}| = \chi'(E)/|G_{\mathbf{q}'}|
\label{eq:irrep-equiv-E}
\end{equation}
Since $\chi(E) = {\rm dim}(\rho)$, where $\chi$ is the character of the representation $\rho$, Eq.~(\ref{eq:irrep-equiv-E}) can be regarded as a dimensionality constraint.
We rule out 23 additional pairs of site-symmetry groups, marked with a D in Table~\ref{tab:pointgrouppairs}, because they do not have irreps that satisfy Eq.~(\ref{eq:irrep-equiv-E}).

We make a finer constraint on dimensionality by looking at specific representations: in particular, we eliminate four additional pairs of point groups, marked by a $D'$ in Table~\ref{tab:pointgrouppairs}, because the specific irreps that satisfy the dimensionality constraint in Eq.~(\ref{eq:irrep-equiv-E}) do not satisfy the necessary condition for irrep-equivalence in Eq.~(\ref{eq:irrep-equiv}),
which we prove in Appendix~\ref{sec:moredimensionality}.

For the remaining pairs of point groups in Table~\ref{tab:pointgrouppairs}, we mark with a W those pairs for which there does not exist a space group with maximal Wyckoff positions whose site-symmetry groups are given by $G_\mathbf{q}$ and $G_{\mathbf{q}'}$.
This eliminates 20 additional pairs. 

This analysis has narrowed our search to only $29$ $(G_\mathbf{q}, G_{\mathbf{q}'})$ pairs that could yield irrep-equivalent EBRs.

In Tables~\ref{tab:diff_q} and Table~\ref{tab:equiv} we list the irrep-equivalent EBRs induced from different sites.
The difference between the two tables is that in Table~\ref{tab:diff_q} the irrep-equivalent EBRs are not homotopically equivalent,
where homotopic equivalence is defined according to Eq.~(\ref{eq:defequivalence}),
while in Table~\ref{tab:equiv}, the irrep-equivalent EBRs are also equivalent.
(Recall that homotopic equivalence is a sufficient condition for irrep-equivalence, but not necessary, as discussed below Eq.~(\ref{eq:defequivalence}).)
To determine whether two irrep-equivalent EBRs are homotopically equivalent, we explicitly checked whether there exists a third intermediate Wyckoff position (on the line connecting the two sites from which the irrep-equivalent EBRs are induced),
such that a band representation induced from the intermediate Wyckoff position is irrep-equivalent to the two EBRs.
If such an intermediate position exists, we deduce that the two EBRs are not only irrep-equivalent, but also homotopically equivalent.

Between Tables~\ref{tab:diff_q} and~\ref{tab:equiv} there are only nine pairs of site-symmetry groups with irreps that induce irrep-equivalent EBRs:
\begin{multline}
(D_2,D_2), (D_3,D_3), (D_4,D_4), (D_6,D_6), \\ (D_{2d}, D_{2d}), (T,T), (O,O), (T_d,T_d), (T,D_2)
\label{eq:diff_site_list_pairs}
\end{multline}
Thus, the $29$ pairs that appear in Table~\ref{tab:pointgrouppairs} provide a necessary, but not sufficient, condition for irrep-equivalence of EBRs induced from different sites.
Interestingly, the last pair in (\ref{eq:diff_site_list_pairs}), $(T,D_2)$, is the only instance where $G_\mathbf{q}$ is not isomorphic to $G_{\mathbf{q}'}$ (for single-valued representations without time-reversal symmetry).
This case was missed in the earlier analysis by BMZ\cite{Bacry1988}; we discuss it in more detail in Sec.~\ref{sec:differences}.

A necessary and sufficient condition for irrep-equivalence could be derived by computing Eq.~(\ref{eq:irrep-equiv}) for each $g\in G$ and $\mathbf{q}, \mathbf{q}'$ that are part of a maximal Wyckoff position, whose site-symmetry groups are given by one of the remaining $(G_\mathbf{q}, G_{\mathbf{q}'})$ pairs in Table~\ref{tab:pointgrouppairs}.
The EBRs induced from irreps of $G_\mathbf{q}$ and $G_{\mathbf{q}'}$ with characters $\chi$ and $\chi'$, respectively, are irrep-equivalent if and only if Eq.~(\ref{eq:irrep-equiv}) is satisfied for all $g\in G$. This is the content of our computational results in Tables~\ref{tab:diff_q} and \ref{tab:equiv} (without time-reversal) and in Tables~\ref{tab:TRdiff_q} and \ref{tab:TRequiv} (with time-reversal).

\subsection{Differences from BMZ}
\label{sec:differences}

Our computational analysis reveals many cases of irrep-equivalence where $\mathbf{q} \neq \mathbf{q}'$, but $G_\mathbf{q}$ and $G_{\mathbf{q}'}$ are isomorphic (diagonal entries in Table~\ref{tab:pointgrouppairs}), as well as two cases without SOC where $G_\mathbf{q}$ and $G_{\mathbf{q}'}$ are not isomorphic (off-diagonal entries in Table~\ref{tab:pointgrouppairs}).
These two cases were missed by BMZ.\cite{Bacry1988}
(Note: BMZ only considered the spinless case. When SOC is included, our tables show many cases where $G_\mathbf{q}$ and $G_{\mathbf{q}'}$ are not the same point group.)

The two cases missed by BMZ occur in space groups $I23$ and $Pn\bar{3}$; in both cases, the site-symmetry group $G_\mathbf{q}$ is isomorphic to $T$ and $G_{\mathbf{q}'}$ is isomorphic to $D_2$.
In $I23$ (No. 197), the irrep-equivalent EBRs are induced from the $T$ irrep on the $2a$ position and $B_{1}$ or $B_{2}$ irrep on the $6b$ position.
In $Pn\bar{3}$ (No. 201), the EBR induced from the $T$ irrep on the $2a$ position and $B_{1}$ or $B_{2}$ irrep on the $6d$ position are irrep-equivalent.
In the Supplemental Material\cite{SMdifferent} we explicitly verify that these cases satisfy Eq.~(\ref{eq:irrep-equiv}).
It remains to determine whether there is an overarching principle that describes why only the case where $G_\mathbf{q}$ is isomorphic to $T$ and $G_{\mathbf{q}'}$ to $D_2$ occurs, out of the several other off-diagonal entries in Table~\ref{tab:pointgrouppairs}.

\section{Conclusion}

In this paper, we have enumerated the irrep-equivalent EBRs with and without TR and SOC.
We have described how the pairs of irrep-equivalent EBRs can give rise to a Berry obstructed atomic limit, which implies that there is a required phase transition (gap closing) between two distinct non-topological phases, which cannot be deduced from their symmetry eigenvalues.

In addition, for two examples, in space groups $F222$ and $P112$, without and with SOC, respectively, we have provided topological invariants that distinguish the irrep-equivalent bands.
We expect that this result can be generalized to all irrep-equivalent EBRs that are not homotopically equivalent.
This hypothesis is intuitive because if a pair of irrep-equivalent EBRs are not homotopically equivalent, then there is an obstruction to deforming them into each other; the obstruction itself would constitute a topological invariant.

Despite the use of topological invariants, the current manuscript has been limited to atomic limit phases: each phase discussed can be described by a Hamiltonian without any momentum-dependence.
However, stable and fragile topological bands can also be irrep-equivalent, either to other topological bands or to trivial bands. 
In future work we plan to extend the present analysis to distinguish topological bands that are hidden from symmetry  labels, as predicted by the theory of topological quantum chemistry.\cite{NaturePaper}

\begin{acknowledgments}
The authors acknowledge Maia Garcia Vergniory, Zhijun Wang, and Benjamin Wieder for helpful conversations while working on earlier publications. 
J. C. acknowledges support from the Flatiron Institute, a division of the Simons Foundation, and the National Science Foundation under grant DMR-1942447. B. B. acknowledges support from the Alfred P. Sloan Foundation, and the National Science Foundation under grant DMR-1945058.
B.A.B. was supported by the the Office of Naval Research (ONR Grant No. N00014-20-1-2303), and was partially supported by the National Science Foundation (EAGER Grant No. DMR 1643312), a Simons Investigator grant (No. 404513), the Packard Foundation,  U.S. Department of Energy (Grant No. DE-SC0016239),  the Schmidt Fund for Innovative Research, the BSF Israel US foundation (Grant No. 2018226), the Gordon and Betty Moore Foundation through Grant No. GBMF8685 towards the Princeton theory program and the NSF-MRSEC (Grant No. DMR-2011750) and acknowledge financial support from the Schmidt DataX Fund at Princeton University. Further support was provided by the NSF-MRSEC No. DMR-1420541
M.I.A. and L.E. are supported by the Government of the Basque Country (project IT1301-19) and the Spanish Ministry of Science and Innovation (PID2019-106644GB-I00).
\end{acknowledgments}

\appendix

\section{Equivalence of band representations}
\label{sec:equivalence}

Recall from Eq.~(\ref{eq:defequivalence}) 
that two band representations
are equivalent iff there exists a unitary matrix-valued function $S(\mathbf{k},\tau,g)$ that interpolates between them as $\tau$ varies from 0 to 1, such that $S(\mathbf{k},\tau,g)$  is smooth in $\mathbf{k}$, continuous in $\tau$, and for all $g\in G$, $\tau\in [0,1]$, $S(\mathbf{k},\tau,g)$ is a band representation. 
We will prove that if two band representations, $S(\mathbf{k},0,g)$ and $S(\mathbf{k},1,g)$, are equivalent, 
 then there exists a unitary matrix $U(\mathbf{k})$, which satisfies:
\begin{equation}
S(\mathbf{k},0,g) = U^\dagger(g\mathbf{k}) S(\mathbf{k},1,g) U(\mathbf{k})
\label{eq:equivalence}
\end{equation}
and which has the periodicity of the BZ, i.e., $U(\mathbf{k}+\mathbf{K}) = U(\mathbf{k})$ for any reciprocal lattice vector $\mathbf{K}$.
$U$ must be Brillouin-zone periodic so that it does not change the boundary conditions of the Hilbert space on which the band representation acts.

We now derive a construction for $U$ in the Hilbert space defined by the set of Wannier functions on which the band representation acts. (In Appendix~\ref{app:unitary}, we present an alternative derivation.)
In real space, we define the localized Wannier functions, $W_{i\alpha}(\mathbf{r} - \mathbf{t},\tau)$, where $i$ indexes a basis vector for the irrep of the site-symmetry group from which the band representation is induced, $\alpha$ indexes a site in the Wyckoff position, $\mathbf{t}$ is a lattice vector and $\tau$ is the parameter that appears in the family of band representations $S(\mathbf{k},\tau,g)$.  
The induced band representation acts on the Fourier transformed functions,
\begin{equation}
a_{i\alpha}(\mathbf{k},\mathbf{r},\tau) = \sum_\mathbf{t} e^{i\mathbf{k} \cdot \mathbf{t} } W_{i\alpha}(\mathbf{r} - \mathbf{t},\tau)
\label{eq:Wannier}
\end{equation}
(c.f. Eq.~(4) of Ref.~\onlinecite{EBRTheory}.)
Because the Hilbert space is fixed as $\tau$ varies, 
the basis of Wannier functions on which the band representation acts evolves according to a unitary transformation, $U_{i\alpha,j\beta}(\mathbf{t},\tau)$, defined by:
\begin{equation}
W_{i\alpha}(\mathbf{r}-\mathbf{t},\tau) = \sum_{j\beta,\mathbf{t}'} U_{i\alpha,j\beta}(\mathbf{t}-\mathbf{t}',\tau) W_{j\beta}(\mathbf{r}-\mathbf{t}',0)
\end{equation}
It follows that
\begin{equation}
a_{i\alpha}(\mathbf{k},\mathbf{r},\tau) = \sum_{j\beta} U_{i\alpha,j\beta} (\mathbf{k},\tau) a_{j\beta}(\mathbf{k},\mathbf{r},0),
\end{equation}
where, suppressing the indices $i\alpha,j\beta$,
\begin{equation}
U(\mathbf{k},\tau) \equiv \sum_\mathbf{t} e^{i\mathbf{k} \cdot \mathbf{t} } U(\mathbf{t},\tau) 
\label{eq:unitary}
\end{equation}
By its definition in Eq.~(\ref{eq:unitary}), $U(\mathbf{k})$ has the periodicity of the BZ.
Further,  since the Fourier transformed Wannier functions transform according to $U$, the band representation transforms according to Eq.~(\ref{eq:equivalence}).
Thus, the matrix $U(\mathbf{k},\tau=1)$ is exactly the matrix $U(\mathbf{k})$ that appears in Eq.~(\ref{eq:equivalence}), which completes the proof.

\section{Tables of irrep-equivalent EBRs}
\label{sec:irrep-equiv-tables}

We present tables of all irrep-equivalent EBRs, with and without time-reversal symmetry.
The results are derived from (and can be checked via) the BANDREP\cite{NaturePaper,GraphDataPaper,GroupTheoryPaper} application on the BCS.

We first consider EBRs without enforcing time-reversal symmetry. In Table~\ref{tab:same_q}, we indicate irrep-equivalent EBRs induced from two distinct irreps of the same site-symmetry group, $G_\mathbf{q}$.
In Table~\ref{tab:diff_q}, we indicate irrep-equivalent EBRs induced from irreps of different site-symmetry groups, $G_\mathbf{q}, G_{\mathbf{q}'}$, of the sites $\mathbf{q}$ and $\mathbf{q}'$, respectively, such that $\mathbf{q}$ and $\mathbf{q}'$ are not part of the same Wyckoff position; excluded from this list are EBRs that are homotopically equivalent, in the sense of Eq.~(\ref{eq:defequivalence}).
The homotopically equivalent EBRs, which are necessarily also irrep-equivalent, are listed in Table~\ref{tab:equiv}.
In Table~\ref{tab:sums}, we list the EBRs which are irrep-equivalent to a sum of two EBRs.

We then move to the time-reversal symmetric EBRs and compute the analogous tables.
In Table~\ref{tab:TRsame_q}, we indicate time-reversal symmetric irrep-equivalent EBRs induced from two distinct irreps of the same site-symmetry group, $G_\mathbf{q}$.
In Table~\ref{tab:TRdiff_q}, we indicate time-reversal symmetric irrep-equivalent EBRs induced from irreps of different site-symmetry groups, $G_\mathbf{q}, G_{\mathbf{q}'}$, of the sites $\mathbf{q}$ and $\mathbf{q}'$, respectively, such that $\mathbf{q}$ and $\mathbf{q}'$ are not part of the same Wyckoff position; 
again, we exclude the homotopically equivalent EBRs from Table~\ref{tab:diff_q} and list them separately in Table~\ref{tab:TRequiv}.
In Table~\ref{tab:TRsums}, we list the time-reversal symmetric EBRs which are irrep-equivalent to a sum of two EBRs.



\section{Proof of Theorem~\ref{th:unitary}: irrep-equivalence is the same as unitary equivalence} 
\label{app:unitary}

Here we prove that two band representations are irrep-equivalent if and only if they are related by a unitary transformation, thus proving Theorem~\ref{th:unitary}.
However, it is important to emphasize -- the main point of the present manuscript -- that unitary equivalence is \textit{not} the same as homotopic equivalence! In other words, given two irrep-equivalent EBRs, there is a unitary transformation that acts on the entire representation (all $\mathbf{k}$ points), but the unitary need not be Brillouin-zone periodic or smooth in $\mathbf{k}$.
(Recall from Appendix~\ref{sec:equivalence} that homotopic equivalence requires the existence of a \emph{smooth} BZ-periodic unitary matrix.)

Following the discussion in Sec.~\ref{sec:irrep-equiv}, we will replace the infinite space groups with their finite Born von Karman counterparts, although we omit the subscript $N$ to avoid clutter.

Let $\rho$ be an $n_q$-dimensional representation of the site-symmetry group, $G_\mathbf{q}$, where $\mathbf{q}$ is a Wyckoff position of multiplicity $n$.
As shown in Eq.~(\ref{eq:blocks}) and discussed in the surrounding text, the matrix form of a band representation consists of $(n\cdot n_q) \times (n\cdot n_q)$ blocks, where each block is labelled by a pair $(\mathbf{k}',\mathbf{k})$; $\mathbf{k}'$ is a row index and $\mathbf{k}$ is a column index. 
Each symmetry operation $h=\{R|\mathbf{v}\}\in G$ maps $\mathbf{k} \mapsto \mathbf{k}' = R\mathbf{k}$.
For each set of columns corresponding to $\mathbf{k}$, there is exactly one non-zero block, which we denote $\rho_G^\mathbf{k}(h)$.
Define $\chi^\mathbf{k}_G(h)= {\rm Tr}\rho^\mathbf{k}_G(h)$.

Recall from Eq.~(\ref{eq:deflittlegroup}) that $G_\mathbf{k}$ denotes the little group of $\mathbf{k}$, which is the set of all space group elements that leave $\mathbf{k}$ invariant.
If $h\in G_\mathbf{k}$, then $h$ leaves $\mathbf{k}$ unchanged and the non-zero block corresponds to $\mathbf{k}=R\mathbf{k} = \mathbf{k}'$.
The collection of blocks $\rho_G^{\mathbf{k} = R\mathbf{k}}(h)$, for all $h\in G_\mathbf{k}$, gives a representation of $G_\mathbf{k}$.
We define irrep-equivalence by:
\begin{defn}
Two band representations, $\rho_G$ and $\rho'_G$, of a space group, $G$, are irrep-equivalent if for each $\mathbf{k}$ and each $h\in G_\mathbf{k}$, $\chi^\mathbf{k}_G(h) = (\chi')^\mathbf{k}_G(h)$.
\label{def:irrep-equiv}
\end{defn}

We define the character of the band representation (as opposed to the character at a particular $\mathbf{k}$) by summing over the diagonal elements at all $\mathbf{k}$.
This is why it is important to utilize the finite Born von Karman groups defined in Sec.~\ref{sec:irrep-equiv-chars}:
the character of an infinite-dimensional representation is ill-defined because the sum over diagonal elements may diverge. (In particular, the trace of the identity element will always diverge.)
In a finite space group, the character of a band representation can be defined by:
\begin{equation}
\chi_G(h) \equiv {\rm Tr} \rho_G(h) = \sum_{\mathbf{k} | h\in G_\mathbf{k} } \chi_G^\mathbf{k}(h),
\label{eq:charbandrep}
\end{equation}
where the second equality follows because the trace is a sum over diagonal elements, which are only non-zero when $R\mathbf{k} = \mathbf{k}$, i.e., $h\in G_\mathbf{k}$.

We now present an example to show how Eq.~(\ref{eq:charbandrep}) works.
Let $G$ be the space group generated by inversion and translations.
Consider a finite space group where, for some even $N$, the translations $\mathbf{t}_{1,2,3}^N$, are identified with the identity element, so that $k_{1,2,3}$ is a multiple of $\frac{2\pi}{N}$, for a total of $N^3$ points in the first BZ.
Let $\rho_G$ be a band representation induced from a one-dimensional irrep of one of the maximal Wyckoff positions of $G$.
Now let $h=\{ E| \mathbf{0} \} $ in Eq.~(\ref{eq:charbandrep}): since $\{ E| \mathbf{0} \}$ is in the little group of all $\mathbf{k}$, $\chi_G( h  )= \sum_\mathbf{k} \chi_G^\mathbf{k}( h )  =N^3 $, since $\chi_G^\mathbf{k}( h )=1$.
This sum would diverge in the full space group where $N\rightarrow \infty$.
Thus, the finite group is necessary so that the characters $\chi_G(h)$ are well-defined for all choices of $h$.
As a second example, let $h$ be the inversion symmetry operation in Eq.~(\ref{eq:charbandrep}).
Since inversion is only in the little group of $\mathbf{k}$ when $k_{1,2,3}\in \{ 0,\pi\}$ (the time-reversal-invariant-momenta, or TRIM, points), $\chi_G(h)= \sum_{\mathbf{k}\in \text{TRIM}} \chi_G^\mathbf{k}(h)$, where the sum on the RHS is finite for any choice of $N$, but only contains all TRIM points when $N$ is even.
This second example shows that, as noted in Sec.~\ref{sec:irrep-equiv-chars}, it is necessary to choose $N$ such that the high-symmetry points of the infinite space group and those of the finite group are identical: if $N$ was odd, then the only TRIM point would be $(0,0,0)$ and the character of $\chi_G(h)$ would be incomplete.

We are now ready to prove that two band representations are irrep-equivalent by Def.~\ref{def:irrep-equiv} if and only if they are related by a unitary transformation.
In the ``only if'' direction, if two band representations, $\rho_G, \rho'_G$, with characters $\chi_G, \chi'_G$ are irrep-equivalent, then, by Def.~\ref{def:irrep-equiv}, $\chi_G^\mathbf{k}(h) = (\chi')_G^\mathbf{k}(h) $ for each $\mathbf{k}$ and $ h\in G_\mathbf{k}$.
From Eq.~(\ref{eq:charbandrep}), it follows that $\chi_G(h) = \chi'_G(h)$, for all $h$.
Consequently, $\rho_G$ and $\rho'_G$ 
are related by a unitary transformation (using the fact that two finite-dimensional representations with the same characters are related by a unitary transformation).\cite{Serre}

In the other direction, consider two band representations, $\rho_G^{(1)}$ and $\rho_G^{(2)}$, that are related by a unitary transformation, $U$.
$U$ must be block diagonal in $\mathbf{k}$ because different $\mathbf{k}$ specify different irreps that transform differently under translation, i.e., they acquire a different phase $e^{-i\mathbf{k}\cdot \mathbf{t}}$.
Since $U$ is block diagonal in $\mathbf{k}$, the blocks $\rho_G^{(1),\mathbf{k}}(h)$ and $\rho_G^{(2),\mathbf{k}}(h)$
are related by a unitary transformation.
Consequently, $\chi_G^\mathbf{k}(h) = \left( \chi'\right)_G^\mathbf{k}(h)$.
Then by Def.~\ref{def:irrep-equiv}, $\rho_G^{(1)}$ and $\rho_G^{(2)}$ are irrep-equivalent, which completes the proof.

Notice that applying Theorem~\ref{th:unitary} to the case of \emph{equivalent} band representations provides a derivation of the unitary matrix that transforms the band representations, which we derived previously in Appendix~\ref{sec:equivalence}. 
Consider an equivalence $S(\mathbf{k},\tau,g)$ between two band representations, as defined in Eq.~(\ref{eq:defequivalence}).
Since equivalent band representations are necessarily irrep equivalent, and since for each $\tau$, $S(\mathbf{k},\tau,g)$ is a band representation, Theorem~\ref{th:unitary} implies the existence of a family of unitary matrices $U(\mathbf{k},\tau)$ satisfying
\begin{equation}
S(\mathbf{k},\tau,g) = U(g\mathbf{k},\tau)S(\mathbf{k},0,g)U^\dag(\mathbf{k},\tau),\label{eq:equivfamily}
\end{equation}
where $U(\mathbf{k},0)=\mathbb{I}$, the identity matrix. 
Since $S(\mathbf{k},\tau,g)$ is an equivalence, it must be continuous in $\tau$ and smooth in $\mathbf{k}$.
Eq.~(\ref{eq:equivfamily}) then implies that $U(\mathbf{k},\tau)$ must also be continuous in $\tau$ and smooth in $\mathbf{k}$.
Further, $U(\mathbf{k},\tau)$ is BZ-periodic, since equivalent band representations act on the same Hilbert space (with the same boundary conditions).
Thus, $U(\mathbf{k},1)$ is a smooth, periodic unitary transformation that relates $S(\mathbf{k},0,g)$ and $S(\mathbf{k},1,g)$.
This provides an alternative proof of Eq.~(\ref{eq:equivalence}).


\section{Representation theory of finite groups}
\label{app:reptheory}

In this appendix, we provide some of the fundamentals of the representation theory of finite groups.
These can be found in BMZ,\cite{Bacry1988} or, for a more complete reference, the book by Serre.\cite{Serre}
The results in this appendix are not specific to crystallographic groups, therefore we adopt a more general notation.

\subsection{Notation}
Following BMZ,\cite{Bacry1988} throughout this appendix, we use $\chi^{(\alpha)}_G$ to denote the characters of an irrep $\rho^{(\alpha)}$ of a group $G$.
We define the conjugacy class of $g$ in $G$ by: $[g]_G \equiv \{ (g')^{-1} g g' | g' \in G \}$; $| [g]_G|$ denotes the number of elements in $[g]_G$.

\subsection{Orthogonality of characters}

We first review the orthogonality of characters.
The inner product of two characters $\chi^{(\alpha)}$ and $\chi^{(\beta)}$ of irreps $\rho^{(\alpha)}$ and $\rho^{(\beta)}$ of $G$ is defined by:
\begin{equation}
\langle \chi^{(\alpha)}_G  | \chi^{(\beta)}_G  \rangle = \frac{1}{|G|} \sum_{g\in G} \left(\chi^{(\alpha)} (g)\right)^* \chi^{(\beta)} (g)
\label{eq:inner}
\end{equation}
This inner product gives rise to an orthogonality relation:
\begin{equation}
\langle \chi_G^{(\alpha)} | \chi_G^{(\beta)} \rangle = \delta_{\alpha\beta}  
\label{eq:orthoirreps}
\end{equation}
A second orthogonality relation is given by summing over all the irreps in $G$:
\begin{equation}
\sum_{\alpha} \chi_G^{(\alpha)}(g) \left(  \chi_G^{(\alpha)}(h)  \right)^* = \begin{cases} | [g]_G | & \text{ if  } h \in [g]_G \\ 0 &\text{ else} \end{cases}
\label{eq:orthoelements}
\end{equation}

\subsection{Induced characters}
We now review the theory of induced representations.
Induction provides an algorithm to construct a representation of a group, $G$, from a representation of a subgroup, $H\subset G$, as we will shortly describe.

Let $H$ be a subgroup of $G$, with coset decomposition $G = \cup_\alpha g_\alpha H$.
A representation of $G$ with characters, $\chi_G$, subduces to a representation of $H$, denoted $\chi_G \downarrow H$ or $ {\rm Res}^G_H \chi_G$, with characters:
\begin{equation}
(\chi_G\downarrow H )(h) \equiv {\rm Res}_H^G\chi_G(h) \equiv \chi_G(h)
\label{eq:restriction}
\end{equation}
The adjoint of subduction is induction.
A representation of $H$ with characters $\chi_H$ induces a representation in $G$ whose characters are given by:
\begin{equation}
(\chi_H \uparrow G)(g) \equiv 
{\rm Ind}_H^G\chi_H (g) \equiv \sum_{\alpha} \tilde{\chi}_H(g_\alpha^{-1} g g_\alpha)
\label{eq:charsind}
\end{equation}
where we use a tilde to denote:
\begin{equation}
\tilde{\chi}(g) = \begin{cases} \chi(g) &\text{if }g\in H \\ 0 &\text{else}
\end{cases}
\label{eq:tilde}
\end{equation}

The Frobenius reciprocity relation (which we will not prove) says that:
\begin{equation}
\langle {\rm Ind}^G_H \chi_H^{(\rho)} | \chi_G^{(\alpha)} \rangle = \langle \chi_H^{(\rho)}| {\rm Res}_H^G \chi_G^{(\alpha)} \rangle
\label{eq:frobenius}
\end{equation}
where the Greek superscripts indicate an irrep of the corresponding group and the inner product is given by Eq.~(\ref{eq:inner}).
The inner product on the LHS is with respect to the group $G$ (since ${\rm Ind_H^G}\chi_H^{(\rho)}$ and $\chi_G^{(\alpha)}$ are representations of $G$), while the inner product on the RHS is with respect to $H$ (since $\chi_H^{(\rho)}$ and ${\rm Res}_H^G\chi_G^{(\alpha)}$ are representations of $H$).

\section{Necessary and sufficient conditions for irrep-equivalence}
\label{app:irrep-equiv}

In this Appendix, we derive the necessary and sufficient condition for irrep-equivalence in Eq.~(\ref{eq:irrep-equiv}) using the fundamentals of representation theory reviewed in Appendix~\ref{app:reptheory}.

Let $H$ and $K$ be two subgroups of $G$ and let $\chi_H$ and $\chi_K$ be characters corresponding to representations of $H$ and $K$.
If $\chi_H$ and $\chi_K$ induce the same representation in $G$, then for any representation, $\alpha$, of $G$ the characters $\chi_G^{(\alpha)}$ satisfy:
\begin{equation}
\langle {\rm Ind}^G_H \chi_H |\chi_G^{(\alpha)} \rangle = \langle {\rm Ind}^G_K \chi_K |\chi_G^{(\alpha)} \rangle
\end{equation}
Frobenius reciprocity (Eq.~(\ref{eq:frobenius})) yields:
\begin{equation}
\langle \chi_H |{\rm Res}^G_H \chi_G^{(\alpha)} \rangle = \langle \chi_K | {\rm Res}^G_K \chi_G^{(\alpha)} \rangle
\end{equation}
Applying the definition of the inner product in Eq.~(\ref{eq:inner}),
\begin{equation}
\frac{1}{|H|} \sum_{h\in H} \chi^*_H(h) \chi_G^{(\alpha)}(h) = \frac{1}{|K|} \sum_{k\in K} \chi^*_K(k) \chi_G^{(\alpha)}(k)
\end{equation} 
Given an element $g\in G$,
multiplying the whole equation by $\chi_G^{(\alpha)}(g)$ and summing over $\alpha$ using Eq.~(\ref{eq:orthoelements}) yields
\begin{equation}
\frac{1}{|H|} \sum_{h\in H \cap [g]_G} \chi_H(h) = \frac{1}{|K|} \sum_{k\in K \cap [g]_G} \chi_K(k), \forall \, g\in G
\label{eq:irrep-equiv-diff-H}
\end{equation}
If $g$ is not conjugate to an element of $H$ or an element of $K$, then both sides of Eq.~(\ref{eq:irrep-equiv-diff-H}) are zero. 
Thus, we need only consider Eq.~(\ref{eq:irrep-equiv-diff-H}) when
$g$ is conjugate to an element of $H$ or an element of $K$ (or both).
Further, since Eq.~(\ref{eq:irrep-equiv-diff-H}) only depends on the conjugacy class $[g]_G$, rather than on $g$ directly, if Eq.~(\ref{eq:irrep-equiv-diff-H}) is satisfied for all $g\in H\cup K$, then it is satisfied for all $g$ conjugate to an element of $H$ or $K$.
Thus, Eq.~(\ref{eq:irrep-equiv-diff-H}) provides the following necessary and sufficient condition for $\chi_H$ and $\chi_K$ to yield the same induced character:
\begin{equation}
\frac{1}{|H|} \sum_{h\in H \cap [g]_G} \!\!  \chi_H(h) =\frac{1}{|K|}  \sum_{k\in K \cap [g]_G} \!\! \chi_K(k) \,\, \forall \, g\in H\cup K,
\label{eq:irrep-equiv-diff-H-2}
\end{equation}
which is exactly Eq.~(\ref{eq:irrep-equiv}).

Now consider the special case where $H=K$ and $\chi, \chi'$ are characters of two representations $\rho,\rho'$ of $H$.
Then, from Eq.~(\ref{eq:irrep-equiv-diff-H-2}), $\rho,\rho'$ induce the same representation in $G$ if and only if:
\begin{equation}
\sum_{h\in H \cap [g]_G} \!\!  \chi_H(h) = \!\! \sum_{h\in H \cap [g]_G} \!\! \chi'_H(h) \,\, \forall \, g\in H,
\label{eq:irrep-equiv-same-H}
\end{equation}

We now prove:
\begin{theorem} A necessary condition for two distinct representations, $\rho$ and $\rho'$, of $H$ to induce the same representation in $G$ is for two conjugacy classes of $H$ to merge in $G$.
\label{th:merging}
\end{theorem}
\noindent Recall, as defined in Sec.~VB, two conjugacy classes in $H$ merge in $G$ if there exist, $h,h'\in H$ such that $[h]_H\neq [h']_H$ but $[h]_G = [h']_G$.
We now prove Theorem~\ref{th:merging} by contradiction: suppose no conjugacy classes of $H$ merge in $G$.
Then $H\cap [g]_G \subset [g]_H$ because if $g'\in H$ and $g'\in [g]_G$, since no conjugacy classes of $H$ merge in $G$, it must be that $g'\in [g]_H$.
Since, by definition, $[g]_H \subset H\cap [g]_G$, it follows that $H\cap [g]_G = [g]_H$.
Thus, Eq.~(\ref{eq:irrep-equiv-same-H}) can be rewritten as:
\begin{equation}
\sum_{h\in [g]_H} \!\!  \chi_H(h) = \!\! \sum_{h\in [g]_H} \!\! \chi'_H(h)   \,\, \forall \, g\in H,
\label{eq:irrep-equiv-same-H-merge}
\end{equation}
which implies that $\chi(g) = \chi'(g)$ for all $g\in H$ because the characters are invariant over the conjugacy class.
Hence, $\chi = \chi'$, which completes the proof of Theorem~\ref{th:merging}.


\section{Little group characters at $\Gamma$}
\label{app:chars-gamma}

We now show that in a band representation, $\rho \uparrow G$, the representation of the little group at $\Gamma$ is determined by mapping $\rho$ into the point group, $P$, of $G$.
We start by defining a map from $G$ to $P$:
\begin{equation}
	h = \{ R | \mathbf{v} \} \mapsto \bar{h} \equiv R, 
	\label{eq:overbar}
\end{equation}
which maps each element of $G$ to its point group part. 
Under this map, a site-symmetry group, $G_\mathbf{q}$, maps to
\begin{equation}
P_\mathbf{q} \equiv \{ \bar{h} | h \in G_\mathbf{q} \}
\end{equation}
and each representation, $\rho$, of $G_\mathbf{q}$, defines a representation $\bar{\rho}$ of $P_\mathbf{q}$ by $\bar{\rho}(\bar{h}) = \rho(h)$, whose characters satisfy:
\begin{equation}
\bar{\chi}(\bar{h}) = \chi(h).
\label{eq:defptrep}
\end{equation}
It follows that the little group character at $\Gamma$ is given by:
\begin{align}
\chi^\Gamma_G (h) &\equiv \sum_\alpha \tilde{\chi}(g_\alpha^{-1} \{ E| -\mathbf{t}_{\alpha\alpha}(h) \} h g_\alpha) \nonumber\\
&= \sum_\alpha \tilde{\bar{\chi}}(\bar{g}_\alpha ^{-1}  \bar{h} \bar{g}_\alpha) \nonumber\\
&= (\bar{\chi} \uparrow P)(\bar{h})
\end{align}
where the first line is exactly the definition of the little group character (Eq.~(\ref{eq:chars})) at $\Gamma$, where $\mathbf{k} = \mathbf{0}$;
the second line drops the translation parts of all space group elements following Eq.~(\ref{eq:defptrep});
and the third line follows from the definition of an induced representation (Eq.~(\ref{eq:charsind})).

Since a necessary condition for irrep-equivalence is for two EBRs to have the same little group irreps at $\Gamma$, an immediate consequence of this result is:
\begin{cor}
A necessary condition for two irreducible representations $\rho$ and $\rho'$ of $G_\mathbf{q}$ to induce irrep-equivalent EBRs in $G$ is that the representations $\bar{\rho}$ and $\bar{\rho}'$ of $P_\mathbf{q}$ induce the same representation in $P$.
\label{cor:induceptgroup}
\end{cor}

We now return to the point groups that we ruled out for irrep-equivalence in Secs.~\ref{sec:theorems} and \ref{sec:samesite}.
Theorem~\ref{th:merging} says that a necessary condition for two distinct irreps of $P_\mathbf{q}$ to induce the same representation of $P$ is that two conjugacy classes with respect to $P_\mathbf{q}$ merge in $P$.
Since the 16 point groups in (\ref{eq:ptgrpsnomerge}) do not have different conjugacy classes with elements in the same crystallographic class, the conjugacy classes with respect to these point groups will not merge in $P$; hence, when $P_\mathbf{q}$ is one of the point groups in (\ref{eq:ptgrpsnomerge}), distinct representations of $P_\mathbf{q}$ induce distinct representations of $P$.
It follows from Corollary~\ref{cor:induceptgroup} that when $G_\mathbf{q}$ is isomorphic to one of the point groups in (\ref{eq:ptgrpsnomerge}), distinct irreps of $G_\mathbf{q}$ will yield EBRs with distinct representations at $\Gamma$ and hence will not be irrep-equivalent.

Furthermore, using the tables on the POINT application of the BCS,\cite{Bilbao3} we have checked that when $P_\mathbf{q}$ is one of the four point groups in (\ref{eq:conjmergenoequiv}), distinct \textit{irreps} always induce distinct representations of $P$, although different \textit{reducible} representations can induce the same representation in $P$ (which is why these groups are not listed in (\ref{eq:ptgrpsnomerge}).)
Consequently, if $G_\mathbf{q}$ is equivalent to one of the point groups in (\ref{eq:ptgrpsnomerge}), distinct irreps of $G_\mathbf{q}$ will induce EBRs with distinct representations at $\Gamma$, which, consequently, will not be irrep-equivalent.


\section{Pairs $(G_\mathbf{q}, G_{\mathbf{q}'})$ marked $D'$ in Table~\ref{tab:pointgrouppairs}.}
\label{sec:moredimensionality}

We used Eq.~(\ref{eq:irrep-equiv-E}) to eliminate pairs of $(G_\mathbf{q}, G_{\mathbf{q}'})$ as candidates for irrep-equivalence if they had no irreps of the compatible dimension; such pairs are marked by a D in Table~\ref{tab:pointgrouppairs} (if they are not already marked by an X).
Here we explain why the additional pairs marked by a D' in Table~\ref{tab:pointgrouppairs} can also be eliminated as candidates for irrep-equivalence based on a combination of dimensionality and zero characters.
\begin{description}
\item[$G_\mathbf{q} = T_d, G_{\mathbf{q}'} = D_4$ or $G_\mathbf{q} = T_d, G_{\mathbf{q}'} = C_{4v}$ ] Character tables of $T_d, D_{4}$ and $C_{4v}$ are shown in Tables~\ref{table:tdchars} and \ref{table:d4chars}. Using $|T_d|=24, |D_4| = |C_{4v} | = 8$, the dimensionality constraint in Eq.~(\ref{eq:irrep-equiv-E}) requires ${\rm dim}(\rho)/{\rm dim (\rho') } = 3$.
This constraint is only satisfied if $\rho$ is one of the three-dimensional irreps, $T_1$ or $T_2$, of $T_d$.
Now let $g$ be the $S_4^+$ rotoreflection in $G_\mathbf{q}$.
Since in the $T_{1,2}$ irreps, $\chi(g) \neq 0$, Corollary~\ref{cor:pointgroupsnonpolarnonzero} immediately applies and enforces that no EBRs induced from these groups will be irrep-equivalent.
%
%
\item[$G_\mathbf{q} = D_{2d}, G_{\mathbf{q}'} = O$ or $G_\mathbf{q} = D_{2d}, G_{\mathbf{q}'} = T_d$] The logic is identical to the previous case. The characters of $D_{2d}$, $O$ and $T_d$ are in Tables~\ref{table:tdchars} and \ref{table:d4chars}.
Using $|D_{2d}|=8$ and $|O|=|T_d| = 24$ the dimensionality constraint in Eq.~(\ref{eq:irrep-equiv-E}) requires ${\rm dim}(\rho)/{\rm dim (\rho') } = 1/3$.
This constraint is only satisfied if $\rho$ is one of the one-dimensional irreps, $A_1,A_2,B_1$ or $B_2$, of $D_{2d}$.
Now let $g$ be the $S_4^+$ rotoreflection in $G_\mathbf{q}$.
Since in the one-dimensional irreps of $D_{2d}$, $\chi(g)  \neq 0$, Corollary~\ref{cor:pointgroupsnonpolarnonzero} immediately applies, enforcing that no EBRs induced from these groups will be irrep-equivalent.
\end{description}

\begin{table}
\begin{tabular}{c|ccccc}
$\rho\, \backslash\, T_d$ & $[E]$ & $[C_{3}]$ & $[C_{2}]$ & $[S_{4} ]$ & $[m]$ \\
$\rho\, \backslash\, O$ & $[E]$ & $[C_3]$ & $[C_2]$ & $[C_4]$ & $[C_2']$ \\
\hline 
$A_1$ & $1$ & $1$ & $1$ & $1$ & $1$\\
$A_2$ & $1$ & $1$ & $1$ & $-1$ & $-1$\\
$E$ & $2$ & $-1$ & $2$ & $0$ & $0$\\
$T_1$ & $3$ & $0$ & $-1$ & $1$ & $-1$\\
$T_2$ & $3$ & $0$ & $-1$ & $-1$ & $1$
\end{tabular}
\caption{Character table for $T_d$ (conjugacy classes in first row) and $O$ (conjugacy classes in second row). Notation follows Ref.~\onlinecite{PointGroupTables}.}\label{table:tdchars}
\end{table}

\begin{table}
\begin{tabular}{c|ccccc}
$\rho \, \backslash\, D_4$ & $[E]$ & $[C_2]$ & $[C_4]$ & $[C_2']$ & $[C_2'']$\\
$\rho \, \backslash\, D_{2d}$ & $[E]$ & $[C_2]$ & $[S_4]$ & $[C_2']$ & $[m]$\\
$\rho \, \backslash\, C_{4v}$ & $[E]$ & $[C_2]$ & $[C_4]$ & $[m_v]$ & $[m_d]$\\
\hline
$A_1$ & $1$ & $1$ & $1$ & $1$ & $1$\\
$A_2$ & $1$ & $1$ & $1$ & $-1$ & $-1$\\
$B_1$ & $1$ & $1$ & $-1$ & $1$ & $-1$\\
$B_2$ & $1$ & $1$ & $-1$ & $-1$ & $1$\\
$E$ & $2$ & $-2$ & $0$ & $0$ & $0$\\
\end{tabular}
\caption{Character table for $D_4$ (conjugacy classes in first row), $D_{2d}$ (conjugacy classes in second row), and $C_{4v}$ (conjugacy classes in third row). Notation follows Ref.~\onlinecite{PointGroupTables}}\label{table:d4chars}
\end{table}

\clearpage

%
%
%
%

\setlength{\tabcolsep}{6pt}
\setlength{\extrarowheight}{1pt}

\LTcapwidth=\textwidth
\begin{longtable*}{cccccc}
\multicolumn{6}{c}%
{\tablename\ \thetable\ } \\
SG & $\mathbf{q}$ & $G_\mathbf{q}$ & Irreps\\
\hline
\endfirsthead
\multicolumn{6}{c}%
{\tablename\ \thetable\ -- \textit{Cont.}} \\
SG & $\mathbf{q}$ & $G_\mathbf{q}$ & Irreps\\
\hline
\endhead
\hline \multicolumn{6}{c}{\textit{Continued}} \\
\endfoot
\hline
\endlastfoot
26 & $b$ & $C_s$ & ${}^1\! \bar{E}, {}^2\! \bar{E}$ \\26 & $a$ & $C_s$ & ${}^1\! \bar{E}, {}^2\! \bar{E}$ \\27 & $d$ & $C_2$ & ${}^1\! \bar{E}, {}^2\! \bar{E}$ \\27 & $c$ & $C_2$ & ${}^1\! \bar{E}, {}^2\! \bar{E}$ \\27 & $b$ & $C_2$ & ${}^1\! \bar{E}, {}^2\! \bar{E}$ \\27 & $a$ & $C_2$ & ${}^1\! \bar{E}, {}^2\! \bar{E}$ \\36 & $a$ & $C_s$ & ${}^1\! \bar{E}, {}^2\! \bar{E}$ \\37 & $b$ & $C_2$ & ${}^1\! \bar{E}, {}^2\! \bar{E}$ \\
\end{longtable*}

\LTcapwidth=\textwidth
\begin{longtable*}{ccccccc}
\multicolumn{7}{c}%
{\tablename\ \thetable\ } \\
SG & $\mathbf{q}$ & $G_\mathbf{q}$ & $\rho$ & $\mathbf{q}'$ & $G_{\mathbf{q}'}$ & $\rho'$\\
\hline
\endfirsthead
\multicolumn{7}{c}%
{\tablename\ \thetable\ -- \textit{Continued}} \\
SG & $\mathbf{q}$ & $G_\mathbf{q}$ & $\rho$ & $\mathbf{q}'$ & $G_{\mathbf{q}'}$ & $\rho'$\\
\hline
\endhead
\hline \multicolumn{7}{r}{\textit{Continued}} \\
\endfoot
\hline
\endlastfoot
16 & $d$ & $D_2$ & $\bar{E}$ & $h$ & $D_2$ & $\bar{E}$ \\16 & $c$ & $D_2$ & $\bar{E}$ & $h$ & $D_2$ & $\bar{E}$ \\16 & $b$ & $D_2$ & $\bar{E}$ & $h$ & $D_2$ & $\bar{E}$ \\16 & $a$ & $D_2$ & $\bar{E}$ & $h$ & $D_2$ & $\bar{E}$ \\16 & $f$ & $D_2$ & $\bar{E}$ & $g$ & $D_2$ & $\bar{E}$ \\16 & $e$ & $D_2$ & $\bar{E}$ & $g$ & $D_2$ & $\bar{E}$ \\16 & $b$ & $D_2$ & $\bar{E}$ & $g$ & $D_2$ & $\bar{E}$ \\16 & $a$ & $D_2$ & $\bar{E}$ & $g$ & $D_2$ & $\bar{E}$ \\
\end{longtable*}

\newpage

\LTcapwidth=\textwidth
\begin{longtable*}{ccccccc}
\multicolumn{7}{c}%
{\tablename\ \thetable\ } \\
SG & $\mathbf{q}$ & $G_\mathbf{q}$ & $\rho$ & $\mathbf{q}'$ & $G_{\mathbf{q}'}$ & $\rho'$\\
\hline
\endfirsthead
\multicolumn{7}{c}%
{\tablename\ \thetable\ -- \textit{Continued}} \\
SG & $\mathbf{q}$ & $G_\mathbf{q}$ & $\rho$ & $\mathbf{q}'$ & $G_{\mathbf{q}'}$ & $\rho'$\\
\hline
\endhead
\hline \multicolumn{7}{r}{\textit{Continued}} \\
\endfoot
\hline
\endlastfoot
16 & $g$ & $D_2$ & $\bar{E}$ & $h$ & $D_2$ & $\bar{E}$ \\16 & $f$ & $D_2$ & $\bar{E}$ & $h$ & $D_2$ & $\bar{E}$ \\16 & $e$ & $D_2$ & $\bar{E}$ & $h$ & $D_2$ & $\bar{E}$ \\16 & $d$ & $D_2$ & $\bar{E}$ & $g$ & $D_2$ & $\bar{E}$ \\16 & $c$ & $D_2$ & $\bar{E}$ & $g$ & $D_2$ & $\bar{E}$ \\16 & $d$ & $D_2$ & $\bar{E}$ & $f$ & $D_2$ & $\bar{E}$ \\16 & $b$ & $D_2$ & $\bar{E}$ & $f$ & $D_2$ & $\bar{E}$ \\16 & $c$ & $D_2$ & $\bar{E}$ & $e$ & $D_2$ & $\bar{E}$ \\
\end{longtable*}

\null 
\vspace{1in}

\LTcapwidth=\textwidth
\begin{longtable*}{ccccc}
\multicolumn{5}{c}%
{\tablename\ \thetable\ } \\
SG & $\mathbf{q}$ & $G_\mathbf{q}$ & $\rho$ & Summed EBRs: $(\mathbf{q}_i, G_{\mathbf{q}_i}, \rho_i)$\\
\hline
\endfirsthead
\multicolumn{5}{c}%
{\tablename\ \thetable\ -- \textit{Continued}} \\
SG & $\mathbf{q}$ & $G_\mathbf{q}$ & $\rho$ & Summed EBRs: $(\mathbf{q}_i, G_{\mathbf{q}_i}, \rho_i)$\\
\hline
\endhead
\hline \multicolumn{5}{r}{\textit{Continued}} \\
\endfoot
\hline
\endlastfoot
64 & $e$ & $C_2$ & ${}^1\! \bar{E}$ & $(b,C_{2h},{}^1\! \bar{E}_u),(b,C_{2h},{}^2\! \bar{E}_g)$ \\64 & $e$ & $C_2$ & ${}^1\! \bar{E}$ & $(b,C_{2h},{}^1\! \bar{E}_g),(b,C_{2h},{}^2\! \bar{E}_u)$ \\64 & $e$ & $C_2$ & ${}^1\! \bar{E}$ & $(a,C_{2h},{}^1\! \bar{E}_u),(a,C_{2h},{}^2\! \bar{E}_g)$ \\64 & $e$ & $C_2$ & ${}^1\! \bar{E}$ & $(a,C_{2h},{}^1\! \bar{E}_g),(a,C_{2h},{}^2\! \bar{E}_u)$ \\64 & $e$ & $C_2$ & ${}^2\! \bar{E}$ & $(b,C_{2h},{}^1\! \bar{E}_u),(b,C_{2h},{}^2\! \bar{E}_g)$ \\64 & $e$ & $C_2$ & ${}^2\! \bar{E}$ & $(b,C_{2h},{}^1\! \bar{E}_g),(b,C_{2h},{}^2\! \bar{E}_u)$ \\64 & $e$ & $C_2$ & ${}^2\! \bar{E}$ & $(a,C_{2h},{}^1\! \bar{E}_u),(a,C_{2h},{}^2\! \bar{E}_g)$ \\64 & $e$ & $C_2$ & ${}^2\! \bar{E}$ & $(a,C_{2h},{}^1\! \bar{E}_g),(a,C_{2h},{}^2\! \bar{E}_u)$ \\
\end{longtable*}

\newpage

\LTcapwidth=\textwidth
\begin{longtable*}{cccccc}
\multicolumn{6}{c}%
{\tablename\ \thetable\ } \\
SG & $\mathbf{q}$ & $G_\mathbf{q}$ & Irreps\\
\hline
\endfirsthead
\multicolumn{6}{c}%
{\tablename\ \thetable\ -- \textit{Continued}} \\
SG & $\mathbf{q}$ & $G_\mathbf{q}$ & Irreps\\
\hline
\endhead
\hline \multicolumn{6}{r}{\textit{Continued}} \\
\endfoot
\hline
\endlastfoot
90 & $b$ & $D_2$ & $B_2, B_3$ \\90 & $a$ & $D_2$ & $B_2, B_3$ \\97 & $d$ & $D_2$ & $B_2, B_3$ \\100 & $b$ & $C_{2v}$ & $B_1, B_2$ \\101 & $b$ & $C_{2v}$ & $B_1, B_2$ \\101 & $a$ & $C_{2v}$ & $B_1, B_2$ \\102 & $a$ & $C_{2v}$ & $B_1, B_2$ \\105 & $b$ & $C_{2v}$ & $B_1, B_2$ \\
\end{longtable*}

\newpage

\LTcapwidth=\textwidth
\begin{longtable*}{ccccccc}
\multicolumn{7}{c}%
{\tablename\ \thetable\ } \\
SG & $\mathbf{q}$ & $G_\mathbf{q}$ & $\rho$ & $\mathbf{q}'$ & $G_{\mathbf{q}'}$ & $\rho'$\\
\hline
\endfirsthead
\multicolumn{7}{c}%
{\tablename\ \thetable\ -- \textit{Continued}} \\
SG & $\mathbf{q}$ & $G_\mathbf{q}$ & $\rho$ & $\mathbf{q}'$ & $G_{\mathbf{q}'}$ & $\rho'$\\
\hline
\endhead
\hline \multicolumn{7}{r}{\textit{Continued}} \\
\endfoot
\hline
\endlastfoot
3 & $c$ & $C_2$ & ${}^1\! \bar{E} {}^2\! \bar{E}$ & $d$ & $C_2$ & ${}^1\! \bar{E} {}^2\! \bar{E}$ \\3 & $b$ & $C_2$ & ${}^1\! \bar{E} {}^2\! \bar{E}$ & $d$ & $C_2$ & ${}^1\! \bar{E} {}^2\! \bar{E}$ \\3 & $a$ & $C_2$ & ${}^1\! \bar{E} {}^2\! \bar{E}$ & $d$ & $C_2$ & ${}^1\! \bar{E} {}^2\! \bar{E}$ \\3 & $b$ & $C_2$ & ${}^1\! \bar{E} {}^2\! \bar{E}$ & $c$ & $C_2$ & ${}^1\! \bar{E} {}^2\! \bar{E}$ \\3 & $a$ & $C_2$ & ${}^1\! \bar{E} {}^2\! \bar{E}$ & $c$ & $C_2$ & ${}^1\! \bar{E} {}^2\! \bar{E}$ \\3 & $a$ & $C_2$ & ${}^1\! \bar{E} {}^2\! \bar{E}$ & $b$ & $C_2$ & ${}^1\! \bar{E} {}^2\! \bar{E}$ \\5 & $a$ & $C_2$ & ${}^1\! \bar{E} {}^2\! \bar{E}$ & $b$ & $C_2$ & ${}^1\! \bar{E} {}^2\! \bar{E}$ \\6 & $a$ & $C_s$ & ${}^1\! \bar{E} {}^2\! \bar{E}$ & $b$ & $C_s$ & ${}^1\! \bar{E} {}^2\! \bar{E}$ \\
\end{longtable*}

\null \vspace{1in}

\LTcapwidth=\textwidth
\begin{longtable*}{ccccccc}
\multicolumn{7}{c}%
{\tablename\ \thetable\ } \\
SG & $\mathbf{q}$ & $G_\mathbf{q}$ & $\rho$ & $\mathbf{q}'$ & $G_{\mathbf{q}'}$ & $\rho'$\\
\hline
\endfirsthead
\multicolumn{7}{c}%
{\tablename\ \thetable\ -- \textit{Continued}} \\
SG & $\mathbf{q}$ & $G_\mathbf{q}$ & $\rho$ & $\mathbf{q}'$ & $G_{\mathbf{q}'}$ & $\rho'$\\
\hline
\endhead
\hline \multicolumn{7}{r}{\textit{Continued}} \\
\endfoot
\hline
\endlastfoot
81 & $c$ & $S_4$ & ${}^1\! E {}^2\! E$ & $d$ & $S_4$ & ${}^1\! E {}^2\! E$ \\81 & $a$ & $S_4$ & ${}^1\! E {}^2\! E$ & $b$ & $S_4$ & ${}^1\! E {}^2\! E$ \\82 & $c$ & $S_4$ & ${}^1\! E {}^2\! E$ & $d$ & $S_4$ & ${}^1\! E {}^2\! E$ \\82 & $a$ & $S_4$ & ${}^1\! E {}^2\! E$ & $b$ & $S_4$ & ${}^1\! E {}^2\! E$ \\85 & $a$ & $S_4$ & ${}^1\! E {}^2\! E$ & $b$ & $S_4$ & ${}^1\! E {}^2\! E$ \\86 & $a$ & $S_4$ & ${}^1\! E {}^2\! E$ & $b$ & $S_4$ & ${}^1\! E {}^2\! E$ \\88 & $a$ & $S_4$ & ${}^1\! E {}^2\! E$ & $b$ & $S_4$ & ${}^1\! E {}^2\! E$ \\111 & $b$ & $D_{2d}$ & $E$ & $d$ & $D_{2d}$ & $E$ \\
\end{longtable*}

\newpage

\LTcapwidth=\textwidth
\begin{longtable*}{ccccc}
\multicolumn{5}{c}%
{\tablename\ \thetable\ } \\
SG & $\mathbf{q}$ & $G_\mathbf{q}$ & $\rho$ & Summed EBRs: $(\mathbf{q}_i, G_{\mathbf{q}_i}, \rho_i)$\\
\hline
\endfirsthead
\multicolumn{5}{c}%
{\tablename\ \thetable\ -- \textit{Continued}} \\
SG & $\mathbf{q}$ & $G_\mathbf{q}$ & $\rho$ & Summed EBRs: $(\mathbf{q}_i, G_{\mathbf{q}_i}, \rho_i)$\\
\hline
\endhead
\hline \multicolumn{5}{r}{\textit{Continued}} \\
\endfoot
\hline
\endlastfoot
21 & $k$ & $C_2$ & ${}^1\! \bar{E} {}^2\! \bar{E}$ & $(d,D_2,\bar{E}),(d,D_2,\bar{E})$ \\21 & $k$ & $C_2$ & ${}^1\! \bar{E} {}^2\! \bar{E}$ & $(c,D_2,\bar{E}),(d,D_2,\bar{E})$ \\21 & $k$ & $C_2$ & ${}^1\! \bar{E} {}^2\! \bar{E}$ & $(b,D_2,\bar{E}),(d,D_2,\bar{E})$ \\21 & $k$ & $C_2$ & ${}^1\! \bar{E} {}^2\! \bar{E}$ & $(a,D_2,\bar{E}),(d,D_2,\bar{E})$ \\21 & $k$ & $C_2$ & ${}^1\! \bar{E} {}^2\! \bar{E}$ & $(c,D_2,\bar{E}),(c,D_2,\bar{E})$ \\21 & $k$ & $C_2$ & ${}^1\! \bar{E} {}^2\! \bar{E}$ & $(b,D_2,\bar{E}),(c,D_2,\bar{E})$ \\21 & $k$ & $C_2$ & ${}^1\! \bar{E} {}^2\! \bar{E}$ & $(a,D_2,\bar{E}),(c,D_2,\bar{E})$ \\21 & $k$ & $C_2$ & ${}^1\! \bar{E} {}^2\! \bar{E}$ & $(b,D_2,\bar{E}),(b,D_2,\bar{E})$ \\
\end{longtable*}


\bibliography{EBRWilsonBib}
\end{document}


\title{Supplemental Material: Topology invisible to eigenvalues in obstructed atomic limits}

\author{Jennifer Cano}
\affiliation{Department of Physics and Astronomy, Stony Brook University, Stony Brook, New York 11974, USA}
\affiliation{Center for Computational Quantum Physics, Flatiron Institute, New York, New York 10010, USA}
\author{L. Elcoro}
\affiliation{Department of Condensed Matter Physics, University of the Basque Country UPV/EHU, Apartado 644, 48080 Bilbao, Spain}
\author{M.~I.~Aroyo}
\affiliation{Department of Condensed Matter Physics, University of the Basque Country UPV/EHU, Apartado 644, 48080 Bilbao, Spain}
\author{B. Andrei Bernevig}
\affiliation{Department of Physics, Princeton University, Princeton, New Jersey 08544, USA}
\author{Barry Bradlyn}
\affiliation{Department of Physics and Institute for Condensed Matter Theory, University of Illinois at Urbana-Champaign, Urbana, IL, 61801-3080, USA}

\maketitle

\section{EBRs in $F222$ induced from the $4c$ and $4d$ positions are distinguishable}
\label{sec:F222-cd}

In this section, we prove that the irrep-equivalent EBRs induced from the $4c$ and $4d$ positions in $F222$ can be distinguished by the topological invariant $n$ defined by Wilson loops in \eqBerry of the main text, which we repeat here for convenience:
\begin{equation}
e^{\frac{2\pi i n}{4}} = W_{\mathbf{g}_1}e^{-\frac{i}{4}\int_\Sigma \Omega \cdot d\Sigma},
\label{eq:22-BerrySM}
\end{equation}
where $\Omega = \nabla \times \mathbf{A}$ is the Berry curvature and $\Sigma$ is a region whose boundary is $\ell$, the loop traced by putting the vectors $\mathbf{g}_i$ and $-\mathbf{g}_1 - \mathbf{g}_2 - \mathbf{g}_3 $ end-to-end; see Fig.~2 in the main text.
Let $\mathbf{q}=\left( \frac{1}{4}, \frac{1}{4}, \frac{1}{4} \right)$ and $\mathbf{q}'=\left( \frac{1}{4}, \frac{1}{4}, \frac{3}{4}\right)$ denote sites in the $4c$ and $4d$ positions, respectively.
The BANDREP application\cite{NaturePaperSM} on the BCS shows that if $\rho$ is an irrep of $G_{\mathbf{q}}$ and $\rho'$ is the corresponding irrep of $G_{\mathbf{q}'}$, 
then the induced band representations $\rho\uparrow G$ and $\rho' \uparrow G$ are irrep-equivalent.
However, following the identical logic as when comparing the irrep-equivalent EBRs induced from the $4a$ and $4b$ positions in the main text, the two EBRs induced from the $4c$ and $4d$ positions are not equivalent because the unitary that relates them is not BZ-periodic.

We now show that the two EBRs induced from the $4c$ and $4d$ positions can also be distinguished by Eq.~(\ref{eq:22-BerrySM}).
Specifically, an EBR induced from a representation of $G_{\mathbf{q} = \left( \frac{1}{4}, \frac{1}{4}, \frac{1}{4} \right)}$ can be continuously deformed to have delta-function localized Wannier functions at $\mathbf{q}$ in each unit cell, which yields $\Omega=0$ and $W_{\mathbf{g}_1} = e^{i\mathbf{g}_1 \cdot \mathbf{q}} = i$, which implies that $n=1$ on the LHS of Eq.~(\ref{eq:22-BerrySM}).
On the other hand, an EBR induced from a representation of $G_{\mathbf{q}' = \left( \frac{1}{4}, \frac{1}{4}, \frac{3}{4}\right)}$ can be continuously deformed to have $\Omega=0$ and $W_{\mathbf{q}_1}= e^{i\mathbf{g}_1 \cdot \mathbf{q}'}=-i$, which implies that $n=3$ on the LHS of Eq.~(\ref{eq:22-BerrySM}).
Thus, we have shown that Eq.~(\ref{eq:22-BerrySM}) completely determines the Wyckoff position of a single-valued EBR in $F222$.


\section{Example of evolution of equivalent band representations: Rice-Mele chain}
\label{sec:RiceMele}
As an example of equivalent band representations, we consider the one-dimensional Su-Schrieffer Heeger (Rice-Mele) chain\cite{ssh1979SM,RiceMeleSM} with inversion symmetry, highlighted in Ref.~\onlinecite{NaturePaperSM}. Recall that a one-dimensional chain with inversion symmetry and Bravais lattice vector $\mathbf{t}_1$ has two maximal Wyckoff positions, denoted $1a$ and $1b$, with coordinates $\mathbf{q}_{1a}=0$ and $\mathbf{q}_{1b}=\frac{1}{2}\mathbf{t}_1$, respectively. The site symmetry groups of the two positions are both isomorphic to the point group $C_s$, with
\begin{align}
G_{1a}&=\langle \{I|0\}\rangle, \\
G_{1b}&=\langle \{I|1\}\rangle.
\end{align}
These groups each have two representations, denoted $A$ and $B$, corresponding to $s$ and $p$ orbitals, respectively. In the $A$ representation the single nontrivial generator is represented by $+1$, while in the $B$ representation it is represented by $-1$. We will construct an explicit equivalence between the band representations
\begin{equation}
\rho_a=(A\oplus B)_{1a}\uparrow G
\end{equation}
and
\begin{equation}
\rho_b=(A\oplus B)_{1b}\uparrow G.
\end{equation}
Following the induction procedure of Ref.~\onlinecite{EBRTheorySM}, we can obtain the representation matrices for $g=\{I|0\}$ ($I$ indicates inversion; we use $E$ for the identity element) in the band representations $\rho_{1a}$ and $\rho_{1b}$. We find using \eqblocks in the main text:
\begin{align}
\rho^\mathbf{k}_{a}(\{ I | 0 \} ) &=\left(\begin{array}{cc}
1 & 0 \\
0 & -1
\end{array}\right)\\
\rho^\mathbf{k}_{b}(\{ I | 0 \} ) &=e^{ik}\left(\begin{array}{cc}
1 & 0 \\
0 & -1
\end{array}\right),
\end{align}
where the basis for the matrices $\rho_{1a}(g)$ and $\rho_{1b}(g)$ are given in terms of functions centered at the $1a$ and $1b$ Wyckoff positions, respectively (see~(\ref{eq:vec})). Although these two matrices are different, we will show that this difference amounts to a continuous and periodic $\mathbf{k}$-dependent change of basis (gauge transformation). Insodoing, we will explicitly construct the equivalence transformation $S(\mathbf{k},\tau)$ which interpolates between these band representations as a function of $\tau$. 
This exercise will be useful when we revisit the space group $F222$ in Sec.~\ref{app:f222}.

Let us work in the basis of $\rho^\mathbf{k}_{a}$, where a column vector $|u_\mathbf{k}\rangle=(a_\mathbf{k},b_\mathbf{k})$ corresponds to a wavefunction
\begin{equation}
\psi_\mathbf{k}(r)=\sum_n e^{ikn}\left(a_\mathbf{k}f(r-n\mathbf{t}_1)+b_\mathbf{k}q(r-n\mathbf{t}_1) \right),
\label{eq:vec}
\end{equation}
where $f(r)=f(-r)$ and $q(r)=-q(-r)$ are centered at $r=0$. The function $f(r)$ transforms as an $s$-orbital under inversion, while $q(r)$ transforms as a $p$-orbital. Let us work in the tight-binding limit, where  
\begin{align}
\int dr rf^*(r-n\mathbf{t}_1)f(r-m\mathbf{t}_1) &=n\delta_{nm}\label{eq:basisoverlapsame} \\
\int dr rq^*(r-n\mathbf{t}_1)q(r-m\mathbf{t}_1) &= n\delta_{nm},\\
\int dr f^*(r)q(r-n\mathbf{t}_1)&=0,\label{eq:basisoverlaps}\\
\int dr rf^*(r)q(r-n\mathbf{t}_1)&=0.\label{eq:tblimitend}
\end{align}

\begin{figure*}[t]
\includegraphics[width=0.5\textwidth]{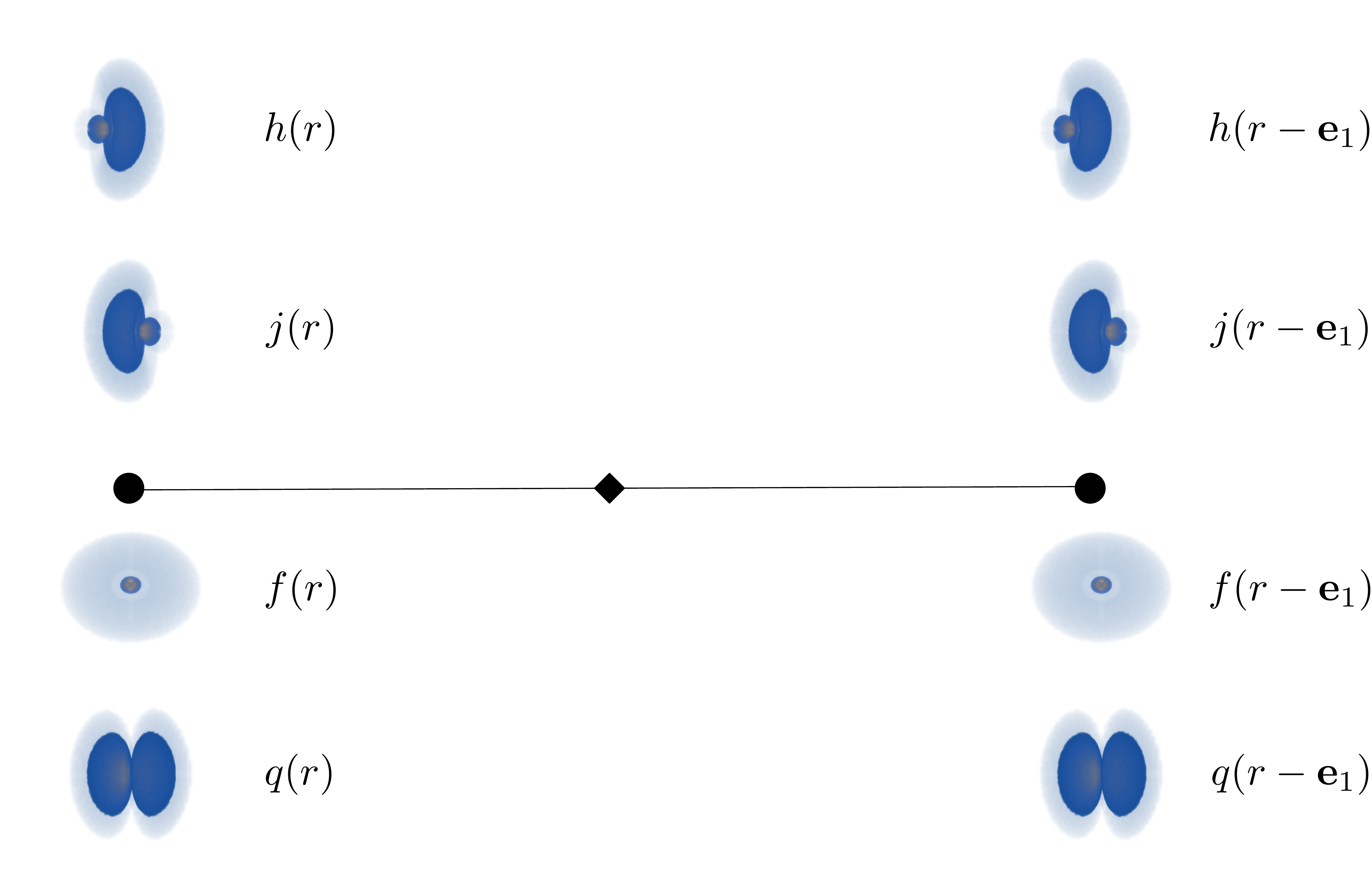}
\caption{Schematic depictions of the basis functions used in defining the equivalence between the $\rho_{a}$ and $\rho_{b}$ band representations. $f(r)$ and $q(r)$ are respectively $s-$orbital and $p-$orbital like functions, centered at the lattice sites of the model (depicted as solid circles). $h(r)$ and $j(r)$ are $sp$-hybrid orbitals given by even-weighted linear combinations of $f(r)$ and $q(r)$. The functions $\tilde{f}(r,\tau)$ and $\tilde{q}(r,\tau)$ are unequal weight superpositions of $sp$ orbitals from neighboring lattice sites, which, as $\tau \rightarrow 1$ have definite parity under inversion about the midpoint of the lattice (indicated by the solid diamond).}\label{fig:bfs}
\end{figure*}

To construct our equivalence, we want to continuously move our basis functions so that they are centered at $r=x$ rather than $r=0$; $0< (x \mod 1) <1/2$ corresponds to the nonmaximal $2c$ Wyckoff position, which has a trivial site symmetry group, $G_{2c}=G_{1a}\cap G_{1b}=\{E|0\}$. Since this site-symmetry group at generic $r$ is trivial, inversion symmetry, $g$, must act by exchanging the basis functions of this band representation (i.e., exchanging the basis function at $r=x$ with its partner at $r=-x$). We can obtain basis functions with this property from $f(r)$ and $q(r)$ in two steps. First, we define a basis of $sp$ hybrid orbitals centered at $x=0$,
\begin{equation}
(h(r),j(r))^T=U(f(r),q(r))^T
\end{equation}
with
\begin{equation}
U=\frac{1}{\sqrt{2}}\left(
\begin{array}{cc}
1 & 1 \\
1 & -1
\end{array}
\right)\label{eq:x0unitary}
\end{equation}
The function $h(r)$ corresponds to an $s+p$ orbital, while $j(r)$ is an $s-p$ orbital. The functions $h(r)$ and $j(r)$ transform into each other under the action of inversion symmetry,
\begin{align}
\{I|0\}h(r)&=j(r)\nonumber \\
\{I|0\}j(r)&=h(r)\label{eq:2c0inv}
\end{align}
Since $h(r)$ and $j(r)$ are centered at $x=0$, we can take them as basis functions for the site-symmetry group $G_{2c}$ as $x\rightarrow 0$. 
(The reason we can find basis functions of the site-symmetry group of $G_{2c}$ as $x\rightarrow 0$ in terms of the basis functions of the site-symmetry group of $G_{1a}$ is because $G_{2c}$ is a subgroup of $G_{1a}$.)
The functions are depicted schematically in Fig.~\ref{fig:bfs}.

Let us now consider the functions $\{h(r), j(r)\}$, which form a basis for the (trivial) group $G_{2c}$ at $x\rightarrow 0$, and $\{h(r-\mathbf{t}_1), j(r-\mathbf{t}_1)\}$, which form a basis for the group $G_{2c}$ at $x\rightarrow 1$. From Eq.~(\ref{eq:2c0inv}), we deduce that both sets form an equivalent representation of $G_{2c}$ (in fact, $G_{2c}$ has only a single representation).  
By taking linear combinations of $\{h(r), j(r)\}$ and $\{h(r-\mathbf{t}_1), j(r-\mathbf{t}_1)\}$, we can thus form a representation of $G_{2c}$ centered at any value $x$. 
In particular, let us consider
\begin{align}
\tilde{f}(r,\tau)&=\cos\frac{\pi \tau }{4} h(r)+\sin\frac{\pi \tau }{4}j(r-\mathbf{t}_1) \\
\tilde{q}(r,\tau )&=\sin\frac{\pi \tau}{4}h(r)-\cos\frac{\pi \tau}{4}j(r-\mathbf{t}_1),\label{eq:tunitary}
\end{align}
for $\tau \in[0,1]$.
The tight binding limit equations Eq.~(\ref{eq:basisoverlapsame}--\ref{eq:tblimitend}) imply that
\begin{align}
\int dr r |h(r-n\mathbf{t}_1)|^2 = \int dr r|j(r-n\mathbf{t}_1)|^2 & = n \\
\int dr r h(r-n\mathbf{t}_1)j(r-m\mathbf{t}_1) &= 0
\end{align}

Using this result, we deduce that the functions $\tilde f(r)$ and $\tilde q(r)$ are centered at
\begin{align}
\langle \tilde{f} |r |\tilde f\rangle &= \int dr\; r |\tilde{f}|^2  = \sin^2 \frac{\pi \tau}{4},\\
\langle \tilde{q} |r | \tilde q \rangle &= \int dr\; r | \tilde{q}|^2 = 1-\sin^2\frac{\pi \tau}{4}
\label{eq:newcenter}
\end{align}
so that as a function of $\tau$, $\tilde{f}(r,\tau)$ and $ \tilde{q}(r,\tau)$ have centers which interpolate between $x=0\mod 1$ at $\tau=0$, and $x=1/2\mod 1$ at $\tau=1$. 
This particular linear combination also makes manifest that $\tilde{f}(r,1)$ and $\tilde{q}(r,1)$ are, respectively, even and odd under the action of $\{I|1\}\in G_{1b}$. 
Note that for all $\tau$, these functions span the representation space of the band representation $\rho_{2c}$, since
the action of $\{I|1\}$ takes a particularly simple form in the basis of $\{\tilde{f},\tilde{q}\}$:
\begin{align}
\{I|1\}\tilde{f}(r,\tau)&=\sin\frac{\pi t}{2}\tilde{f}(r,\tau) - \cos\frac{\pi t}{2} \tilde{q}(r,\tau)\\
\{I|1\}\tilde{q}(r,\tau)&=-\sin\frac{\pi t}{2}\tilde{q}(r,\tau) - \cos\frac{\pi t}{2} \tilde{f}(r,\tau)
\end{align}

Let us now combine these observations to obtain an explicit equivalence between the $\rho_{a}$ and $\rho_{b}$ band representations. Combining the transformations Eq.~(\ref{eq:x0unitary}) and (\ref{eq:tunitary}) and moving to momentum space (where translation by $\mathbf{t}_1$ is represented by $e^{-ik}$), we build the $\tau$- and $k$-dependent transformation,
\begin{equation}
U(k,\tau)=\left(\begin{array}{cc}
\cos\frac{\pi \tau}{4} & e^{-ik}\sin\frac{\pi \tau}{4} \\
\sin\frac{\pi \tau}{4} & -e^{-ik}\cos\frac{\pi \tau}{4}
\end{array}
\right)U,
\end{equation}
where $U$ is defined in Eq.~(\ref{eq:x0unitary}).
This is a continuous family of periodic unitaries, which for all $0<\tau <1$ transforms the basis functions $\{f(r),q(r)\}$ of $\rho_{a}$ into a basis for the band representation $\rho_{2c}$. Furthermore, at $\tau=1$,
\begin{equation}
U(k,1)=\frac{1}{2}\left(
\begin{array}{cc}
1+e^{-ik} & 1-e^{-ik} \\
1-e^{-ik} & 1+e^{-ik}
\end{array}
\right),
\label{eq:RMunitary}
\end{equation}
and 
\begin{equation}
U^\dag(-k,1)\rho_{a}^\mathbf{k}(\{I|0\})U(k,1)=e^{-ik}\sigma_z=\rho_{b}^\mathbf{k}(\{I|0\}),
\end{equation}
showing that $U(k,1)$ transforms $\{f(r),q(r)\}$ into a basis for the band representation $\rho_{b}$. Thus, $U(k,\tau)$ implements the equivalence of the band representations $\rho_{a}$ and $\rho_{b}$ per the definition in \eqdefequivalence in the main text.


\section{Model of a Berry obstructed atomic limit in SG $F222$}\label{app:f222}

In this section, we will explicitly construct a tight-binding model in space group $F222$ which realizes the Berry obstructed atomic limit described in Sec.~III between EBRs induced from the $4a$ and $4b$ positions (defined in \eqbandrepa and \eqdefrho in the main text).
 This construction illustrates a general algorithm for constructing obstructed atomic limits for any equivalence of band representations. To do this, we consider a pair of orbitals at the $4a$ Wyckoff position $\mathbf{q}=(0,0,0)$, and demonstrate a phase transition to a phase with Wannier functions centered at the $4b$ Wyckoff position $\mathbf{q}=(0,0,\half)$. 
The site-symmetry groups are given by:
\begin{equation}
G_\mathbf{q}=\lbrace  \{ E | 000\}, \{C_{2,100}|000\}, \{C_{2,010}|000\}, \{C_{2,001}|000\}\rbrace ,
\end{equation}
and 
\begin{equation}
G_\mathbf{q'}=\lbrace  \{ E | 000\}, \{C_{2,100}|001\}, \{C_{2,010}|001\}, \{C_{2,001}|000\}\rbrace,
\end{equation}
Both $G_\mathbf{q}$ and $G_\mathbf{q'}$ are isomorphic to $D_2$.
As discussed below \eqbandrepb in the main text (and in Ref.~\onlinecite{EBRTheorySM}), the band representations $A_\mathbf{q}\uparrow G$ and $A_\mathbf{q'}\uparrow G$ are irrep-equivalent, as are the band representations $B_{1,\mathbf{q}}\uparrow G$ and $B_{1,\mathbf{q'}}\uparrow G$, 
where $A_\mathbf{q}$ and $B_{1,\mathbf{q}}$ ($A_\mathbf{q'}$ and $B_{1,\mathbf{q}'}$) label the $A$ and $B_1$ irreps of $G_\mathbf{q}$ ($G_{\mathbf{q}'}$).
The characters of the $A$ and $B_1$ representations can be found in the character table of $D_2$ (\tabledchars{} in the main text).

First, we will prove the \emph{equivalence} of band representations
\begin{equation}
\left(A_{\mathbf{q}}\oplus B_{1,\mathbf{q}}\right)\uparrow G \equiv \left(A_{\mathbf{q'}}\oplus B_{1,\mathbf{q'}}\right)\uparrow G.\label{eq:equivf222}
\end{equation}
To do so, note that the intersection group $G_\mathbf{q''}=G_\mathbf{q}\cap G_\mathbf{q'}=\langle\{C_{2,001}|000\}\rangle$ is isomorphic to the point group $C_2$, and is the stabilizer group of $\mathbf{q}''=(0,0,z)$, which is part of the $8g$ Wyckoff position (the $8g$ Wyckoff position has two sites, $\pm \mathbf{q}''$, in the primitive unit cell and eight sites in the conventional unit cell).

Let us consider the representation $\rho\equiv A\uparrow D_2$, induced from the $A$ (trivial) representation of $C_2$ into $D_2$.
Since $C_2$ is an index-$2$ normal subgroup of $D_2$ with only one conjugacy class, we deduce immediately from the character table of $C_2$ (\tablecchars{} in the main text) that 
\begin{equation}
\chi_\rho(C_{2,001})=2,\;\;\chi_\rho(C_{2,100})=\chi_\rho(C_{2,010})=0.
\end{equation}
From the character table of $D_2$ (\tabledchars{} in the main text), it follows that the induced representation $\rho$ of $D_2$ is given by the sum $A\oplus B_1$.
Thus, 
\begin{equation}
A_{\mathbf{q}''} \uparrow G = \left( A_{\mathbf{q}''} \uparrow G_{\mathbf{q}} \right) \uparrow G = \left( A_\mathbf{q} \oplus B_{1,\mathbf{q}} \right) \uparrow G
\label{eq:induceq}
\end{equation}
The same logic follows when induced to $\mathbf{q}'$, i.e., 
\begin{equation}
A_{\mathbf{q}''} \uparrow G = \left(A_{\mathbf{q}''} \uparrow G_{\mathbf{q}'} \right) \uparrow G = \left( A_{\mathbf{q}'} \oplus B_{1,\mathbf{q}'} \right) \uparrow G 
\label{eq:induceqp}
\end{equation}
Equating Eqs.~(\ref{eq:induceq}) and (\ref{eq:induceqp}) proves the equivalence Eq.~\ref{eq:equivf222}.
 (A basis could be constructed following the example in Sec.~\ref{sec:RiceMele}.)

Next, we construct a tight-binding model realizing this equivalence. 
We start with a single $s$ and $p_z$ orbital at the $1a$ position of the unit cell, which transform in the $A_\mathbf{q}\oplus B_{1,\mathbf{q}}$ representation of the site symmetry group $G_\mathbf{q}$. Let us denote the band representation induced from this reducible representation as $\rho_a\equiv\left(A_\mathbf{q}\oplus B_{1,\mathbf{q}}\right)\uparrow G$. Using \eqblocks for the induced representation, we find that the matrix representation is given by
\begin{align}
\rho_a(\{C_{2,100}|000\},\mathbf{k})=\rho_a(\{C_{2,010}|000\},\mathbf{k})&=\sigma_z\\
\rho_a(\{C_{2,001}|000\},\mathbf{k})&=\sigma_0
\end{align}
where the Pauli matrices act in the basis of $(s,p_z)$. A similar analysis of the equivalent band representation $\rho_b\equiv\left(A_\mathbf{q'}\oplus B_{1,\mathbf{q'}}\right)\uparrow G$ induced from an $s$ and $p_z$ orbital at the $1b$ position yields
\begin{align}
\rho_b(\{C_{2,100}|000\},\mathbf{k})=\rho_b(\{C_{2,010}|000\},\mathbf{k})&=e^{-ik_z}\sigma_z\\
\rho_b(\{C_{2,001}|000\},\mathbf{k})&=\sigma_0,
\end{align}
where we have introduced rather suggestively the shorthand $k_z=(\mathbf{t}_2+\mathbf{t}_3-\mathbf{t}_1)\cdot\mathbf{k}$. Noting that for a quasi-1D chain oriented along the $\mathbf{t}_3$ direction, both $C_{2,100}$ and $C_{2,010}$ are indistinguishable from inversion symmetry, we can draw inspiration from the Rice-Mele model described in Sec.~\ref{sec:RiceMele}. As in that case, the representations $\rho_a$ and $\rho_b$ are unitarily equivalent, and related by,
\begin{equation}
\rho_b(g,\mathbf{k})=U^\dag(g\mathbf{k})\rho_a(g,\mathbf{k})U(\mathbf{k})
\end{equation}
where
\begin{equation}
U(\mathbf{k})=\frac{1}{2}\left(
\begin{array}{cc}
1+e^{-ik_z} & 1-e^{-ik_z} \\
1-e^{-ik_z} & 1+e^{-ik_z}
\end{array}
\right)
\end{equation}
is identical to Eq.~(\ref{eq:RMunitary}).
$U$ is the BZ periodic unitary transformation that implements the equivalence $\rho_a\equiv \rho_b$. 
Finally, we define the projector onto the $s$-orbital states at $4a$ (which has one site in the primitive unit cell, and forms the basis for the $A_\mathbf{q}\uparrow G$ band representation),
\begin{equation}
P_0=\frac{1}{2}(1+\sigma_z),
\end{equation}
which acts within the space of orbitals spanning the $A_\mathbf{q}\oplus B_{1,\mathbf{q}}$ site-symmetry representation. Similarly, we form the projector
onto the basis functions of the $A_\mathbf{q'}\uparrow G$ band representation,
\begin{equation}
P_1=U(\mathbf{k})P_0U^\dag(\mathbf{k})=\frac{1}{2}(1+\sigma_z\cos k_z+\sigma_y\sin k_z)
\end{equation} 
acting on the same space. The Hamiltonian
\begin{equation}
H=-tP_1 - \epsilon P_0
\label{eq:f222oal}
\end{equation}
has the property that when $\epsilon>t$ the band representation of the valence band is given by $P_0\rho_aP_0$, while for $t>\epsilon$, the band representation of the valence band is given by $P_1\rho_aP_1=P_0\rho_bP_0$. 
This follows from examining the flat-band limits, $t=0, H=-\epsilon P_0$, and $\epsilon=0,H=-t P_1$, 
and recalling from the main text that $P_0\rho_aP_0$ and $P_0\rho_bP_0$ cannot be related by any periodic unitary transformation, which prevents the band representation matrices from changing without closing the bulk gap. Exploiting the latter statement, we can add any other symmetry allowed terms to Eq.~(\ref{eq:f222oal}) without changing our conclusions here. 

However, unlike in the SSH phase, the transition between the two obstructed atomic limits in this case is generically via an intermediate semimetallic \textit{phase}, since the models are in three dimensions.
In particular, since $F222$ is a noncentrosymmetric group, the intermediate phase between the two obstructed atomic limits will, generically, be a Weyl semimetal. For instance, the fine-tuned critical Hamiltonian
\begin{equation}
H_c=-P_0-P_1
\end{equation}
has a gapless surface $k_z=\pi$. Adding the symmetry-allowed perturbation
\begin{equation}
\delta H = -(\alpha\cos{k_x}+\beta\cos{k_y})\sigma_z- \gamma \sin{k_x}\sin{k_y}\sigma_x,\label{eq:weyl}
\end{equation}
will, as a function of $\alpha,\beta$ and $\gamma$, open gaps in the $k_z=\pi$ plane, leaving behind four gapless Weyl points. The combined Hamiltonian $H_c+\delta H$ has eigenvalues
\begin{widetext}
\begin{equation}
E_{\pm}=-2\pm\sqrt{\sin^2k_z+\gamma^2\sin^2k_x\sin^2k_y+(1+\cos k_z + \alpha\cos k_x + \beta \cos k_y)^2}.
\end{equation}
\end{widetext}
For small values of $|\alpha|< |\beta|$ and non-zero $\gamma$,
this Hamiltonian has four Weyl points at $(k_x, k_y, k_z) = (0, \pm \arccos (-\frac{\alpha}{\beta}), \pi)$ and $(\pi, \pm \arccos(\frac{\alpha}{\beta}), \pi)$. (If $|\alpha| > |\beta|$, then the four Weyl points are at $(\pm \arccos(-\frac{\beta}{\alpha}), 0, \pi)$ and $(\pm \arccos(\frac{\beta}{\alpha}),\pi,\pi)$.)

Finally, we note that for our simple model Eq.~(\ref{eq:f222oal}), we can compute the invariant $n$ in Eq.~(\ref{eq:22-BerrySM}) in both flat band limits; the results hold away from these limits because $n$ is quantized. Since Eq.~(\ref{eq:f222oal}) has no $k_x$ or $k_y$ dependence and is the exactly solvable Su-Schrieffer-Heeger model\cite{ssh1979SM}, we have immediately that (c.f. for instance Ref.~\onlinecite{NaturePaperSM}), $\Omega=0$ and
\begin{align}
W_\mathbf{g_1}=\begin{cases}
1, & |t|<|\epsilon| \\
-1, & |t|>|\epsilon|
\end{cases}
\end{align}
Combining these, we see from Eq.~(\ref{eq:22-BerrySM}) that when $|t|<|\epsilon|$, $n=0 \mod 4$, while when $|t|>|\epsilon|$, $n=2 \mod 4$. Furthermore, since $n$ is an integer, its value will not change when we add terms such as Eq.~(\ref{eq:weyl}), provided no gaps close.

To conclude, we note that the procedure used to construct the obstructed atomic limit here is quite general, and can be applied to any equivalence of band representations. The key idea is to construct a homotopy between two projectors related by the unitary transformation which implements the equivalence of representations. The fact that the band representations are equivalent ensures that the resulting sum of projectors gives a local Hamiltonian, due to the analyticity and periodicity of the unitary transformation.


\section{Two cases missed by BMZ}
\label{sec:different-site-exs}

Here we explicitly verify that the irrep-equivalent EBRs missed by BMZ\cite{Bacry1988SM} in SGs $I23$ and $Pn\bar{3}$ are, in fact, irrep-equivalent.

Recall from \thirrepequiv{} that band representations induced from a representation of $G_\mathbf{q}$ with character $\chi$ and a representation of $G_{\mathbf{q}'}$ with character $\chi'$ 
are irrep-equivalent if and only if \eqirrepequiv{} is satisfied for every $g\in G_\mathbf{q} \cup G_{\mathbf{q}'}$; for convenience, we repeat \eqirrepequiv{} here:
\begin{equation}
\frac{1}{|G_\mathbf{q}|} \sum_{h\in G_\mathbf{q} \cap [g]_G } \chi(h) = 
\frac{1}{|G_{\mathbf{q}'}|} \sum_{h' \in G_{\mathbf{q}'} \cap [g]_G } \chi' (h'),
\label{eq:irrep-equivSM}
\end{equation}
where $[g]_G \equiv \{ (g')^{-1} g g' | g'\in G \}$ denotes the conjugacy class of $g$ in $G$.

\subsection{$I23$}

The group $I23$ (No. 197) is a body-centered cubic, generated by $\{ C_{2,001} | \mathbf{0} \}$, $\{ C_{3,111} | \mathbf{0} \}$ and the lattice translations $\{ E | \bar{\frac{1}{2}}, \frac{1}{2}, \frac{1}{2} \}$, $\{ E | \frac{1}{2}, \bar{\frac{1}{2}}, \frac{1}{2} \}$, and $\{ E | \frac{1}{2}, \frac{1}{2}, \bar{\frac{1}{2}} \}$.
The site-symmetry group of $\mathbf{q} = (0,0,0)$ (part of the $2a$ position) is isomorphic to $T$:
\begin{equation}
G_\mathbf{q} = \langle  \{ C_{3,111} | \mathbf{0} \}, \{ C_{2,001} | \mathbf{0} \} \rangle,
\label{eq:I23-q-gens}
\end{equation}
where the elements enclosed by angled brackets are the generators of the group; the character table for $T$ is in Table~\ref{table:Tchars}.
The site-symmetry group of $\mathbf{q}'=\left(0,\frac{1}{2},\frac{1}{2}\right)$ (part of the $6b$ position) is isomorphic to $D_2$:
\begin{equation}
G_{\mathbf{q}'} = \langle \{ C_{2,100} | 011 \} , \{ C_{2,010} | 001 \} , \{ C_{2,001} | 010 \}  \rangle
\label{eq:I23-qp-gens}
\end{equation}
The character table for $D_2$ is in \tabledchars{}.

\begin{table}[b]
\begin{tabular}{c|cccc}
$\rho$ & $[E]$ & $[C_{2}]$ & $[C_{3}]$ & $[C_{3}^{-1}]$ \\ 
\hline
$A$ & $1$ & $1$ & $1$ & $1$\\
$^1E$ & $1$ & $1$ & $\omega$ & $\omega^2$\\
$^2E$ & $1$ & $1$ & $\omega^2$ & $\omega$\\
$T$ & $3$ & $-1$ & $0$ & $0$
\end{tabular}
\caption{Character table for $T$; $\omega = e^{2\pi i/3}$.}\label{table:Tchars}
\end{table}

To evaluate Eq.~(\ref{eq:irrep-equivSM}), we need to know which conjugacy classes of the site-symmetry groups merge in $G$. 
Since $G_\mathbf{q}$ is isomorphic to the full point group of $G$, no conjugacy classes of $G_\mathbf{q}$ merge in $G$.
The conjugacy classes of $\{ C_{2,010}|001\}$ and $\{ C_{2,001} | 010\}$ in $G_{\mathbf{q}'}$ do merge in $G$, which can be  explicitly checked by conjugating the elements of $G_{\mathbf{q}'}$ by the three-fold rotations of the form $\{ C_{3,111} | (t_x,t_y,t_z)\}$, where $(t_x,t_y,t_z)$ is a lattice vector:
\begin{widetext}
\begin{align}
\{ C_{3,111} | (t_x,t_y,t_z) \}^{-1} \{ C_{2,100} | (0,1,1) \} \{ C_{3,111} | (t_x,t_y,t_z) \} &= \{ C_{2,001} | (-2t_y+1,-2t_z+1,0) \} 
\label{eq:197-xz}\\
\{ C_{3,111} | (t_x,t_y,t_z) \}^{-1} \{ C_{2,010} | (0,0,1) \} \{ C_{3,111} | (t_x,t_y,t_z) \} &= \{ C_{2,100} | (0,-2t_z+1,-2t_x) \} 
\label{eq:197-yx}\\
\{ C_{3,111} | (t_x,t_y,t_z) \}^{-1} \{ C_{2,001} | (0,1,0) \} \{ C_{3,111} | (t_x,t_y,t_z) \} &= \{ C_{2,010} | (1-2t_y,0,-2t_x) \}
\label{eq:197-zy}
\end{align}
\end{widetext}
Eqs.~(\ref{eq:197-xz}) -- (\ref{eq:197-zy}) prove the following statements about conjugacy classes, which we will utilize momentarily:
\begin{itemize}
\item $\{ C_{2,001} | 010 \}$ and $\{ C_{2,010} | 001 \} $ are conjugate in $G$ (using Eq.~(\ref{eq:197-zy}) with $(t_x,t_y,t_z) = ( \frac{1}{2}, \frac{1}{2}, \bar{\frac{1}{2}} )$, which is a lattice vector of the body-centered cubic lattice.)
\item $\{ C_{2,100} | 011\}$ is not conjugate to either $ \{ C_{2,001} | 010 \}$ or $\{ C_{2,010} | 001 \} $ in $G$ (using Eqs.~(\ref{eq:197-xz}) and (\ref{eq:197-yx}))
\item $ \{ C_{2,100} | 011 \} $ is conjugate to $\{C_{2,001} | \mathbf{0} \} \in G_\mathbf{q}$ (using Eq.~(\ref{eq:197-xz}) with $(t_x,t_y,t_z) = ( \bar{\frac{1}{2}}, \frac{1}{2}, \frac{1}{2})$)
\item $\{ C_{2,010} | 001 \} $ and $\{ C_{2,001} | 010 \} $ are not conjugate to elements of $G_\mathbf{q}$ (using Eqs.~(\ref{eq:197-yx}) and (\ref{eq:197-zy}))
\end{itemize}


We can now explicitly verify that the EBRs induced from the $T$ irrep of $G_\mathbf{q}$
and the $B_1$ or $B_2$ irreps of $G_{\mathbf{q}'}$
are irrep-equivalent by checking Eq.~(\ref{eq:irrep-equivSM}) for each element of $G_\mathbf{q}$ and $G_{\mathbf{q}'}$:
\begin{description}
\item[$g=\{ C_{3,111} | \mathbf{0} \} $ or $\{ C_{3,111} | \mathbf{0} \}^{-1} $] There are no elements of $G_{\mathbf{q}'}$ conjugate to $g$. Hence, the RHS of Eq.~(\ref{eq:irrep-equivSM}) is zero. Furthermore, in the $T$ irrep of $G_\mathbf{q}$, the character $\chi(g)=0$; since we have established that no conjugacy classes of $G_\mathbf{q}$ merge in $G$ (since $G_\mathbf{q}$ is isomorphic to the full point group of $G$), the LHS of Eq.~(\ref{eq:irrep-equivSM}) is also zero by \thnomerge.
\item[$g=\{ C_{2,001} | \mathbf{0} \} $] On the left-hand side of Eq.~(\ref{eq:irrep-equivSM}), $\chi(g) = -1$ in the $T$ representation; since there are three elements in $[g]_{G_\mathbf{q}}$ and $|G_{\mathbf{q}}| = 12$, 
\begin{equation}
\frac{1}{|G_\mathbf{q}|} \sum_{h\in G_\mathbf{q} \cap [g]_G } \chi(h) = \frac{1}{12}\cdot 3 \cdot (-1) = -\frac{1}{4}
\end{equation}
We proved (Eqs.~(\ref{eq:197-xz})--(\ref{eq:197-zy})) that $g$ is conjugate to $\{ C_{2,100} | 011 \} \in G_{\mathbf{q}'}$, but not to any other element of $G_{\mathbf{q}'}$.
Thus, only $\{ C_{2,100} | 011 \}$ enters the sum on the RHS of Eq.~(\ref{eq:irrep-equivSM}).
The character of $\{ C_{2,100} | 011 \}$ in either the $B_1$ or $B_2$ irrep of $D_2$ is $-1$ (see \tabledchars{}).
Since $|G_{\mathbf{q}'}| = 4$ and the conjugacy class of $\{ C_{2,100} | 011 \}$ in $G_{\mathbf{q}'}$ has only one element, the RHS is also equal to $-\frac{1}{4}$.
\item[$g=\{ C_{2,100} | 011 \} $] Same as $g=\{ C_{2,001} | \mathbf{0} \}$, since they are conjugate in $G$.
\item[$g=\{ C_{2,010} | 001 \} $] Eqs.~(\ref{eq:197-xz})--(\ref{eq:197-zy}) proved that no elements of $G_\mathbf{q}$ are conjugate to $g$. Hence, the LHS of Eq.~(\ref{eq:irrep-equivSM}) is zero. Eqs.~(\ref{eq:197-xz})--(\ref{eq:197-zy}) also proved that the conjugacy classes of $\{ C_{2,010} | 001 \} $ and $\{ C_{2,001} | 010 \}$ merge in $G$; hence, the RHS of Eq.~(\ref{eq:irrep-equivSM}) is proportional to the sum:
\begin{equation}
\chi(\{ C_{2,010} | 001 \} ) + \chi(\{ C_{2,001} | 010 \}) = 0,
\end{equation}
where the equality follows by using the characters of either the $B_1$ or $B_2$ irrep of $D_2$ (\tabledchars{}.)
\item[$g=\{ C_{2,001} | 010 \} $] Same as $g=\{ C_{2,010} | 001 \} $, since they are conjugate in $G$.
\end{description}

\subsection{$Pn\bar{3}$}

The group $Pn\bar{3}$ (No. 201) is a primitive cubic group generated by $\{ C_{2,001} | \frac{1}{2} \frac{1}{2}0 \}$, $\{ C_{3,111} | \mathbf{0} \}$, inversion, and lattice translations (here we are using origin choice 2\cite{Bilbao2SM}).
The site-symmetry group of $\mathbf{q} = (\frac{1}{4},\frac{1}{4},\frac{1}{4} )$ (part of the $2a$ position) is isomorphic to $T$:
\begin{equation}
G_\mathbf{q} = \langle  \{ C_{3,111} | \mathbf{0} \}, \{ C_{2,001} | \frac{1}{2}\frac{1}{2} 0 \} \rangle,
\label{eq:Pn3-q-gens}
\end{equation}
where the elements enclosed by angled brackets are the generators of the group; the character table for $T$ is in Table~\ref{table:Tchars}.
The site-symmetry group of $\mathbf{q}'=\left( \frac{1}{4}, \frac{3}{4}, \frac{3}{4} \right)$ (part of the $6d$ position) is isomorphic to $D_2$:
\begin{equation}
G_{\mathbf{q}'} = \langle \{ C_{2,100} | 0\frac{3}{2}\frac{3}{2} \} , \{ C_{2,010} | \frac{1}{2}0\frac{3}{2} \} , \{ C_{2,001} | \frac{1}{2}\frac{3}{2}0 \}  \rangle
\label{eq:Pn3-qp-gens}
\end{equation}
The character table for $D_2$ is in \tabledchars{}.

As in the previous section,
to evaluate Eq.~(\ref{eq:irrep-equivSM}), we determine which conjugacy classes of the site-symmetry groups merge in $G$.
No conjugacy classes of $G_\mathbf{q}$ merge in $G$ because the only point group element in $G$ that is not in $G_\mathbf{q}$ is inversion symmetry, which does not merge any conjugacy classes.
No conjugacy classes of $G_{\mathbf{q}'}$ merge in $G$, either, which can be explicitly checked by conjugating the elements of $G_{\mathbf{q}'}$ by the three-fold rotations and six-fold rotoreflections, analogous to Eqs.~(\ref{eq:197-xz})--(\ref{eq:197-zy}) (conjugating by inversion does not change the conjugacy class):
\begin{widetext}
\begin{align}
\{ C_{3,111} | (t_x,t_y,t_z) \}^{-1} \{ C_{2,100} | (0,\frac{3}{2},\frac{3}{2}) \} \{ C_{3,111} | (t_x,t_y,t_z) \} &= \{ C_{2,001} | (-2t_y+\frac{3}{2}, -2t_z + \frac{3}{2},0) \} 
\label{eq:201-xz}\\
\{ C_{3,111} | (t_x,t_y,t_z) \}^{-1} \{ C_{2,010} | (\frac{1}{2},0,\frac{3}{2}) \} \{ C_{3,111} | (t_x,t_y,t_z) \} &= \{ C_{2,100} | (0,-2t_z+\frac{3}{2},-2t_x+\frac{1}{2}) \} 
\label{eq:201-yx}\\
\{ C_{3,111} | (t_x,t_y,t_z) \}^{-1} \{ C_{2,001} | (\frac{1}{2},\frac{3}{2},0) \} \{ C_{3,111} | (t_x,t_y,t_z) \} &= \{ C_{2,010} | (\frac{3}{2}-2t_y,0, \frac{1}{2}-2t_x) \}
\label{eq:201-zy} \\
\{ S_{6,111}^+ | (t_x,t_y,t_z) \}^{-1} \{ C_{2,100} | (0,\frac{3}{2},\frac{3}{2}) \} \{ S_{6,111}^+ | (t_x,t_y,t_z) \} &= \{ C_{2,001} | (2t_y-\frac{3}{2}, 2t_z -\frac{3}{2},0) \} 
\label{eq:201-xz-i}\\
\{ S_{6,111}^+ | (t_x,t_y,t_z) \}^{-1} \{ C_{2,010} | (\frac{1}{2},0,\frac{3}{2}) \} \{ S_{6,111}^+ | (t_x,t_y,t_z) \} &= \{ C_{2,100} | (0,2t_z-\frac{3}{2},2t_x-\frac{1}{2}) \} 
\label{eq:201-yx-i}\\
\{ S_{6,111}^+ | (t_x,t_y,t_z) \}^{-1} \{ C_{2,001} | (\frac{1}{2},\frac{3}{2},0) \} \{ S_{6,111}^+ | (t_x,t_y,t_z) \} &= \{ C_{2,010} | (-\frac{3}{2}+2t_y,0, -\frac{1}{2}+2t_x) \}
\label{eq:201-zy-i}
\end{align}
\end{widetext}
(Recall, following Sch\"onflies notation, that $S_n^\pm$ denotes an $n$-fold rotoreflection.\cite{PointGroupTablesSM})
Eqs.~(\ref{eq:201-xz}) -- (\ref{eq:201-zy-i}) yield the following information:
\begin{itemize}
\item $\{ C_{2,010} | \frac{1}{2}0\frac{3}{2} \}$ and $\{ C_{2,001} | \frac{1}{2}\frac{3}{2}0 \}$ are conjugate in $G$ (use Eq.~(\ref{eq:201-zy-i}) with $(t_x,t_y,t_z) = (1,1,0)$)
\item The conjugacy class of $\{ C_{2,100} | 0\frac{3}{2}\frac{3}{2} \}$ in $G_{\mathbf{q}'}$ does not merge with any other conjugacy class of $G_{\mathbf{q}'}$ in $G$
\item $\{ C_{2,001} | \frac{1}{2}\frac{1}{2}0\} \in G_\mathbf{q}$ is conjugate to $\{C_{2,100} | 0\frac{3}{2}\frac{3}{2} \} \in G_{\mathbf{q}'}$ (use Eq.~(\ref{eq:201-xz-i}) with $(t_x,t_y,t_z) = (0,1,1)$)
\end{itemize}

We now explicitly verify that the EBRs induced from the $T$ irrep of $G_\mathbf{q}$ and the $B_{1}$ or $B_{2}$ irrep of $G_{\mathbf{q}'}$ are irrep-equivalent by checking Eq.~(\ref{eq:irrep-equivSM}) for each element of $G_\mathbf{q} \cup G_{\mathbf{q}'}$:

\begin{description}
\item[$g=\{ C_{3,111} | \mathbf{0} \} $ or $\{ C_{3,111} | \mathbf{0} \}^{-1}$] Exactly as in the previous section, there are no elements of $G_{\mathbf{q}'}$ conjugate to $g$. Hence, the RHS of Eq.~(\ref{eq:irrep-equivSM}) is zero. Furthermore, in the $T$ irrep of $G_\mathbf{q}$, the character $\chi(g)=0$; since we have established that no conjugacy classes of $G_\mathbf{q}$ merge in $G$, the LHS of Eq.~(\ref{eq:irrep-equivSM}) is also zero.
\item[$g=\{ C_{2,001} | \frac{1}{2}\frac{1}{2} 0 \} $] We proved (Eq.~(\ref{eq:201-xz-i})) that $\{ C_{2,001} | \frac{1}{2}\frac{1}{2}0 \}\in G_\mathbf{q}$ is conjugate to $\{ C_{2,100} | 0\frac{3}{2} \frac{3}{2} \} \in G_{\mathbf{q}'}$, but it is not conjugate to any other element of $G_{\mathbf{q}'}$. On the LHS of Eq.~(\ref{eq:irrep-equivSM}), $\chi(g) = -1$ in the $T$ representation. Since there are three elements in $[g]_{G_\mathbf{q}}$ and $|G_{\mathbf{q}}| = 12$, 
\begin{equation}
\frac{1}{|G_\mathbf{q}|} \sum_{h\in G_\mathbf{q} \cap [g]_G } \chi(h) = \frac{1}{12}\cdot 3 \cdot (-1) = -\frac{1}{4}
\end{equation}
On the RHS of Eq.~(\ref{eq:irrep-equivSM}), the character of $\{ C_{2,100} | 011 \}$ in either the $B_1$ or $B_2$ irrep of $D_2$ is $-1$ (see \tabledchars{}.), $|G_{\mathbf{q}'}| = 4$ and the conjugacy class of $\{ C_{2,100} | 011 \}$ in $G_{\mathbf{q}'}$ has only one element; hence the right-hand-size is also equal to $-\frac{1}{4}$.
\item[$g=\{ C_{2,100} | 0\frac{3}{2}\frac{3}{2} \} $] Same as $g=\{ C_{2,001} |\frac{1}{2}\frac{1}{2} 0 \}$, since they are conjugate.
\item[$g=\{ C_{2,010} | \frac{1}{2} 0 \frac{3}{2} \} $] We proved using Eqs.~(\ref{eq:201-xz})--(\ref{eq:201-zy-i}) that no elements of $G_\mathbf{q}$ are conjugate to $g$. Hence the LHS of Eq.~(\ref{eq:irrep-equivSM}) is zero. We also proved using Eq.~(\ref{eq:201-zy-i}) that the conjugacy classes of $g=\{ C_{2,010} | \frac{1}{2}0\frac{3}{2} \} $ and $\{ C_{2,001} | \frac{1}{2}\frac{3}{2}0 \}$ merge in $G$; hence the RHS of Eq.~(\ref{eq:irrep-equivSM}) is proportional to the sum:
\begin{equation}
\chi(\{ C_{2,010} | \frac{1}{2}0\frac{3}{2} \} ) + \chi(\{ C_{2,001} | \frac{1}{2}\frac{3}{2}0 \}) = 0,
\end{equation}
where the equality follows by using the characters of either the $B_1$ or $B_2$ irrep of $D_2$ (\tabledchars{}.)
\item[$g=\{ C_{2,001} | \frac{1}{2}\frac{3}{2}0 \} $] Same as $g=\{ C_{2,010} | \frac{1}{2}0\frac{3}{2} \} $, since they are conjugate.
\end{description}

\bibliography{SMBib}